\documentclass[twocolumn]{aastex631}

\usepackage{amsmath}       
\usepackage{mathtools}
\usepackage{mathrsfs}
\usepackage{xspace}
\usepackage[nointegrals]{wasysym}
\usepackage{enumitem}
\usepackage{chemformula}
\usepackage{rotating}
\usepackage{tablefootnote}
\usepackage{appendix}
\usepackage{float}
\usepackage{textcomp,gensymb}
\usepackage{silence}
\usepackage{float} 
\usepackage[normalem]{ulem}
\usepackage{multirow}

\let\tablewidth\undefined
\usepackage{tabularray}

\def\vplanet{\texttt{\footnotesize{VPLanet}}\xspace}
\def\clima{\texttt{\footnotesize{Clima}}\xspace}
\def\magmoc{\texttt{\footnotesize{MagmOc}}\xspace}

\def\picaso{\texttt{\footnotesize{PICASO}}\xspace}

\def\radheat{\texttt{\footnotesize{RadHeat}}\xspace}

\def\thermint{\texttt{\footnotesize{ThermInt}}\xspace}

\def\SciPy{\texttt{\footnotesize{SciPy}}\xspace}

\begin{document}

\title{A whole-planet model of the Earth without life for terrestrial exoplanet studies}

\correspondingauthor{Samantha Gilbert-Janizek}
\email{samroseg@uw.edu}

\author[0009-0004-8402-9608]{Samantha Gilbert-Janizek}
\affiliation{Department of Astronomy and Astrobiology Program, University of Washington, Box 351580, Seattle, Washington 98195}

\author[0000-0001-6487-5445]{Rory K. Barnes}
\affiliation{Department of Astronomy and Astrobiology Program, University of Washington, Box 351580, Seattle, Washington 98195}

\author[0000-0001-6241-3925]{Peter E. Driscoll}
\affiliation{Earth and Planets Laboratory, Carnegie Institution for Science, Washington, DC 20015, USA}

\author[0000-0002-0413-3308]{Nicholas F. Wogan}
\affiliation{SETI Institute, Mountain View, CA 94043}
\affiliation{NASA Ames Research Center, Moffett Field, CA 94035}

\author[0000-0002-8119-3355]{Avi M. Mandell}
\affiliation{NASA Goddard Space Flight Center, 8800 Greenbelt Road, Greenbelt, MD 20771, USA}
\affiliation{Sellers Exoplanet Environment Collaboration, 8800 Greenbelt Road, Greenbelt, MD 20771, USA}

\author[0000-0002-7961-6881]{Jessica L. Birky}
\affiliation{Astronomy Department, University of Washington, Seattle, WA 98195, USA}

\author[0000-0001-9355-3752]{Ludmila Carone}
\affiliation{Space Research Institute, Austrian Academy of Sciences, Schmiedlstrasse 6, A-8042 Graz, Austria}

\author{Rodolfo Garcia}
\affiliation{Department of Astronomy and Astrobiology Program, University of Washington, Box 351580, Seattle, Washington 98195}

\makeatletter
\let\internallinenumbers\relax
\let\endinternallinenumbers\relax
\makeatother

\begin{abstract}
    As the only known habitable (and inhabited) planet in the universe, Earth informs our search for life elsewhere. Future telescopes like the Habitable Worlds Observatory (HWO) will soon look for life on rocky worlds around Sun-like stars, so it is critical that we understand how to distinguish habitable planets from inhabited planets. However, it remains unknown if life is necessary to maintain a habitable planet, or how all of the components of an evolving planet impact habitability over time.  {As a first step toward answering these questions,} we present a coupled interior-atmosphere evolution model of the Earth without life from 50 Myr to 5 Gyr that reproduces 19 key observations of the pre-industrial Earth after 4.5 Gyr  within estimated measurement uncertainties. We also produce a reflected light spectrum covering the possible wavelength range of HWO. Our findings {suggest} that life {may} not {be} required to maintain {long-term} habitable surface conditions. The model presented here is apt for predicting the long-term habitability of Earth-like exoplanets {by coupling the interior and surface evolution}. By generating realistic reflected light spectra from evolved atmospheric states, this model represents significant progress towards {characterizing the observability of} whole-planet {evolution}, which may ultimately provide a robust abiotic baseline for interpreting biosignature observations with HWO.
\end{abstract}

\section{Introduction}
\label{sec:intro}

Earth is currently the only known inhabited planet in the universe, and geological evidence suggests that it was habitable with liquid water oceans for most of its 4.54 Gyr history \citep{cameron2024evidence}. Owing to its long-term habitability, Earth serves as the template that guides our search for biospheres elsewhere. Consequently, a validated model of the entire planet is valuable for both understanding our own planet and optimizing the search for life on worlds outside the Solar System. {Building on previous studies}, here we present a coupled one-dimensional core-mantle-crust-climate model of the Earth that reproduces 19 key observations of the pre-industrial Earth within estimated  measurement uncertainties. 

An open question regarding Earth's evolution is whether or not it could be habitable {over Gyr time-scales} without life's influence on the environment. The idea that life plays an active role in maintaining conditions favorable to its survival is popularly known as the ``Gaia hypothesis'' \citep{lovelock1974atmospheric}. {Clearly biology plays an important role in Earth's surface environment, which motivates our central} question: is life \textit{required} to maintain {the} {long-term} habitab{ility} {of a rocky} planet? 

As a first step towards answering this question, we simulate the Earth through time without life with the goal of reproducing the habitable conditions of the pre-industrial Earth after 4.5 Gyr. {We define habitability as the conditions required to maintain liquid water on the planetary surface \citep[e.g.,][]{dole1964planets, robinson2018characterizing}.} Ultimately, whether or not life is a requirement for long-term habitability has vast implications for the search for life on exoplanets; if life is not required for habitability, then the detection of habitable conditions alone is insufficient evidence for life. Instead, distinguishing inhabited from habitable and uninhabited worlds will require identifying atmospheric gases or surface features that can only be produced by life.

Developing this capability is critical as next-generation telescopes like the Habitable Worlds Observatory (HWO) prepare to search for life on nearby exoplanets. HWO is a space-based, infrared/optical/ultraviolet direct imaging telescope of 6 meters or larger that will discover and characterize Earth-like planets around Sun-like stars \citep{national2021pathways, habworldstargets, feinberg2026habitableworldsobservatorysconcept}. The proper interpretation of photometry and spectroscopy of the ``pale dots'' detected in the habitable zone (HZ) of nearby stars with HWO is predicated on a complete and self-consistent model of Earth-like planets, defined as rocky planets with approximately Earth's mass and radius \citep{wolf2017constraints} and active interior-atmosphere volatile exchange \citep{rodler2014feasibility, fan2019earth}. This broad definition likely includes planets with diverse compositions \citep{shim2014earth, santerne2018earth}, atmospheric structures \citep{vladilo2013habitable}, stellar hosts \citep{habworldstargets}, instellations \citep{kopparapu2013habitable}, and even habitable planets that never developed life \citep{cockell2014habitable, wordsworth2014abiotic, lopez2019detecting, krissansen2021oxygen}. The atmospheres of ``abiotic Earths'' must be well-quantified to serve as null hypotheses for identifying life's signatures and ruling out its imitators, or ``false positives'', on exoplanets observed with HWO \citep{meadows2018exoplanet, krissansen2021oxygen, constantinou2025comparative}.

Many studies have simulated the co-evolution of the Earth and Sun based on geological evidence \citep{barley2005late, kharecha2005coupled, zerkle2012bistable, charnay2013exploring, arney2016pale, krissansen2018constraining, catling2020archean, wogan2022rapid} and extrapolated these results to predict the conditions of Earth-like planets around other stars \citep{segura2005biosignatures, kaltenegger2007spectral, segura2010effect, frank2014radiogenic, godolt20153d, godolt2016assessing, arney2017pale, gebauer2017evolution,  wolf2017constraints, rugheimer2018spectra, arney2019k, young2023inferring}. The existing work largely models the atmospheric evolution of these systems by coupling climate to photochemistry to achieve atmospheric compositions that are in radiative-convective equilibrium. However, atmospheric evolution cannot be fully understood in isolation from interior processes that regulate volatile cycling through the crust and mantle. 

The emerging concept of ``whole-planet'' modeling was first introduced in a pair of papers \citep{driscoll2013divergent,foley2016whole} that argued long-term habitability depends on evolving interior-atmosphere volatile exchange. \citet{foley2016whole} argue that climate influences whether or not plate tectonics develops, as low surface temperatures promote long-lived weak sheer zones, or ``damage'', in the lithosphere. In turn, plate tectonics facilitates volatile recycling at plate boundaries, which is particularly important for stabilizing the climate via the carbon cycle \citep{walker1981negative, berner1983carbonate}. {The carbon cycle}, therefore, can also affect the cooling rate of the interior {by modulating mantle viscosity and melting, which influence tectonics and heat loss from the mantle}, ultimately influencing the core cooling rate and thus magnetic field generation \citep{foley2016whole}. {Because core convection and magnetic field strength are directly tied to the core's thermal and compositional evolution, the present-day magnetic field provides an independent observational constraint on a planet's internal dynamics, offering a valuable calibration point for whole-planet models.}
Finally, the strength of the planetary magnetic field {has been hypothesized to} affect the rate of atmospheric escape, {potentially playing a role} in the desiccation of Venus and thus its distinct evolutionary path \citep{watson1981dynamics, driscoll2013divergent, gillmann2022long}{;} {the direct causal link between field strength and escape rates, however, remains an active area of research \citep[e.g.][]{gunell2018intrinsic}.} 

Despite its potential importance, many questions about whole planet modeling remain.  \citet{krissansen2021oxygen} developed a rigorously coupled mantle-surface-atmosphere model that includes a magma ocean phase and a transition from a mobile lid to plate tectonics, to explore the long-term carbon and oxygen cycles.  However, they did not focus on the coupled evolution of the mantle and core, instead assuming that the core cooling decays exponentially.  Drawing on the ideas of \citet{nimmo2002does}, \citet{armann2012simulating} argue that the surface and core are interconnected, demonstrating that an episodic stagnant lid can suppress core cooling. 

Building on this work, \citet{garcia2026venus} produced a coupled one-dimensional solar-atmosphere-lithosphere-mantle-core model of Venus that reproduces present-day atmospheric \ch{H2O} and \ch{CO2} abundances and the lack of a core dynamo. Their ability to fully validate the model is limited by the dearth of observations of Venus's surface and interior. Furthermore, as in \citet{driscoll2013divergent}, \citet{garcia2026venus} assume a gray atmospheric model. To calculate the planet's energy balance, \citet{krissansen2021oxygen} use a pre-calculated radiative transfer grid to solve for the wavelength-dependent outgoing long-wave radiation (OLR) from the planet, and use a simple parameterization of albedo, planet-star distance, and stellar luminosity to solve for the absorbed short-wave radiation (ASR). Thus, no whole-planet evolutionary model yet includes radiative transfer calculations to solve for the ASR and OLR balance of the changing planetary climate {as well as the magnetic field evolution of the core.  By adding these features to our whole-planet model we can incorporate additional constraints on its evolution and potential habitability.}

The presence of these interconnected feedback mechanisms make clear that long-term planetary evolution cannot be fully understood by studying individual subsystems in isolation. The stellar environment, atmosphere, surface, mantle, and core form a coupled system where changes in one component may cascade to the whole system over geological timescales. For example, the evolution of atmospheric composition -- the primary observable for missions like HWO -- is mediated not only by photochemistry and climate, but also by outgassing set by mantle convection and water content, which in turn depend on the thermal state of the core and mantle. To correctly interpret observations of Earth-like exoplanets and establish a robust null hypothesis for biosignature detection, we require validated whole-planet models that self-consistently track the co-evolution of all major planetary reservoirs. The pre-industrial Earth, with its relatively well-constrained properties across all subsystems, provides one possible validation target for such models. 

Here we present a one-dimensional, core-mantle-crust-climate model of the abiotic Earth that tracks the evolution of the core-mantle system, geochemistry, atmosphere, incoming solar flux, and solar effective temperature over the age of the Solar System. We use our model to broadly reproduce observed quantities of the pre-industrial Earth. We choose this calibration point as the modern Earth is not in radiative balance due to human-driven climate change. Due to large uncertainties in early Earth conditions, we do not attempt to validate the model on the conditions of the early Earth. Finally, as a proof of concept, we demonstrate that atmospheric states from our model can be used to generate realistic reflected light spectra comparable to observations HWO will obtain of Earth-like exoplanets.

In the following section, we discuss the known properties of the pre-industrial Earth (PIE) to establish our model's validation targets. Then, we describe the whole planet abiotic Earth model in Section \ref{sec:methods}, and present its validated results in Section \ref{sec:results}. We discuss future model upgrades as well as potential applications of our validated model for HWO observations of Earth-like exoplanets in Section \ref{sec:discussion}. Finally, we summarize our results and conclusions in Section \ref{sec:conclusions}.

\section{Properties of the Pre-Industrial Earth}
\label{sec:props}

 The ``pre-industrial'' era defines the period of Earth's history prior to the industrial revolution (c. 1800), when widespread changes to human land-use and atmospheric pollution first arose \citep{hawkins2017estimating}. From records, geological evidence, and modeling reconstructions, the PIE is known to have had lower atmospheric and ocean CO$_2$ concentrations and cooler global average surface temperatures than today. Measurements of these properties with their uncertainties and corresponding sources are summarized in Table \ref{tab:earth-meas}. We report interior properties based on present-day measurements because (1) pre-industrial measurements of these properties are largely nonexistent, and (2) mantle and core properties evolve on a sufficiently slow timescale that any variation from their present-day values should be insignificant. Similarly, we assume that the total mass of the ocean has not varied significantly in the last 300 years.

\vspace{-25pt}

\begin{deluxetable*}{lrc}
\setlength{\tabcolsep}{15pt}
\tablecaption{Selected properties of the pre-industrial Earth.}
\tablehead{
\colhead{Property} & 
\colhead{Value} & 
\colhead{Reference(s)}
}
\label{tab:earth-meas}
\startdata
\cutinhead{Atmosphere}
Atmospheric \ch{CO2} ($p$\ch{CO2}) & 28.4 $\pm$ 0.4 Pa & (1) \\
Atmospheric \ch{H2O} ($p$\ch{H2O}) & $900 \pm 700$ Pa & (2, 3) \\
Global avg. surface temperature ($T_{\mathrm{surf}}$) & 286.9 $\pm$ 0.1 K & (4, 5) \\
\cutinhead{Surface}
Ground albedo ($A_g$) & $0.14 \pm 0.02$ & (6--8) \\
Ocean mass ($w_{\textrm{ocean}}$) & 1.4 $\times 10^{21}$ kg (1 T.O.) & (9) \\
Oceanic (CO$_2$)$_{\mathrm{aq}}$ & $8 \pm 2$ $\mu$mol/kg & (10) \\
Surface Ocean Total Dissolved Inorganic Carbon (DIC) & $2000 \pm 200$ $\mu$mol/kg & (10, 11) \\
Ocean pH & $8.18 \pm 0.05$ & (12) \\
\cutinhead{Interior}
Upper mantle temperature ($T_{\mathrm{UM}}$) & $1587_{-34}^{+161}$ K & (13) \\
Core-mantle boundary temp. ($T_{\mathrm{CMB}}$) & 4000 $\pm$ 200 K & (14) \\
Upper mantle heat flow ($Q_{\mathrm{UM}}$) & 38 $\pm$ 3 TW & (13) \\
Core-mantle boundary heat flow ($Q_{\mathrm{CMB}}$) & 11 $\pm$ 6 TW & (13) \\
{Upper TBL viscosity} ($\nu_{\mathrm{UM}}$)\tablenotemark{a} & $10^{17.38 \pm 1}$ m$^2$ s$^{-1}$ ($10^{20.9 \pm 1}$ Pa$\cdot$s) & (15--18) \\
{Lower TBL viscosity} ($\nu_{\mathrm{LM}}$)\tablenotemark{a} & $10^{17.28 \pm 1}$ m$^2$ s$^{-1}$ ($10^{21.0 \pm 1}$ Pa$\cdot$s) & (15--18) \\
Upper mantle melt fraction ($f_{\mathrm{UM}}$) & $11.5 \pm 3.5$\% & (19) \\
Mantle melt mass flux & $(1.3 \pm 0.8) \times 10^6$ kg/s & (20, 21) \\
Inner core radius ($R_{\mathrm{IC}}$) & 1221 $\pm$ 1 km & (22) \\
Magnetic moment & 80 $\pm$ 4 ZAm$^2$ (1.00 $\pm$ 0.05 E.U.) & (23, 24) \\
Magnetopause radius & 9.10 $\pm$ 0.14 R$_{\oplus}$ (1.00 $\pm$ 0.02 E.U.) & (25) \\
\enddata
\tablenotetext{a}{{Viscosities at the inner edges of the two thermal boundary layers (TBLs) (${\sim}100$ km depth and ${\sim}300$ km above the CMB present-day). Nominal values are the median of central values from references (15--18); the $\pm 1$ dex uncertainty reflects inter-study disagreement. Kinematic values assume $\rho = 3300$ and $5200$ kg m$^{-3}$ at the respective depths.}}
\tablenotetext{}{References: (1) \citet{etheridge1996natural}, (2) \citet{robinson2011earth}, (3) \citet{lustig2023earth}, (4) \citet{hawkins2016connecting}, (5) \citet{hawkins2017estimating}, (6) \citet{solomon2007ipcc}, (7) \citet{kargel2014global}, (8) \citet{genda2016origin}, (9) \citet{charette2010volume}, (10) \citet{feely2001uptake}, (11) \citet{eide2017global}, (12) \citet{jiang2019surface}, (13) \citet{jaupart2015treatise}, (14) \citet{hirose2013composition}, (15) \citet{paulson2005modelling}, {(16) \citet{mitrovica2004new}, (17) \citet{lau2016inferences}, (18) \citet{steinberger2006models},} (19) \citet{gale2013mean}, (20) \citet{cogne2004temporal}, (21) \citet{li2015seismic}, (22) \citet{dziewonski1981preliminary}, (23) \citet{de2010planetary}, (24) \citet{kivelson2014planetary}, (25) {calculated by} \citet{driscoll2013divergent} {using equations from \citet{griessmeier2004effect}}}
\end{deluxetable*}

\subsection{Atmosphere}

Earth retains a 1-bar atmosphere consisting of 78\% N$_2$ \citep{united1976us}. Oxygen, which comprises 21\% of our atmosphere, became a major constituent after the proliferation of photosynthetic life \citep{kump2008rise, farquhar2011geological, lyons2014rise}. The remaining atmosphere consists of ``trace components'', including \ch{CO2} and \ch{H2O}, which are effective greenhouse gases that help regulate planetary climate \citep{tyndall1861xxiii}. Water vapor is the most efficient greenhouse gas \citep{fleming1998historical}, and because warmer air can hold more water vapor, the atmospheric abundance of \ch{H2O} relates to global temperature in a positive feedback loop \citep{held2000water, patel2023increase}. {However, water vapor's short atmospheric residence time of $\sim$10 days \citep{trenberth1998atmospheric} -- set by condensation and precipitation -- means that \ch{H2O} cannot itself act as a long-term climate forcing. Instead, the non-condensing greenhouse gases such as \ch{CO2}, with atmospheric lifetimes of centuries to millennia, set the temperature structure that determines the equilibrium water vapor abundance \citep{lacis2010atmospheric}.} Below we detail measurements of Earth's atmospheric \ch{CO2} and \ch{H2O} abundances, as well as global average surface temperature. 

A record of atmospheric \ch{CO2} levels has been pieced together by analyzing air trapped in polar ice sheets. \citet{etheridge1996natural} analyzed air bubbles in Antarctic ice cores dating back to 1000 C.E., finding that pre-industrial atmospheric \ch{CO2} abundances were around 280 ppm with annual fluctuations of $\sim$4 ppm. In Earth's 1-bar atmosphere (101325 Pa), this abundance is equivalent to a partial pressure of 28.4 $\pm$ 0.4 Pa. 

The abundance of \ch{H2O} in our atmosphere varies with latitude, with higher abundances in the humid tropics \citep{mockler1995water, allan2022global}, as well as altitude, with much higher abundances below the atmospheric cold trap \citep{brewer1949evidence, randel2019diagnosing}. The abundance and distribution of water vapor is also seasonally variable, but the resulting change to the global atmosphere is small, on the order of $\pm1 \times 10^{15}$ kg \ch{H2O}, or less than 0.02\% of the total atmospheric mass \citep{trenberth1987global}. Satellite and sounding measurements have provided anchor points for 1-D models to map the vertical distribution of water vapor throughout Earth's atmosphere; for example, \citet{robinson2011earth} analyzed archived satellite data corresponding to the dates of the EPOXI Earth-flyby mission in 2008, extracting water vapor abundance profiles at latitudes of 0$\degree$, 19$\degree$, 41$\degree$, and 66$\degree$ \citep[][their Fig. 4]{lustig2023earth}. Each latitude profile represents the globally averaged vertical profile for all HEALPix bins with the corresponding north and south latitudes. The mean water vapor profile includes a near-surface water vapor volume mixing ratio of 0.0089, with near-surface abundances of 0.0158 and 0.00182 for the mean profiles acquired at 0$\degree$ and 66$\degree$, respectively \citep{lustig2023earth}. Treating the equatorial and mid-latitude averages as upper and lower bounds, these measurements correspond to a present-day partial pressure of $900\pm 700$ Pa, a large uncertainty that reflects the strong latitude-dependency of near-surface water vapor abundances. 

Earth's global average surface temperature can be traced through time directly via historical records, and indirectly by estimating evolving radiative forcings, using global climate model simulations, and finally by compiling proxy evidence from tree-rings, corals, and ice cores. \citet{hawkins2017estimating} combine all of these approaches to estimate that pre-industrial temperatures were 0.55--0.80${\degree}$C cooler than the global average surface temperature from 1986 to 2005. Establishing the corresponding 1986-2005 baseline as 14.4$\degree$C \citep{hawkins2016connecting}, we therefore assume the pre-industrial global average surface temperature to be 286.9 $\pm$ 0.1 K \citep{hawkins2017estimating}.

\subsection{Surface Properties}

To first order, Earth's surface includes exposed land, oceans, and polar ice caps. These three surface types, along with atmospheric water vapor clouds, contribute to the overall brightness or albedo of the planet, regulating planetary absorption of stellar radiation. In addition to \ch{H2O}, the oceans contain dissolved inorganic carbon (DIC) in the form of aqueous \ch{CO2}, carbonate (\ch{CO3^2-}), and bicarbonate (\ch{HCO3^-}), which together mediate the overall pH of the marine environment. 

Earth sustains a surface liquid water ocean of mass $1.4 \times 10^{21}$ kg, or by definition, 1 terrestrial ocean (T.O.) \citep{charette2010volume}. Earth's vast oceans are a major carbon sequestration reservoir. Atmospheric \ch{CO2} dissolves into the oceans as aqueous \ch{CO2} and reacts with water to form carbonic acid, which dissociates into \ch{H^+} and \ch{HCO3^-}, which itself dissociates into \ch{H^+} and \ch{CO3^2-} \citep{pilson2012introduction}. 

Studies of ocean carbon isotopic fractionation reveal that the pre-industrial oceanic carbon budget is largely abiotic. Since photosynthetic respiration preferentially uses the \ch{^{12}C} isotope \citep{o1981carbon}, ocean water enriched in heavier \ch{^{13}C} indicates abiotic dissolved inorganic carbon \citep{quay2003changes}. Isotopic fractionation studies suggest non-human biological activity accounts for only $\sim$2\% of the total DIC in the oceans \citep{carroll2022attribution}. Though marine life pumps dissolved carbon from the near-surface water column to the deep ocean \citep{le2019pathways}, gravitational sinking dominates carbon export in modern oceans \citep{boyd2019multi, nowicki2022quantifying, siegel2023quantifying}. Recent work using models to reconstruct pre-industrial ocean chemistry supports near-surface, pre-industrial DIC concentrations of $2000 \pm 200$ $\mu$mol/kg \citep{feely2001uptake, eide2017global}, an aqueous CO$_2$ concentration of $8 \pm 2$ $\mu$mol/kg \citep{feely2001uptake}, and a global average surface ocean pH of $8.18 \pm 0.05$ \citep{jiang2019surface}.

Planetary surface albedo, a linear combination of albedo contributions from different surface types and clouds, affects climate by regulating the Earth's absorbed short-wave radiation from the Sun. Based on data from spectral libraries \citep{aster_spectral_library_1999, clark_usgs_spectral_2003}, previous planetary modeling work assumes that the albedos of seawater, rock, and ice are largely wavelength independent \citep{kaltenegger2007spectral, driscoll2013divergent}, where ocean albedo $A_{oc}$ $= 0.1$, rock albedo $A_{r}$ $ = 0.17$, and ice albedo $A_{i}$ $= 0.6$. The ocean covers 71\% of Earth's surface area \citep{genda2016origin}, and the remainder comprises land with seasonal variations in ice coverage. 

Satellite measurements have been used to quantify a perennial surface ice area of $16.01$ -- $16.04 \times 10^6$ km$^2$ \citep{solomon2007ipcc, kargel2014global}, or a perennial ice surface fraction of $3.4$\% that implies a remaining ice-free land surface fraction of 25.6\%. On Earth, most of the ice-less land is covered by vegetation \citep{chu2019fractional}, with 54\% grasslands ($A\sim0.4$) and 30\% forests ($A\sim0.1$). Since we are investigating an abiotic planet without life of any kind, we will assume that the surface of the abiotic Earth would instead be dominated by rock. Calculating the total expected ground albedo ($A_g$) as a linear combination of surface fraction and albedo contributions, we thus calculate $A_g$ = $0.14$.  Seasonal changes in Earth's albedo imply variations of $\pm 0.02$ \citep{bartman1980time}.

In reality, Earth's \textit{total} albedo also receives major contributions from atmospheric water vapor clouds, as satellite measurements suggest Earth's total albedo is closer to $A_T = 0.29$ \citep{wielicki1996clouds, loeb2009toward}. This total albedo is $\sim$0.15 brighter than the composite ground albedo because clouds are highly reflective, and they partially block the path of stellar radiation to the ground \citep{ramanathan1989cloud, harrison1990seasonal}. For an albedo representing an average of thick and thin clouds, \citet{de2010planetary} and \citet{driscoll2013divergent} use $A_{cl} = 0.89$.  

\subsection{Interior}

Beneath a thin crustal layer, Earth has a convecting solid mantle, molten outer core, and solid iron inner core. Here we describe how various measurements and models have been used to establish values (with uncertainties) for key core and mantle properties of the Earth, including temperature, heat flow, viscosity, upper mantle melt fraction, and magnetic moment. {These properties are not independent: mantle viscosity governs convective vigor and plate speed, which in turn set the rates of volatile outgassing and regassing that control atmospheric \ch{CO2} and \ch{H2O} content. As discussed in Section \ref{sec:intro}, the magnetic moment and magnetopause radius provide an additional, independent observational constraint on the planet's evolving internal heat budget, since core convection and magnetic field generation are directly tied to the thermal and compositional evolution of the core-mantle system. These interior properties control the dynamical processes that ultimately shape the planet's evolving surface habitability.}

The mantle geotherm describes how the laterally-averaged temperature of the mantle changes with depth. The temperature of the mantle at various depths may be estimated using petrological constraints on the temperature of mid-ocean ridge basalts \citep{kinzler1992primary} and the identification of several important phase changes: the 440 km deep olivine-wadsleyite phase change \citep{katsura2004olivine}, the 660 km deep post-spinel phase transition, \citep{katsura2003post}, and the inner core boundary \citep{alfe2002composition, labrosse2003thermal, jaupart2015treatise}. Using the depth of these phase changes as anchor points, previous studies use isentropic profiles combined with thermal boundary layers to describe a continuous mantle geotherm \citep[e.g.,][]{katsura2004olivine, driscoll2014thermal}. Synthesizing these measurements and models has resulted in the following literature estimates for the temperature of the upper mantle (at the base of the lithosphere) and the core-mantle boundary: $T_{\mathrm{UM}} = 1587_{-34}^{+161}$ K \citep{jaupart2015treatise} and $T_{\mathrm{CMB}} = 4000 \pm 200$ K \citep{hirose2013composition}, respectively.

Over geological time the mantle undergoes solid state convection, transporting heat from the CMB to the surface. The viscosity of the mantle, an important quantity for determining the vigor of mantle convection and heat transfer, is expected to vary significantly between the upper and lower mantle due to variations in temperature, pressure, and composition  \citep{jeanloz1986temperature, boehler1996melting, jaupart2015treatise}. {Mantle viscosity is not merely a descriptive feature of the present-day Earth: it is a material transport property that plays a central role in controlling the long-term evolution of the interior.  Water content and heat reduce mantle viscosity, enhancing convective vigor and plate speed, which accelerates the rates of volatile regassing and degassing that set atmospheric composition over time \citep{seales2020deep, garcia2026venus}. We describe how viscosity enters our convection and plate-speed parameterization in Section \ref{sec:methods}.}

{The radial viscosity structure of the mantle has been inferred from glacial isostatic adjustment (GIA) observations, joint inversions of GIA and mantle convection data, and geodynamically optimized profiles \citep{paulson2005modelling, mitrovica2004new, steinberger2006models, lau2016inferences}. Although the inferred viscosity profiles differ in places, the depths corresponding to the two thermal boundary-layer edges yields per-study central values of [$10^{20.5}$--$10^{21.25}$] Pa$\cdot$s at ${\sim}100$~km depth and [$10^{20}$--$10^{22}$] Pa$\cdot$s at ${\sim}300$~km above the CMB. We adopt the median of the per-study central values as our nominal targets, $\eta_{\mathrm{UM}} = 10^{20.9}$ Pa$\cdot$s and $\eta_{\mathrm{LM}} = 10^{21.0}$ Pa$\cdot$s, with an adopted uncertainty of $\pm 1$ order of magnitude. This envelope reflects the disagreement found among independent inversions arising from differing ice histories, tomographic models, regularization, and treatment of deep radial structure. Indeed, the inter-study variability often exceeds the formal posterior width of any single study. Some \textit{bulk} lower-mantle viscosity estimates ($\gtrsim 10^{22}$ Pa$\cdot$s) exceed our adopted estimate, likely because they do not account for temperature in the lower mantle thermal boundary layer approaching its melting temperature, which will significantly decrease the local viscosity.  In fact,   models that explicitly resolve the structure in the lower-most mantle do show viscosity decreasing around the lower mantle thermal boundary layer \citep{mitrovica2004new, steinberger2006models}. Assuming densities of $\rho_{\mathrm{UM}} = 3300$ kg/m$^3$ and $\rho_{\mathrm{LM}} = 5200$ kg/m$^3$ at the respective boundary-layer depths, these correspond to nominal kinematic viscosities of $\nu_{\mathrm{UM}} \approx 2.4 \times 10^{17}$ m$^2$/s and $\nu_{\mathrm{LM}} \approx 1.9 \times 10^{17}$ m$^2$/s.}

Although surface measurements constrain the upper mantle heat flow to $Q_{\mathrm{UM}} = 38 \pm 3$ TW \citep{jaupart2015treatise}, 
 the core-mantle boundary (CMB) heat flow can not be measured directly. However, the temperature jump across the lower mantle thermal boundary layer can be constrained by extrapolating the mantle geotherm down and the inner core boundary (ICB) temperature up to the CMB. This {approach}, combined with the estimated viscosity of the lower mantle, typically gives an estimate of $Q_{\mathrm{CMB}}$ to be $11 \pm 6$ TW \citep{jaupart2015treatise}. Thermal evolution models of the Earth, therefore, {should} be within these margins of uncertainty to be considered accurate.  

The subduction of Earth's tectonic plates drives the upwelling of ambient mantle at mid-ocean ridges, facilitating the cycling of volatiles between the solid interior and atmosphere over geological timescales. The fraction of upwelling mantle that melts and its rate of flow determine the melt mass flux, which also plays a role in heat transport and can have a direct impact on planetary thermal evolution. The extent of melting of mid-ocean ridge basalts (MORB) has been estimated by compiling various measurements of basalt chemistry as a function of axial depth, and averaging the data over large distances to account for small-scale variability in basalt composition \citep{klein1987global}. This analysis suggests that the average extent of MORB melting is 8 -- 15\% \citep{gale2013mean}.

Measurements of crustal production rates at plate boundaries probe mantle melting by providing estimates on the upward mass flux of mantle melt. \citet{cogne2004temporal} use direct measurements of currently visible seafloor surfaces bounded by pairs of geophysical isochrones in the Atlantic, Indian, Antarctic, Pacific, Nazca, and Cocos ocean basins to estimate a seafloor production rate of 5 -- 15 km$^3$/year. \citet{li2015seismic} use three-dimensional seismic tomography to measure magmatic accretion at the Southwest Indian Ridge, measuring a crustal production rate of $\sim$20 km$^3$/year, providing an upper bound on mantle melt mass flux. Assuming an upper mantle density of $\sim${3300} kg/m$^3$, a crustal production rate range of 5 -- 20 km$^3$/year corresponds to a mantle melt mass flux of $(1.3 \pm 0.8) \times 10^6$ kg/s (values below zero not permitted). The amount of heat transported by the melt has been estimated to be $\sim$1.5 TW \citep{nakagawa2012influence}.

Earth's magnetic field is generated by convection in its liquid iron outer core.  Core convection is primarily generated compositionally through the solidification of the solid iron inner core, where light elements are rejected into the liquid, thereby decreasing the local density of the fluid and driving fluid flow. The radius of the inner core has been measured to high precision by seismographs, yielding $R_\textrm{IC} =1221 \pm 1$ km \citep{dziewonski1981preliminary}. 
 
 Key measurements of the terrestrial magnetic field include the dipolar magnetic moment, which quantifies the strength of the core field, and magnetopause radius, which is the boundary between the terrestrial magnetic field and that of the solar wind. Measurements of Earth's magnetic field suggest a terrestrial magnetic moment of 80 ZAm$^2$ \citep{de2010planetary, kivelson2014planetary}, which we define as 1 Earth unit. Similarly, {using the standoff distance equation from \citet{griessmeier2004effect},}  \citet{driscoll2013divergent} {calculate a magnetopause radius of}  $\sim$9.1 Earth radii \citep{driscoll2013divergent}, which we also define as 1 Earth unit. {This calculated value is consistent with previous measurements from geosynchronous satellites which suggested a plausible magnetopause radius range between 9--10 $R_{\oplus}$ \citep{shue1998magnetopause}}. While the {values} {for both the magnetic field moment and magnetopause radius} lack formal uncertainty estimates, we adopt conservative validation tolerances of $\pm0.015$ Earth units for the magnetopause radius and $\pm0.05$ Earth units for magnetic moment. These $\leq5$\% tolerances effectively account for known variability in these quantities: the location of the magnetopause varies with solar wind conditions, while the magnetic dipole moment exhibits secular variation.

\section{Methods} \label{sec:methods}

\begin{figure*}[htb!]
    \centering
\includegraphics[width=0.75\linewidth, 
    trim=13cm 5cm 21cm 5cm,clip]
    {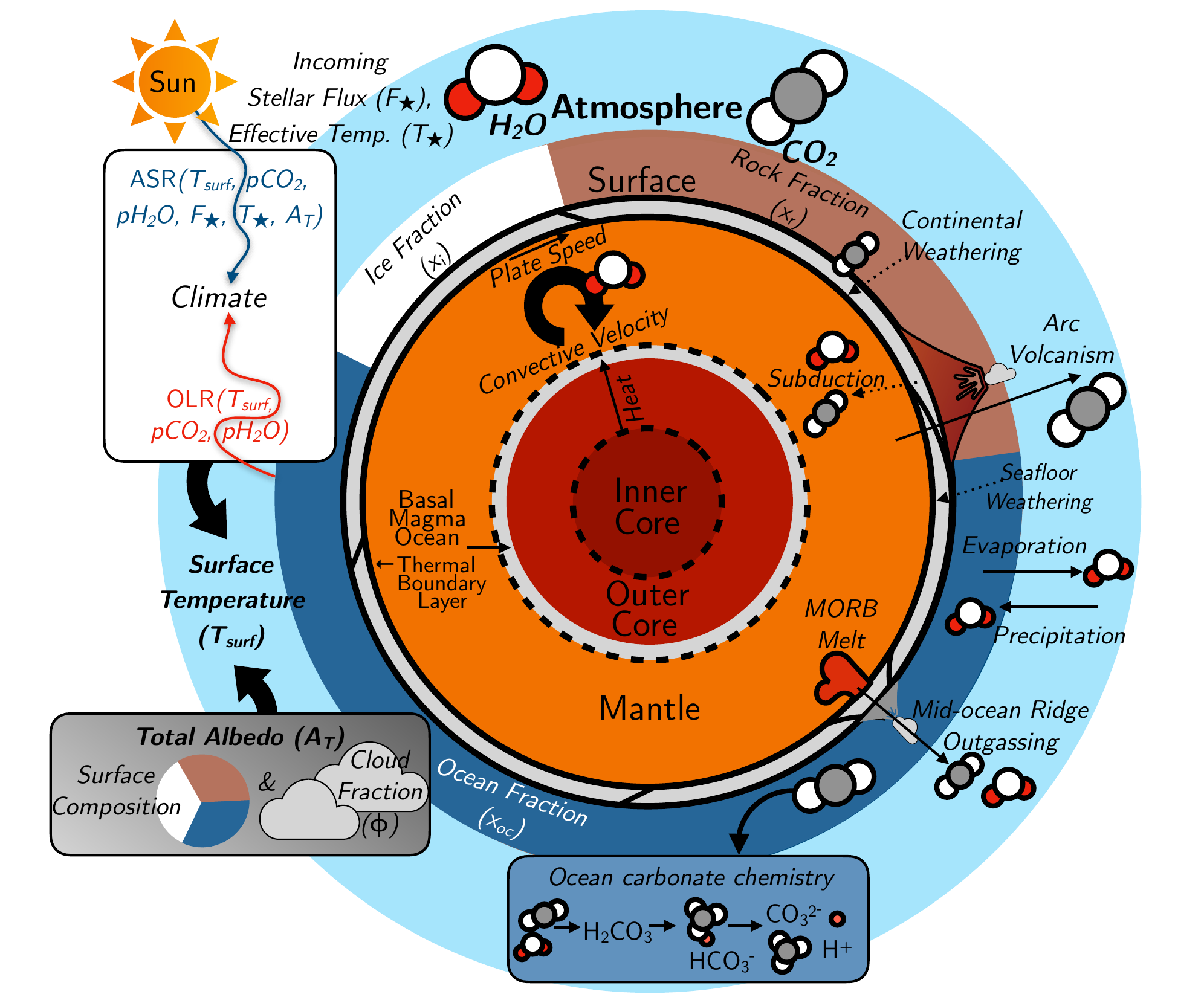}   
    \caption{A schematic of the key processes included in our abiotic Earth model. Remotely observable quantities (surface temperature, albedo, and atmospheric H$_2$O and CO$_2$) are in bold. Volatile cycling is diagrammed in more detail in Figure \ref{fig:schematic}. \label{fig:model-diagram}}
\end{figure*}

\begin{figure*}
    \centering
    \includegraphics[width=\linewidth, 
    trim=9cm 4cm 2cm 0cm,clip]{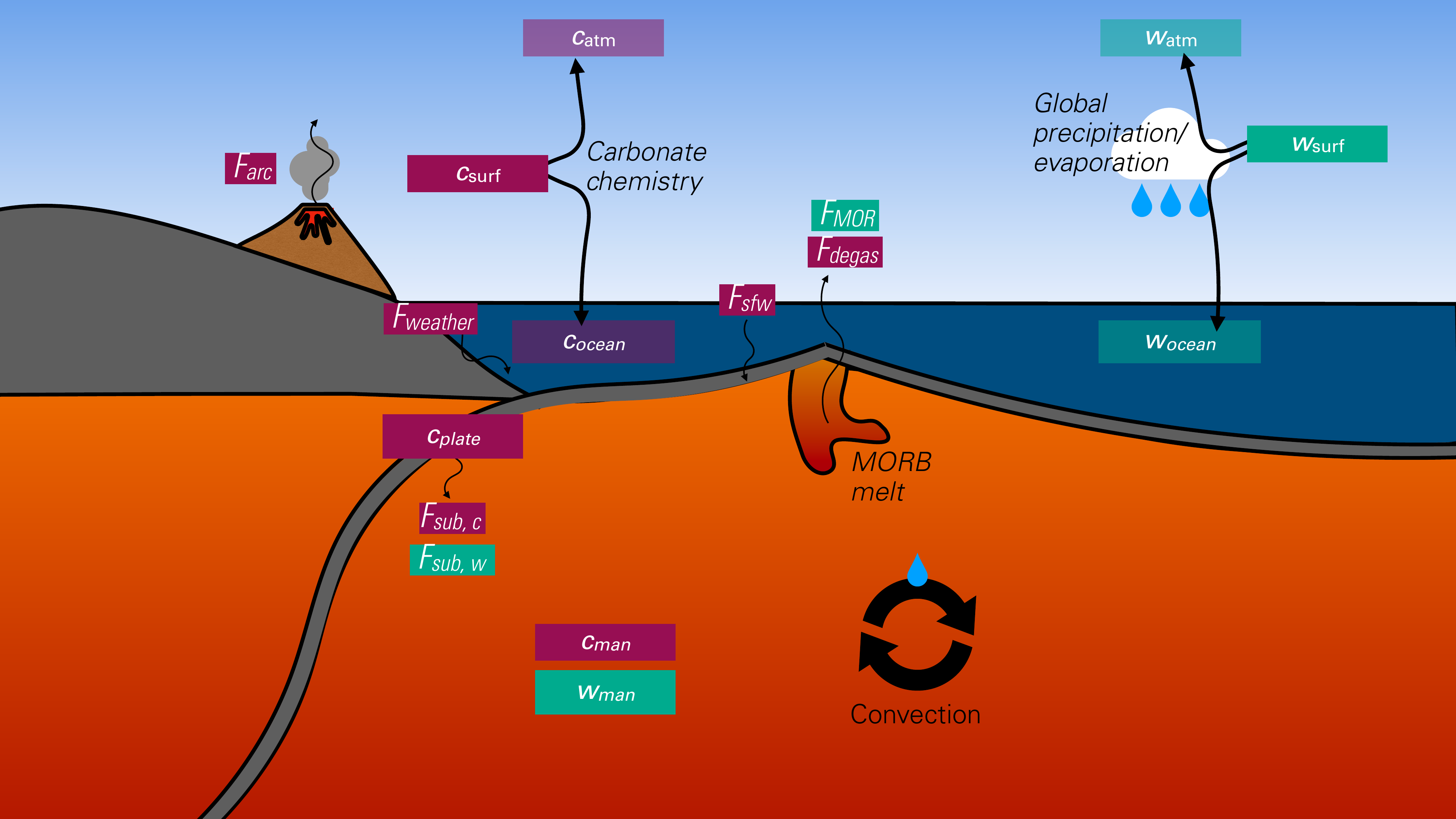}
    \caption{A schematic showing the atmospheric sources and sinks that define the new volatile cycling processes in \vplanet's \thermint module, inspired by Figure 1 in \citet{foley2015role}. Fluxes related to \ch{CO2} are color-coded magenta, and fluxes related to \ch{H2O} are color-coded teal. We incorporate the climate-stabilizing carbon cycle from \citet{foley2015role}, tracking the mass of CO$_2$ across the mantle ($c_{\textnormal{man}}$), plate ($c_{\textnormal{plate}}$), and surface ($c_{\textnormal{surf}}$) reservoirs. We include the deep water cycle from \citet{seales2020deep}, and a precipitation/evaporation parameterization from \citet{driscoll2013divergent}, which allow us to track the mass of H$_2$O across the mantle ($w_{\textnormal{man}}$) and surface ($w_{\textnormal{surf}}$) reservoirs. At each timestep, surface carbon and water are partitioned into the atmosphere ($c_{\textnormal{atm}}$, $w_{\textnormal{atm}}$) and the ocean ($c_{\textnormal{ocean}}$, $w_{\textnormal{ocean}}$) by enforcing equilibrium at the air-water boundary. This schematic does not depict modeled ocean chemistry.}
    \label{fig:schematic}
\end{figure*}

Our fully-coupled model builds on \vplanet \citep{barnes2020vplanet, garcia2026venus} by incorporating published models of the carbon cycle \citep{foley2015role}, the deep water cycle \citep{seales2020deep}, the surface hydrological cycle \citep{driscoll2013divergent}, and climate \citep{wogan2025open} for planets with plate tectonics. Our abiotic Earth model components are summarized in the schematic in Figure \ref{fig:model-diagram}, with a more detailed schematic of \ch{CO2} and \ch{H2O} cycling between the interior, surface, and atmosphere shown in Figure \ref{fig:schematic}.

In this section we describe the {assumptions and} governing equations used to upgrade \vplanet's \thermint module. Any equations and physical parameters not detailed here are implemented as described by \citet[][Appendix K]{barnes2020vplanet}. Model constants are defined in Appendix \ref{sec:constants}.

\subsection{Model Assumptions}
\label{sec:assumptions}

{We make several simplifying assumptions about interior and atmosphere processes for model tractability.  For the interior, we assume that the model is initialized immediately following the solidification of the magma ocean. Since this brief phase (1--10 Myr) likely dictated the initial composition of the atmosphere \citep{elkins2008linked, hamano2013emergence, lammer2018origin, stueeken2020mission, chao2021lava, carone2025co2}, we choose a set of appropriate initial conditions that apply after the bulk of the mantle has solidified.} 

{We also assume active plate tectonics throughout Earth's history, consistent with Earth-moon geology that suggests plate tectonics began as early as the middle Hadean \citep{maruyama2018initiation}. A number of alternative hypotheses for Earth's tectonic evolution exist{, including} {transitions} from a stagnant lid mode to modern plate tectonics during the Archean \citep{condie2016great}, {from} sluggish to active lid convection {\citep{al2024coupled},} {or a slowdown in plate velocity due to compositional buoyancy and lithospheric stiffening \citep{korenaga2006archean}.} {Since} we validate our model only against pre-industrial Earth {properties}, assessing the tectonic regime during the Hadean and Archean is outside the scope of this work.}  

{For the atmosphere, {as in previous work \citep{krissansen2021oxygen}, we assume that the atmosphere maintains a constant 1 bar of \ch{N2}, since our 1-bar, abiotic atmospheres lack photosynthetic oxygen.} Since abiotic methane sources generate fluxes several orders of magnitude lower than biological production \citep{etiope2013abiotic, wogan2020abundant, thompson2022case}, we focus our model on \ch{CO2} and \ch{H2O} as the primary greenhouse gases regulating climate on an Earth without life, with a discussion of reduced species in Section \ref{sec:discch4}.} Additionally, {we assume atmospheric escape is negligible, including magnetically mediated escape processes, in our model. The present-day escape rate is small:  observed hydrogen loss rates are negligible relative to Earth's volatile inventories over the model integration time and the tropopause cold trap keeps most of the water vapor inventory close to the surface \citep{kasting1984response, tian2013atmosphere, gronoff2020atmospheric}.} 

{Extending this assumption to earlier epochs, when the strength of the geodynamo may have differed, requires further justification, which we base on two considerations. First, the effect of a weaker paleofield on escape is uncertain: a planetary magnetosphere may suppress some escape channels (sputtering and ion pickup) while enabling others (polar cap and cusp outflow), resulting in a non-monotonic relationship between magnetic field strength and total escape rate \citep{gunell2018intrinsic, gronoff2020atmospheric}. Second, the strongest empirical constraint on Earth's integrated escape history, the xenon isotope record, admits interpretations ranging from the loss of an ocean's worth of hydrogen to little escape at all (see Section~\ref{sec:earthcomp}). Given these open questions, we choose zero escape as a reasonable limiting case for Earth, and leave coupled escape processes as future work (Section~\ref{sec:futurework}).} 

{Finally, we acknowledge that due to large uncertainties in early Earth conditions, we validate the model only against properties of the pre-industrial Earth. We therefore use observed pre-industrial continental weathering rates as our reference, acknowledging that these rates likely reflect some degree of biological enhancement by land vegetation and soil microbial communities. However, the quantitative contribution of biology to silicate weathering rates remains poorly constrained, with field estimates of the biotic enhancement factor spanning nearly two orders of magnitude \citep[e.g.,][]{dessert2003basalt, cochran1996promotion}, and no true abiotic control site exists on Earth today \citep{wild2022contribution}. We discuss addressing these uncertainties in Section \ref{sec:discussion}.}

\subsection{Core, Mantle, and Basal Magma Ocean}
\label{sec:interior}

We use the \vplanet core-mantle interior module, \thermint, and the radiogenic heating module, \radheat, as described by \citet{barnes2020vplanet}. We make two key modifications to the nominal model: (1) a basal magma ocean (BMO) layer, which  treats radiogenic heating contributions from both the core and the BMO in one combined term prior to present-day solidification, and (2) a mantle viscosity influenced by the water content in the mantle as described in \citet{garcia2026venus}. The mantle water content is controlled by the deep water cycle and is described in Section \ref{sec:water}. 
 
Earth's mantle may have solidified from the middle out \citep{boukare2025solidification}, where a prolonged basal magma ocean (BMO) would have solidified slowly and become enriched in incompatible elements, including radiogenic species of K, Th, and U \citep{labrosse2007crystallizing}. {With potential enrichment as high as $\sim$200 times the bulk silicate Earth \citep{labrosse2007crystallizing}, the BMO may have provided a stable, thermally insulating layer that slowed core cooling. However, the existence of the BMO is not yet widely accepted \citep{jackson2010evidence, campbell2012evidence, carlson2014did}. For example, \citet{campbell2012evidence} note that the measured heat flux from upper mantle plumes is inconsistent with a radioactively-enriched layer in the lower mantle.} To capture the thermal effect of an enriched BMO we define a new effective BMO + core radiogenic heating rate, $Q_{\mathrm{rad,BMO}+C} = Q_{\mathrm{rad,BMO}} + Q_{\mathrm{rad,}C}$ that folds in the radiogenic heating of the BMO.  With this modification the core energy balance in \thermint becomes:

\begin{equation}
\dot{T_C} = \frac{Q_{\mathrm{rad,BMO}+C} - Q_{\mathrm{BMO}}}{M_Cc_C - \frac{dM_{IC}}{dT_C} (L_{H} + E_G)}, 
\label{eq:derivfinal}
\end{equation}

\noindent where $M$ denotes mass, $c_C$ denotes the core specific heat capacity, $\dot{T}$ denotes the evolving reservoir temperature, and $Q$ denotes heat flow. The subscripts indicate the relevant reservoir, where $C$, CMB, and BMO represent the core, core-mantle-boundary, and basal magma ocean layer, respectively. Thus, $Q_{\mathrm{rad, BMO}}$ represents the radiogenic heat production in the BMO, and $Q_{\mathrm{rad,}C}$ represents the radiogenic heat production in the core. Additionally, $\frac{dM_{IC}}{dT_C}$ is the derivative of the inner core mass with respect to core temperature, and $L_{H}$ and $E_G$ give the latent and gravitational energy released at the inner core boundary per unit mass. For the full derivation, see Appendix \ref{sec:BMOcalc}. 

This approach approximates the effect of a BMO enriched in radiogenic elements without modeling its solidification. In reality, as the mantle gradually solidifies downward towards the core and the liquid BMO shrinks, it becomes further enriched in radiogenic elements  \citep{labrosse2007crystallizing}. This effectively maintains a higher concentration of radiogenic elements outside the core and provides a stable thermally insulating layer.

To match estimates of upper mantle heat flow within uncertainties, we find that our model requires approximately 3.5 TW of combined core and lower mantle radiogenic power by 4.5 Gyr. Given that core radiogenic heating today is thought to be significantly lower than this amount \citep{watanabe2014abundance, blanchard2017solubility, faure2020uranium}, the bulk of the heating must come from the BMO. To assess whether this excess heating is physically plausible we calculate the total present-day heat budget from a thin, enriched BMO layer. 

\citet{labrosse2007crystallizing} estimate that if the early Earth's magma ocean gradually solidified downward toward the core, incompatible radiogenic elements would be expelled from solid mantle minerals and concentrated in the remaining liquid BMO. They find the primordial BMO could be enriched in the main radioactive elements (\ch{^{40}K}, \ch{^{232}Th}, \ch{^{235}U}, and \ch{^{238}U}) by 200 times the concentrations in the bulk silicate Earth (see their Figure 4). Assuming the BMO is a spherical shell with thickness {4} km and that it is slightly denser than the solid mantle due to its relative enrichment in iron (6000 kg m$^{-3}$), {we find a present-day combined heating budget of 3.67 TW due to \ch{^{40}K}, \ch{^{232}Th}, \ch{^{235}U}, and \ch{^{238}U}. This calculation demonstrates that attributing 3.5 TW of core heating to a radiogenically enriched, thin, long-lived BMO is physically plausible.}     

\subsection{Carbon Cycle} \label{sec:carbon}

The model conserves the mass of CO$_2$ over geological time as it cycles between three reservoirs: the mantle ($c_{\textnormal{man}}$), plate ($c_{\textnormal{plate}}$), and surface ($c_{\textnormal{surf}}$), which includes the ocean ($c_{\textnormal{ocean}}$), and atmosphere ($c_{\textnormal{atm}}$) reservoirs. At each time step, we calculate the following fluxes as in \citet{foley2015role}: arc volcanic ($F_{\textnormal{arc}}$), continental weathering ($F_{\textnormal{weather}}$), seafloor weathering ($F_{\textnormal{sfw}}$), degassing ($F_{\textnormal{degas}}$), and subduction ($F_{\textnormal{sub,c}}$) (see Figure \ref{fig:schematic}). The fluxes are balanced between the planet's  CO$_2$ reservoirs according to the following equations \citep{foley2015role}:

\begin{equation}
    {\dot{c}_{\textnormal{plate}}} = \frac{F_{\textnormal{weather}}}{2} + F_{\textnormal{sfw}} - F_{\textnormal{sub,c}},
\end{equation}

\begin{equation}
    {\dot{c}_{\textnormal{man}}} = \chi_{\mathrm{sub,c}}F_{\textnormal{sub,c}} - F_{\textnormal{degas}}, 
\end{equation}

\noindent where {$\chi_{\mathrm{sub,c}}$} is a {factor} that {represents} the {subduction efficiency}. The continental weathering flux is divided by a factor of 2 to account for the fact that half of the carbon initially sunk into the continental plate is returned to the atmosphere when carbonates form on the seafloor \citep{berner1983carbonate, foley2015role}. Finally, we have

\begin{equation} \label{eq:carbsurf}
  \begin{aligned}
    \dot{c}_{\textnormal{surf}} & = F_{\textnormal{arc}} + F_{\textnormal{degas}} - \frac{F_{\textnormal{weather}}}{2} - F_{\textnormal{sfw}},\\
  \end{aligned}
\end{equation}

\noindent where by mass conservation $c_{\textnormal{surf}} = c_{\textnormal{atm}} + c_{\textnormal{ocean}}$ (see Section \ref{sec:oceanchem} for a detailed description of partitioning \ch{CO2} into the atmosphere and ocean).

We define the fluxes according to \citet{foley2015role}, with some minor changes. The subduction flux is given by

\begin{equation}
    F_{\textnormal{sub}} = 2 c_{\textnormal{plate}} v L \frac{d_{\textnormal{melt}}}{V_{\textnormal{man}}},
\end{equation}

 \noindent where $v$ is the speed of the tectonic plates (see below), $L$ is the present-day length {of the ocean} trench{es}, $d_{\textnormal{melt}}$ is the depth to the base of the upper mantle melt region (see \citet[][Appendix K]{barnes2020vplanet}), and $V_{\textnormal{man}}$ is the volume of the mantle. The $d_{\textnormal{melt}}/V_{\textnormal{man}}$ term represents division by the effective area of the mantle melt. The arc volcanic flux is subsequently defined as 
 
 \begin{equation}
     F_{\textnormal{arc}} = (1 - \chi_{\mathrm{sub,c}}) F_{\textnormal{sub}},
 \end{equation}
 
\noindent where {$1 - \chi_{\mathrm{sub,c}}$} is the fraction of subducting carbon that degasses. The degassing flux is therefore given by

 \begin{equation}
     F_{\textnormal{degas}} = 2 f_{\mathrm{UM}} \chi_{\mathrm{degas,c}} {c_{\textnormal{man}}}  \frac{v L d_{\textnormal{melt}}}{V_{\textnormal{man}}},
 \end{equation}

\noindent where $f_{\textrm{UM}}$ is the upper mantle melt fraction,{$\chi_{\mathrm{degas,c}}$} is the fraction of the {melted} mantle that degasses, and $V_{\textnormal{man}}$ is the volume of the mantle. We include the additional $f_{\textrm{UM}}$ term to ensure that the degassing flux is tightly coupled to the size of the melt layer. {As in \citet{foley2015role}, we assume that the length of the ocean ridges is equivalent to that of the trenches.} We note that the upper mantle melt fraction describes partial melting of the upwelling mantle, and not the whole-mantle volumetric melt fraction sometimes used in models that parameterize the solidification of the magma ocean \citep[e.g.,][]{elkins2008linked, krissansen2021oxygen}. 

The arc and ridge degassing fluxes are the \ch{CO2} sources in the surface (atmosphere + ocean). Both terms include efficiency factors {$\chi_{\mathrm{sub,c}}$} and {$\chi_{\mathrm{degas,c}}$}. We calibrate these factors to give a total carbon outgassing flux similar to present-day observations of $\sim$7.5 Tmol/yr \citep{catling2017atmospheric}, as well as PIE $p$\ch{CO2} values.

Prior to subduction, seafloor and continental weathering deposit carbon into the oceanic plate. The seafloor and continental weathering fluxes strongly influence the $p$\ch{CO2} evolution in our model. \citet{foley2015role} gives the seafloor weathering flux as

\begin{equation}
F_{\textnormal{sfw}} = F_{\textnormal{sfw},0} \left(\frac{v}{v_0}\right)\left(\frac{p\ch{CO2}}{p\ch{CO2}_{,0}}\right)^{\alpha},   
\end{equation}

\noindent where $F_{\textnormal{sfw},0}$ is the present-day seafloor weathering flux, $v_0$ is the present-day speed of the tectonic plates, $x_{oc}$ gives the ocean surface fraction, and $x_{oc,0}$ represents the present-day ocean surface fraction of 0.71. Lastly, $p$\ch{CO2} is the partial pressure of \ch{CO2}, $p$\ch{CO2}$_{,0}$ is its present-day value, and $\alpha$ quantifies the dependence of basalt carbonation on atmospheric \ch{CO2}. {\citet{krissansen2018constraining} showed that seafloor weathering should include a strong temperature dependence to reproduce the Cenozoic ${}^{87}$Sr/${}^{86}$Sr record. Incorporating an Arrhenius temperature relation from \citet{krissansen2018constraining}, our seafloor weathering equation becomes:}

\begin{align}
\label{eq:sfw}
F_{\textnormal{sfw}} = {} & F_{\textnormal{sfw},0} \left(\frac{v}{v_0}\right)\left(\frac{p\ch{CO2}}{p\ch{CO2}_{,0}}\right)^{\alpha} \nonumber \\
& \times \exp\left(\frac{E_\mathrm{bas}}{R_g}\left(\frac{1}{T_{\mathrm{pore,0}}} - \frac{1}{T_{\mathrm{pore}}}\right)\right),
\end{align}

\noindent {where $E_{\mathrm{bas}}$ = 92 kJ/mol is the empirically derived activation energy for basalt weathering \citep{coogan2015alteration}, $R_g$ is the universal gas constant, $T_{\mathrm{pore}}$ is the pore-space temperature, and $T_{\mathrm{pore,0}}$ is its modern-day value. The pore-space temperature is directly related to deep-ocean temperatures ($T_{\mathrm{D}}$) \citep{coogan2015alteration}, which in turn are related to global mean surface temperature ($T_{\mathrm{surf}}$, see Section \ref{sec:temp}). By fitting empirical data of global mean surface temperature and deep ocean temperature, \citet{krissansen2018constraining} give $T_{\mathrm{pore}}$ = $T_{\mathrm{D}} + 9$ and $T_{\mathrm{D}} = a_{\mathrm{grad}}T_{\mathrm{surf}} + b_{\mathrm{int}}$, with $a_{\mathrm{grad}}$ = 1.02 and $b_{\mathrm{int}}$ = -16.7 as the best-fit parameters.}

Finally, \citet{foley2015role} derives a continental weathering flux that accounts for supply-limited behavior at high $p$\ch{CO2} such that

\begin{equation}
\begin{aligned}
   F_{\textnormal{weather}} &= F_{\textnormal{weather},s} \times \biggl\{1 - \exp\biggl[-\frac{F_{\textnormal{weather},0}}{F_{\textnormal{weather},s}} \frac{x_{r}}{x_{l,0}} \\
    &\qquad \times \biggl(\frac{p\ch{CO2}}{p\ch{CO2}_{,0}}\biggr)^{\beta_1} \biggl(\frac{P_{\textnormal{sat}}}{P_{\textnormal{sat},0}}\biggr)^{a_1} \\
    &\qquad \times \exp\biggl(\frac{E_a}{R_g}\biggl(\frac{1}{T_{\textnormal{surf},0}}-\frac{1}{T_{\textnormal{surf}}}\biggr)\biggr) \biggr]\biggr\},
\end{aligned}
\end{equation}

\noindent where $F_{\textnormal{weather},s}$ is the supply limit to weathering, $F_{\textnormal{weather},0}$ is the present-day silicate weathering rate, $\beta_1$ is the partial pressure of CO$_2$ ($p$\ch{CO2}) scaling parameter for silicate weathering, $P_{\textnormal{sat}}$ is the saturation vapor pressure, $P_{\textnormal{sat},0}$ is the present-day saturation vapor pressure, $a_1$ is the saturation vapor pressure scaling parameter for silicate weathering, $E_a$ is the activation energy of the weathering reaction, and $T_{\textnormal{surf},0}$ is the present-day surface temperature. The equation also includes a term for scaling the evolving exposed rock fraction, $x_r$, at a given time step by the present-day land fraction, $x_{l,0}$. The treatment of evolving surface composition is described in further detail in Section \ref{sec:albedo}.

The supply limit to weathering represents the maximum achievable flux given the available weatherable rock on the surface of the planet. We calculate the supply limit using the relation $F_{\textnormal{weather},s} = (A_{\textnormal{Earth}} x_{\textnormal{r}} E_{\textnormal{max}} f_{cc} \rho_r)/{\bar{m}_{cc}}$, where $E_{\textnormal{max}}$ is the maximum erosion rate, $f_{cc}$ is the fraction of Mg, Ca, K, and Na in the continental crust, $\rho_r$ is the density of the regolith, and ${\bar{m}_{cc}}$ is the average molar mass of Mg, Ca, K, and Na \citep{foley2015role}. {Mg, Ca, K, and Na are relevant as these are the weatherable components of felsic continental crust \citep{west2012thickness}.} This relation gives the supply-limit in units of mol/s, which we convert to kg/s by multiplying by the molar mass of CO$_2$, $\bar{m}_c$. The saturation vapor pressure is related to its present-day value and the surface temperature of the planet according to the relation

\begin{equation}
P_{\textnormal{sat}} = P_{\textnormal{sat},\textnormal{ref}} \exp{\left[-\frac{\bar{m}_w L_w}{R_g} \left(\frac{1}{T_{\textnormal{surf}}} - \frac{1}{T_{\textnormal{sat},\textnormal{ref}}}\right) \right]},
\end{equation}

\noindent where $\bar{m}_w$ is the molar mass of water, $L_w$ is the latent heat of water, and $T_{\textnormal{sat},0}$ is the reference saturation vapor temperature \citep{driscoll2013divergent}. 

The fluxes $F_{\textnormal{arc}}$, $F_{\textnormal{sfw}}$, $F_{\textnormal{degas}}$, and $F_{\textnormal{sub,c}}$ are related to the speed of the tectonic plates, $v$. We implement the \citet{seales2020deep} parameterization, which relates $v$ to the mantle Rayleigh number:

\begin{equation}
v = \frac{a_2 \kappa}{2 (R_\mathrm{man} - R_c)} \left(\frac{Ra}{Ra_{\textnormal{crit}}}\right)^{2\beta_2},
\end{equation}

\noindent where $a_2$ and $\beta_2$ are scaling terms, $\kappa$ is the thermal diffusivity, $R_\mathrm{man}$ is the radius of the mantle, $R_c$ is the radius of the core, $Ra$ is the mantle Rayleigh number (see \citet{barnes2020vplanet}, Appendix K), and $Ra_\textnormal{crit} = 660$ is the critical Rayleigh number.

\subsection{Water Cycle} \label{sec:water}

We track the mass of water as it cycles between the mantle ($w_{\textnormal{man}}$) and the surface ($w_{\textnormal{surf}}$), which includes the ocean ($w_{\textnormal{ocean}}$), and the atmosphere ($w_{\textnormal{atm}}$), via a subduction flux ($F_{\textnormal{sub,w}}$) and a mid-ocean ridge degassing flux ($F_{\textnormal{MOR}}$). The fluxes balance according to the following system of differential equations:

\begin{equation}
    {\dot{w}_{\textnormal{man}}} = F_{\textnormal{sub,w}} - F_{\textnormal{MOR}},
\end{equation}

\noindent and

\begin{equation}
\dot{w}_{\textnormal{surf}} = F_{\textnormal{MOR}} - F_{\textnormal{sub,w}},
\end{equation}

\noindent where mass conservation requires that $w_{\textnormal{surf}} = w_{\textnormal{ocean}} + w_{\textnormal{atm}}$.

The abundance of water in the mantle impacts the viscosity of the upper mantle and thus the efficiency of heat transfer via the Arrhenius viscosity. \citet{garcia2026venus} give the Arrhenius component of the viscosity as

\begin{equation}
    \begin{aligned}
    \nu_{\textnormal{Arr}} = & \\ \nu_{\textnormal{ref}}  {\exp{\biggl[\frac{E_{a,\textnormal{man}}}{T_{\textnormal{UM}}R_g} - \frac{\Delta{E_\textnormal{H$_2$O}}(w_\textnormal{man}/M_\textnormal{man})}{T_{\textnormal{UM}}R_g}\biggr]}},
    \end{aligned}
\end{equation}

\noindent where $\nu_{\textnormal{ref}}$ is the reference viscosity of the upper mantle, $E_{a, \textnormal{man}}$ is the viscosity activation energy of the upper mantle, $\Delta{E_\textnormal{H$_2$O}}$ is the depression of mantle viscosity activation energy due to water, $M_{\textnormal{man}}$ is the mass of the mantle, and $T_{\textnormal{UM}}$ is the temperature of the upper mantle. The viscosity of the upper mantle {(within $\sim$100 km depth of the upper thermal boundary layer)} is then calculated by dividing the Arrhenius viscosity term by the upper mantle viscosity melt factor \citep{garcia2026venus}. {The viscosity of the lower mantle (within $\sim$300 km depth of the CMB) is calculated by scaling the upper mantle viscosity by a fixed viscosity ratio, a standard closure in parameterized thermal evolution models \citep{driscoll2014thermal, barnes2020vplanet, patovcka2020minimum}.} The melt viscosity factor is calculated following \citet{driscoll2015tidal}. First, we define the normalized melt fraction:
\begin{equation}
    \Phi = \frac{f_{\mathrm{UM}}}{\phi_s},
\end{equation}

\noindent where $f_{\mathrm{UM}}$ is the upper mantle melt fraction and $\phi_s = 0.8$ is the reference melt fraction parameter. Next, we calculate the melt geometry factor:

\begin{equation}
    F = (1 - \xi) \cdot \mathrm{erf}\left[\frac{\sqrt{\pi}}{2(1-\xi)} \Phi \left(1 + \Phi^{\gamma}\right)\right],
\end{equation}

\noindent where the viscosity-melt reduction coefficient $\xi = 5 \times 10^{-4}$, the viscosity-melt reduction exponent $\gamma = 6.0$, and $\mathrm{erf}$ is the error function. Finally, the melt viscosity reduction factor is given by:

\begin{equation}
    \eta_{\mathrm{melt}} = \frac{1 + \Phi^{\delta}}{(1 - F)^{B\phi_s}},
\end{equation}

\noindent where the viscosity-melt reduction exponent $\delta = 6.0$ and the viscosity-melt reduction parameter $B = 2.5$ \citep{costa2009model}. {The mantle viscosity depends on melt fraction through the parameterization of \citet{costa2009model}, which we adopt because it is constructed to be smooth and monotonic across the full range of melt fractions traversed by our model.   The more general multiphase framework developed by \citet{keller2019continuum} spanning the full range of melt fractions adopt this same relation as their reference effective-viscosity closure for basalt–olivine, showing that it reproduces the experimentally-constrained exponential melt-weakening at low melt fraction \citep{hirth2003rheology} while transitioning smoothly across the disaggregation threshold toward the solid limit—unlike diffusion-creep closures \citep{takei2009viscous, rudge2018viscosities} or the Einstein–Roscoe relation, which drop to zero or introduce unrealistically sharp transitions at their critical fractions.}

To calculate the fluxes of water into and out of the mantle, we assume the definitions of $F_{\textnormal{sub,w}}$ and $F_{\textnormal{MOR}}$ given by \citet{seales2020deep} so that

\begin{equation}
    F_{\textnormal{sub,w}} = f_h \rho_{\mathrm{man}}  D_{\textnormal{hydr}} S \chi_{\mathrm{regas,w}},
\end{equation}

\noindent where $f_h$ is the mass fraction of water in the serpentinized layer, $\rho_{\mathrm{man}}$ is the density of the mantle, $D_{\textnormal{hydr}}$ is the thickness of the serpentinized layer, $S$ is the spreading rate of the tectonic plates, and {$\chi_{\mathrm{regas,w}}$} is the regassing efficiency factor. The thickness of the serpentinized layer is related to surface temperature $T_{\textnormal{surf}}$ according to the relation $D_{\textnormal{hydr}} = ({k\left|{T_{\textnormal{surf}} - 973}\right|})/{q_m}$, where $k$ is the thermal conductivity and $q_m$ is the heat flux through the upper mantle (see \citet{barnes2020vplanet}, Appendix K). Note that this parameterization of the subduction flux assumes that there is always a sufficient supply of water to drive the serpentinizing reaction. The spreading rate is related to the speed of the plates as $2 L v$, where the factor of $2L$ represents bidirectional spreading. 

The mid-ocean ridge degassing flux is thus given by

\begin{equation}
    F_{\textnormal{MOR}} = \rho_{\mathrm{man}} f_{\textnormal{UM}} X_{\textnormal{melt}} d_{\textnormal{melt}} S \chi_{\mathrm{degas,w}},
\end{equation}

\noindent where $X_{\textnormal{melt}}$ is the abundance of water in the erupting magma and {$\chi_{\mathrm{degas,w}}$} is the degassing efficiency factor. The parameters $f_{\textnormal{UM}}$ and $d_{\textnormal{melt}}$ are calculated as described by \citet[][Appendix K]{barnes2020vplanet}, while the fraction of water in the melt can be found by taking $X_{\textnormal{melt}} = {\chi_m}/[{D_{\textnormal{H}_2 \textnormal{O}} + f_{\textnormal{UM}}(1 - D_{\textnormal{H}_2 \textnormal{O}})}]$, where $\chi_m$ is the mass fraction of water in the mantle ($w_{\textnormal{man}}$ / $M_\mathrm{man}$) and $D_{\textnormal{H}_2 \textnormal{O}}$ is the bulk distribution coefficient \citep{seales2020deep}. 

The mass of \ch{H2O} on the surface (atmosphere + ocean) results from the balance of the mid-ocean ridge degassing sources and the subduction sinks at the plate boundaries. In our model, both terms include degassing and regassing efficiency factors, {$\chi_{\mathrm{degas,w}}$} and {$\chi_{\mathrm{regas,w}}$}, respectively. We tune these factors to give the PIE mass of water in the oceans and atmosphere after 4.5 Gyr, and to give a degassing flux on the order of present-day observations, $\sim$95 Tmol/yr \citep{catling2017atmospheric}.

\subsection{Atmospheric Constituents}
\label{sec:atmcomponents}

Our Earth-like atmospheres consist of \ch{N2}, \ch{H2O}, and \ch{CO2}. Atmospheric \ch{H2O} and \ch{CO2} are allowed to vary based on the balance of sources and sinks from volatile cycling{, while \ch{N2} is kept at a constant 1 bar (see \ref{sec:assumptions}).} In the model framework, we treat both \ch{H2O} and \ch{CO2} as combined surface reservoirs. At each time step, we partition each gas between the atmosphere and the ocean by enforcing instantaneous equilibrium. Thus, by mass conservation we require that $c_{\textnormal{surf}} = c_{\textnormal{ocean}}+c_{\textnormal{atm}}$, and $w_{\textnormal{surf}} = w_{\textnormal{ocean}}+w_{\textnormal{atm}}$. Here, we describe the partitioning of water, while the partitioning of \ch{CO2} is described in the following subsection. 



To maintain hydrostatic equilibrium at the air-water boundary, at each time step we partition water between the atmosphere and the ocean by initially calculating the mass of water in the atmosphere ($w_{\textnormal{atm}}$) in excess of the amount of moisture the air can hold at a given time step \citep{driscoll2013divergent}. Since the Earth's oceans and atmosphere achieve equilibrium on the order of days, whereas \thermint processes evolve on geological timescales, we assume the flow of water between the atmosphere and the ocean occurs instantaneously in the model. At each time step, we calculate the saturation vapor pressure \citep{foley2015role} as a function of $T_{\textnormal{surf}}$, and we then multiply by a factor of $RH \times (M_{\textnormal{atm,T}}/P_{\textnormal{T}}) \times ({\bar{m}_w}/{\bar{w_{atm}}})$ to obtain ${m_{\textnormal{moist}}}$, the maximum mass of water that the air can hold, where $RH$ represents the tuneable global average relative humidity. 

When $w_{\textnormal{atm}}$ $> {m_{\textnormal{moist}}}$, the excess water mass is subtracted from $w_{\textnormal{atm}}$ and added to $w_{\textnormal{ocean}}$. In the inverse case when $w_{\textnormal{atm}}$ $< {m_{\textnormal{moist}}}$, to maintain equilibrium between the air and the ocean, we subtract the mass difference from $w_{\textnormal{ocean}}$ and add it to $w_{\textnormal{atm}}$. To obtain $p$\ch{H2O} in Pa, we convert $w_{\textnormal{atm}}$ to moles of \ch{H2O}, divide by the total moles in the atmosphere, and multiply by $P_{\textnormal{T}}$. This simplified 0-D approximation provides the initial atmospheric \ch{H2O} partial pressure needed as input to the climate calculation (Section \ref{sec:temp}), where it is subsequently refined using the vertically-resolved atmospheric structure. 

\subsection{Ocean Chemistry}
\label{sec:oceanchem}
The total reservoir of carbon in the ocean (total dissolved inorganic carbon, or DIC) is the sum of the aqueous [\ch{CO2}], [\ch{CO3^2-}], and [\ch{HCO3^-}]. Since we enforce equilibrium at the air-ocean boundary, we assume that all ocean chemistry values are representative of surface waters. Initially, atmospheric \ch{CO2} {dissociates in} ocean water following Henry's law \citep{pilson2012introduction}, which quantifies the solubility of a gas according to its partial pressure above a liquid so that,

\begin{equation}
    (\ch{CO2})_{\mathrm{aq}} = H \times p\ch{CO2},
\end{equation}

\noindent where (\ch{CO2})$_{\mathrm{aq}}$ gives the aqueous concentration of CO$_2$ in units of mol/kg \ch{H2O}, $H$ is Henry's law coefficient in units of mol/kg/atm, and $p$\ch{CO2} is the partial pressure of atmospheric \ch{CO2}. We obtain $c_{\textnormal{ocean}}$ by multiplying (\ch{CO2})$_{\mathrm{aq}}$ by the total mass of water in the ocean, $w_{\textrm{ocean}}$, and the molar mass of \ch{CO2}. By definition, $c_{\textnormal{atm}} = ({p\ch{CO2}}/{P_{\textnormal{T}}}) \times ({\bar{m}_c}/{\bar{m}_{atm}}) \times M_{\textnormal{atm, T}}$, where $P_{\textnormal{T}}$ is the total pressure, $\bar{m}_{atm}$ is the average molar mass of the atmosphere, and $M_{\textnormal{atm, T}}$ is the total mass of the atmosphere. Substituting $p\ch{CO2} = {\ch{CO2_{\textnormal{aq}}}}/{H}$ into the mass conservation constraint, we solve a quadratic equation to obtain (\ch{CO2})$_{\mathrm{aq}}$ based on the integration of Equation \eqref{eq:carbsurf} at each time step. Finally, to obtain $p$\ch{CO2} in Pa, we take the (\ch{CO2})$_{\mathrm{aq}}$ solution and divide by the Henry's law coefficient, and convert from atm to Pa.

Next, aqueous CO$_2$ and H$_2$O combine to make carbonic acid in the following equilibrium reaction:
\begin{equation}
(\ch{CO2})_{\mathrm{aq}} + \ch{H2O <=>[a][b] H2CO3},
\end{equation}
\noindent where the forward reaction (a) is governed by the rate constant $k_a$, and the reverse reaction (b) is governed by the rate constant $k_b$. Depending on the ocean pH, carbonic acid readily {dissociates} into H$^+$ and bicarbonate in the following equilibrium reaction
\begin{equation}
\ch{H2CO3 <=>[c][d] H^+ + HCO3^-},
\end{equation}
\noindent with $k_c$ and $k_d$ for the forward and reverse reactions, respectively. Finally, bicarbonate may further dissolve into H$^+$ and a carbonate ion
\begin{equation}
\ch{HCO3^- <=>[e][f] H^+ + CO3^2-},
\end{equation}
\noindent where $k_e$ and $k_f$ give the dissolution and combination rate constants, respectively. 

As in \cite{schwieterman2019rethinking} and \cite{krissansen2021oxygen}, we assume that calcite is saturated so that
\begin{equation}
\label{eq:carb}
    [\ch{CO3^2-}] = \frac{\Omega_{\textnormal{cal}} \times K_{sp} (T_{\textnormal{surf}})}{[\ch{Ca^2+}]},
\end{equation}
\noindent where $\Omega_{\textnormal{cal}}$ is the unit-less global ocean calcite saturation factor, $K_{sp}$ is the calcite solubility product, and [$\ch{Ca^2+}$] is the concentration of calcite in the oceans in mol/kg. To calculate solubility self-consistently with surface temperature, we use the temperature and salinity-dependent solubility product formula given by \citet{millero2005chemical}. As in \citet{schwieterman2019limited}, for all solubility parameterizations we assume that ocean salinity remains constant at 35 parts per thousand. We also assume that the temperature of the ocean is equivalent to the global average surface temperature, as we assume that the ocean and atmosphere equilibrate faster than our model's geological timescales. 

Using Equation (\ref{eq:carb}), a saturation factor, and a calcite concentration, we can calculate the carbonate concentration in the ocean. \citet{schwieterman2019limited} assumed $\Omega_{\textnormal{cal}}=1$, but calcite may need to be supersaturated ($\Omega_{\textnormal{cal}}\sim10$--$20$) for abiotic precipitation to occur. Furthermore, over Earth's history, the calcite concentration has varied from [\ch{Ca^{2+}}] $=10^{-2}$ to $3 \times 10^{-1}$ mol/kg \citep{halevy2017geologic}. We therefore tune $\Omega_{\textnormal{cal}}/[\ch{Ca^2+}]$ to match the observed total dissolved inorganic carbon in the surface oceans for pre-industrial conditions (see Table \ref{tab:constants}). 

This tuning approach requires justification, as our model represents an Earth without oxygenic photosynthesis or complex skeletal organisms, analogous to conditions during the Archean and early Proterozoic eons. In the modern ocean, carbonate precipitation is predominantly biogenic \citep{Chave1970, Bialik2022}, though \citet{Bialik2022} estimate that abiotic aragonite precipitation accounts for approximately 15\% of CO$_2$ efflux in oligotrophic regions where surface waters remain supersaturated with respect to calcium carbonate. However, evidence supports that abiotic carbonate precipitation processes, while kinetically slower than biogenic precipitation, can achieve similar steady-state DIC concentrations over geological timescales. Laboratory experiments show that abiotic calcite precipitates at rates two to five orders of magnitude slower than biotic calcite \citep{carpenter1992srmg}, but given billion-year timescales, kinetic limitations become less constraining. 

Furthermore, geological evidence supports widespread abiotic carbonate precipitation in Earth's early history. The Precambrian geological record shows extensive carbonate platform deposition \citep{Grotzinger1989, Grotzinger1993, Grotzinger2000}, including meter-thick beds of fibrous calcite and aragonite precipitated directly on the Archean seafloor \citep{Sumner1996}. \citet{Sumner1996} argued that high rates of abiotic precipitation in Archean oceans were facilitated by elevated Fe$^{2+}$ concentrations in anoxic seawater, which paradoxically promoted aragonite precipitation while inhibiting calcite nucleation. This geological evidence demonstrates that abiotic processes can generate substantial carbonate deposits, supporting our assumption that calcite saturation is achievable without biology.

We find that the optimal $\Omega_{\textnormal{cal}}/[\ch{Ca^2+}]$ to match pre-industrial DIC is $408.1$ kg/mol. Recent spatial mapping of global calcite saturation of present-day oceans shows an average $\Omega_{\textnormal{cal}} = 4.3$ \citep{shaik2025advanced}, and the present-day [\ch{Ca^2+}] is 10.28 mmol/kg \citep{halevy2017geologic}, corresponding to a present-day ratio of $\Omega_{\textnormal{cal}}/[\ch{Ca^2+}] = 398$. Conservatively estimating that a supersaturation factor of $\Omega_{\textnormal{cal}} = 20$ is required for abiotic carbonate precipitation, our calibrated $\Omega_{\textnormal{cal}}/[\ch{Ca^2+}]$ corresponds to a calcite concentration of $49$ mmol/kg. We discuss the possible limitations of this tuning approach in Section \ref{sec:calcite}.

We model this chemical system using the following equations adopted from \citet{pilson2012introduction}:
\begin{equation}
\label{eq:k1}
    k_e = \frac{[\ch{H^+}][\ch{HCO3^-}]}{(\ch{CO2})_{\mathrm{aq}}},
\end{equation}

\noindent and 

\begin{equation}
\label{eq:k2}
    k_f = \frac{[\ch{H^+}][\ch{CO3^2-}]}{[\ch{HCO3^-}]}.
\end{equation}

Since $k_e/k_f = {[\ch{HCO3^-}]^2}/({[\ch{CO3^2-}][\ch{CO2}]})$, we can combine these rate constant equations to solve for the concentration of bicarbonate:

\begin{equation}
\label{eq:bicarb}
    [\ch{HCO3^-}] = \sqrt{(\ch{CO2})_{\mathrm{aq}} [\ch{CO3^2-}] \times \frac{k_e}{k_f}}.
\end{equation}

Finally, we can solve for the concentration of $[\ch{H^+}]$ by taking

\begin{equation}
    [\ch{H^+}] = \frac{K_{sp} [\ch{HCO3^-}]}{[\ch{CO3^2-}] },
\end{equation}
\noindent where pH $= -\log_{10}{[\ch{H^+}]}$.

 \subsection{Surface Albedo} \label{sec:albedo}

The surface albedo of the planet in our model consists of contributions from the ground and atmospheric water vapor clouds. Based on previous work \citep{driscoll2013divergent}, we simulate the evolving total planetary surface albedo $A_T$ by allowing the ground albedo $A_g$ and cloud fraction $\phi$ to vary with time so that

\begin{equation}
    A_{T} = A_{g} (1 - \phi) + A_{cl} \phi.
\end{equation}

\noindent We assume that our Earth-like water vapor clouds have an albedo of $A_{cl} = 0.89$, which represents an average albedo of thick and thin clouds \citep{de2010planetary, driscoll2013divergent}. We assume a constant fraction of the atmospheric H$_2$O reservoir condenses into clouds, and tune the wavelength-independent cloud opacity pressure to ensure we match the PIE surface temperature at 4.5 Gyr. As in \citet{driscoll2013divergent}, we then calculate the cloud reflectivity fraction using the two-stream Eddington approximation,

\begin{equation}
    \phi = \frac{\gamma (1 - e^{-2\beta_{\tau, cl}}) }{1 - \gamma^2 e^{-2\beta_{\tau, {cl}}}},
\end{equation}

\noindent with $\tau_{cl} = 4800$ Pa as the constant cloud opacity pressure and scattering constants

\begin{equation}
\gamma = \frac{\beta -2(1 - \bar{w})}{\beta +2(1 - \bar{w})}
\end{equation}

\noindent and 

\begin{equation}
    \beta = \sqrt{3(1 - \bar{w})(1 - \bar{w}g_w)},
\end{equation}
 \noindent where $\bar{w}$ is the single-scattering albedo $1 - \bar{w} = 1\times10^{-7}$ and $g_w=0.74$ is the asymmetry factor \citep{hashimoto2001predictions}. Given this {simplified} cloud model {cannot account for more complex effects like radiative heating due to clouds and wavelength-dependent reflectivity}, we tune {the cloud opacity pressure} to match pre-industrial Earth {temperature}. {Thus}, we will \textit{not} attempt to match the pre-industrial Earth cloud {fraction} with the model.  

The ground albedo $A_g$ consists of contributions from surface rock ($x_{r}$), ocean ($x_{oc}$), and ice ($x_{i}$) fractions, where

\begin{equation}
    A_g = x_{r} A_{r} + x_{oc} A_{oc} + x_{i} A_{i}.
\end{equation}

\noindent We take the albedo of rock to be $A_r = 0.17$ \citep{kaltenegger2007spectral, driscoll2013divergent}, the albedo of ocean water to be $A_{oc} = 0.1$ \citep{de2010planetary, driscoll2013divergent}, and the albedo of ice to be $A_i = 0.6$ \citep{kaltenegger2007spectral, driscoll2013divergent}. We assume a conical ocean basin {as in \citet{driscoll2013divergent},} with a slope angle of $\theta_{oc} = 0.06188 \degree$ calibrated to give $x_{oc} = 0.71$ when $w_\mathrm{ocean} = 1.4 \times 10^{21}$ kg. To model ice growth, we include the simple relation from \citet{driscoll2013divergent}, which gives $x_i = x_{i,\mathrm{max}} \times (1 - x_{oc}) \times ({x_{oc}}/{x_{oc,0}})$, where $x_{i,\mathrm{max}} = 0.12$ is the maximum ice fraction, $(1 - x_{oc})$ is the fraction of land, and $x_{oc,0}$ is the present-day ocean fraction of 0.71. 

Previous work has established that habitability requires only some amount of liquid water on the planetary surface \citep{kasting1993habitable}, and that global average surface temperatures can be meaningfully compared to spatially-resolved three-dimensional climate states when assessing habitability thresholds \citep{charnay2013exploring, arney2016pale}. \citet{charnay2013exploring} demonstrated through 3D general circulation modeling that complete glaciation (e.g., a ``snowball'' Earth) likely does not occur until global average surface temperatures fall below 240 K, owing to spatial heterogeneity in ice coverage and meridional heat transport that maintains liquid water reservoirs at the equator. We therefore allow $x_{i} = 1.0$ when the global average surface temperature is $\leq$ 240 K, though like \citet{driscoll2013divergent} our model does not currently handle transitions between global glaciation phases and $x_{i,\mathrm{max}}$.  

\subsection{Climate} \label{sec:temp}

Planetary surface temperature is related to surface and atmospheric characteristics as well as the radiation received from the star. We determine surface temperature by calculating thermal equilibrium between the absorbed short-wave radiation (ASR) and the outgoing long-wave radiation (OLR). OLR is a function of surface temperature and the atmospheric abundances of water vapor and CO$_2$, while ASR is additionally a function of the total albedo, incoming stellar flux, and effective stellar temperature. 

To calculate ASR and OLR, we use the \clima module of the \texttt{Photochem} software package \citet{wogan2025open}. \clima uses standard two-stream methods \citep{toon1989rapid} to solve the radiative transfer equations, incorporating opacity tables to simulate continuum UV absorption, Rayleigh scattering, and collision-induced absorption, while accounting for line absorption with the correlated-k method. This approach explicitly captures the wavelength-dependent absorption and scattering processes that gray atmosphere models cannot. \clima has been benchmarked against other community radiative transfer codes including SOCRATES \citep{wolf2022exocam}, ExoRT \citep{wolf2022exocam}, SMART \citep{meadows1996ground}, and the radiative transfer code used by \citet{kopparapu2013habitable}, demonstrating good agreement across a range of atmospheric compositions and stellar spectra. Furthermore, the code reproduces the observed pressure-temperature profiles of Venus, Earth, Mars, Jupiter and Titan \citep{wogan2025open}.

To resolve 1-D atmospheric structure, we use the mode of \clima that assumes an isothermal stratosphere above a convective troposphere with a pseudo-moist adiabat. The stratospheric temperature is estimated via the skin temperature for an assumed bond albedo, $A_b$. As in any 1-D climate model, \clima has a limited ability to self-consistently calculate the radiative impact of clouds. As a result, we follow previous work \citep[e.g.,][]{kopparapu2013habitable} and ``paint'' them on the surface by incorporating them into the total surface albedo, which is an input to the radiative transfer calculation (see Section \ref{sec:albedo}).

Using the stellar evolution module in \vplanet \citep{baraffe2015new}, we calculate the evolving incoming stellar flux ($F_{\star}$) and temperature ($T_{\star}$) as a function of time. To determine the top-of-atmosphere solar flux, \clima calculates a black-body distribution corresponding to $T_{\star}$. This ensures that ASR accounts for how the evolving Sun impacts wavelength-dependent atmospheric opacity. Rather than calling the full radiative transfer calculation at every timestep, which would be computationally prohibitive for billion-year integrations, we pre-compute a grid of OLR and ASR values as functions of surface temperature, atmospheric \ch{CO2} partial pressure, the total \ch{H2O} surface inventory, stellar flux, stellar effective temperature, and surface albedo. 

The grid spans surface temperatures from 100 K to 500 K (41 points, spaced at 10 K intervals), \ch{CO2} partial pressures from $10^{-6}$
 to $10^{4}$ bar (21 points, logarithmically spaced at 0.5 dex intervals), \ch{H2O} total surface pressures from $10^{-3}$ to $10^{4}$
bars (15 points, logarithmically spaced at 0.5 dex intervals), stellar flux from 800 to 2000 W/m$^2$
(13 points, spaced at 100 W/m$^2$ intervals), surface albedoes of 0.0, 0.1, 0.3, 0.5, 0.7, and 0.9 (6 discrete values), and stellar effective temperatures from 5000 to 6000 K (5 points, spaced at 250 K intervals). At each model timestep, we pass the current values of $T_{\mathrm{surf}}$, $p$\ch{CO2}, total surface water inventory (in bars), stellar flux ($F_{\star}$), $A_{\mathrm{T}}$, and stellar effective temperature ($T_{\star}$) to a multidimensional linear interpolation routine that returns the corresponding OLR and ASR from the pre-computed grid. The complete grid contains approximately 9.7 million pre-computed atmospheric states.

We then use Brent's method \citep{brent2013algorithms} to iteratively solve for the equilibrium surface temperature where ASR equals OLR, using a bracket of $[100, 600]$ K and a tolerance of $2 \times 10^{-12}$. Once convergence is achieved, we pass the solved $T_{\textnormal{surf}}$ back to \vplanet to update all temperature-dependent parameters. This grid-based approach maintains the approximate accuracy of full radiative transfer calculations ($<\sim1$ K error) while reducing computational cost by approximately three orders of magnitude, enabling efficient exploration of long-term planetary evolution.

In \clima, we parameterize the \ch{H2O} inventory as a pressure column (in bars) representing the total surface water reservoir  ($w_{\textnormal{surf}}$ converted to an equivalent pressure). \clima internally partitions the combined surface reservoir between atmospheric vapor (limited by saturation vapor pressure at each vertical level) and condensed phases, capturing the coupled radiative-thermodynamic effects of water vapor feedback. Once a steady-state climate is computed, we extract the atmospheric composition using \clima's vertically-resolved solution. We extract the surface water vapor mixing ratio and the total surface pressure from the interpolated grid output, which represent \clima's self-consistent partitioning of water between the atmosphere and ocean when accounting for temperature stratification throughout the atmospheric column. This more physically accurate partitioning refines our initial surface-only estimates described in Sections \ref{sec:atmcomponents}. We recalculate $w_{\textnormal{atm}}$, $w_{\textnormal{ocean}}$, $c_{\textnormal{atm}}$ and $c_{\textnormal{ocean}}$ to match \clima's atmospheric state, ensuring that the water distribution used in subsequent carbon cycle and thermal evolution calculations remains consistent with the radiative transfer solution. 

\subsection{Reflected Light Spectra}

We use the \picaso code \citep{batalha2019exoplanet} to compute synthetic HWO reflected light spectra (0.2-2.0 $\mu$m) for the true Earth and the abiotic Earth's atmospheric state after 4.5 Gyr. We use a set of $R = 15{,}000$ resampled opacities archived on Zenodo \citep{Opacities2025}, derived from a similar set of HITEMP \citep{rothman2010hitemp} and HITRAN \citep{gordon2017hitran2016} opacities used by \clima for climate calculations (Section \ref{sec:temp}). The opacity database accounts for line absorption from the following relevant species: \ch{CO2}, \ch{H2O}, \ch{O2}, \ch{O3}, and \ch{CH4}. Relevant CIA partners include \ch{CO2}-\ch{CO2}, \ch{H2O}-\ch{H2O}, \ch{H2O}-\ch{N2}, \ch{N2}-\ch{N2}, \ch{N2}-\ch{O2}, and \ch{O2}-\ch{O2}. \picaso also includes Rayleigh scattering for all specified molecules.

We compute two reflected light spectra: (1) a realistic modern Earth spectrum using temperature and mixing ratio profiles from the Intercomparison of Radiation Codes in Climate Models (ICRCCM) case 62, which represents the averaged mid-latitude Earth during the summer months \citep{lincowski2018evolved}, and (2) an abiotic Earth spectrum using the \vplanet atmospheric state after 4.5 Gyr, which includes only \ch{CO2} and \ch{H2O} as greenhouse gases along with 1 bar of \ch{N2}. For both spectra, we input an expected surface albedo of 0.14 (without clouds) and include a cloud deck. We model clouds as a single gray deck extending from 0.6 to 0.7 bar with an optical depth of 10, a single-scattering albedo of 0.99, and an asymmetry parameter of 0.85. We compute both cloudy and cloud-free spectra and weight them by a cloud fraction of 50\% to approximate Earth's observed cloud coverage. Both calculations assume solar illumination at quadrature (90$\degree$ phase angle) and are binned to $R = 140$. 

The ICRCCM Earth reference profile has been re-interpolated onto the \vplanet pressure grid to enable direct comparison. The \vplanet grid has a surface pressure of 1.016 bar compared to the ICRCCM value of 0.989 bar, representing a difference of approximately 0.03 bar, or 3\% higher surface pressure. The top-of-atmosphere (TOA) extends to $1.1 \times 10^{-5}$ bar in the \vplanet grid versus $1 \times 10^{-7}$ bar in the original ICRCCM profile. The number of atmospheric layers differs as well, with 51 layers in the \vplanet grid compared to 63 layers in the original ICRCCM profile. These differences are minor -- the 3\% surface pressure difference causes negligible changes in atmospheric column density, and the higher TOA cutoff removes only the uppermost stratosphere and mesosphere, which are optically thin and contribute minimally to the spectrum. The temperature-pressure profile, water vapor distribution, and other trace gas abundances have been linearly interpolated in log-pressure space to preserve the atmospheric structure of the original ICRCCM profile. 

\subsection{Initial Conditions}
\label{sec:IC}

Our model requires the initialization of key parameters, including those describing the initial conditions of the interior and the surface volatile inventories. Here we briefly describe how we initialized our model at 50 Myr; i.e., following the magma ocean phase. Most other initial values are calibrated to match the relevant measurements of the PIE. Though the model evolution is largely insensitive to the initial \ch{CO2} and \ch{H2O} inventories, we base the total initial masses of volatiles on previous work.

To compute a plausible evolutionary model for Earth, we optimized 16 initial conditions and constants of the interior (given in Table \ref{tab:input_parameters}) to reproduce Earth's present-day interior properties (given in Table \ref{tab:earth-meas}). To perform the optimization, we assumed a Gaussian likelihood function based on the mean and uncertainty of each output parameter, and minimized the negative log-likelihood using the Nelder-Mead simplex algorithm \citep{Nelder1965} implemented in \texttt{SciPy} \citep{jones2001scipy}. Due to the multi-modal nature of likelihood space, we initialized the optimization at 200 different starting locations uniformly sampled over a wide parameter range, and selected the maximum likelihood fit out of all runs that converged. 

As in \citet{seales2020deep} we initialize our model with 2 terrestrial oceans (TOs) of water. From \citet{sleep2001carbon}, we assume that Earth's total initial \ch{CO2} inventory is $1.1 \times 10^{21}$ kg. Initially, to partition the volatiles between the interior and surface reservoirs, we follow estimations of the primordial Earth atmosphere immediately following the solidification of the magma ocean. \citet{elkins2008linked} predicts primordial Earth \ch{CO2}--\ch{H2O} atmospheres ranging from 90--3350 bars. Consistent with the lower end of these {model} estimates, we initialize the surface with 5 bars of \ch{CO2} and 530 bars of \ch{H2O} (2 T.O.) for a combined primordial atmosphere of 535 bars. 

Though the magma ocean phase may conclude with a higher surface inventory of \ch{CO2}, through initial model testing we found that the $p$\ch{CO2} evolution is largely independent of the initial atmospheric carbon inventory. This model behavior occurs because {our} seafloor weathering flux (Equation [\ref{eq:sfw}]) {is assumed to be} directly dependent on plate speed and $p$\ch{CO2} relative to the present-day, which {may} be very large immediately following the magma ocean phase{, though primordial Earth conditions remain uncertain}. As a result, basaltic seafloor weathering is found to be efficient enough to draw nearly all of the \ch{CO2} out of the atmosphere. Therefore, we initialize the model with 5 bars to maximize model stability -- all \ch{CO2} reservoirs must initially contain non-zero mass, and the remaining \ch{CO2} is split between the plate and the mantle.

In summary, our model couples interior thermal evolution (\thermint, \radheat), volatile cycling (carbon and water), ocean chemistry, and climate (\clima) into a unified framework. We track \ch{CO2} and \ch{H2O} masses across mantle, ocean, and atmospheric reservoirs with fluxes controlled by plate tectonics and weathering. Key model features include a radiogenically enriched BMO, water-dependent mantle viscosity, calcite-saturated ocean chemistry, and evolving planetary surface albedo from ground and cloud contributions. We initialize the model as described above, then evolve the system forward 4.5 Gyr to match pre-industrial Earth properties.

\section{Results}
\label{sec:results}

In this section, we show that the whole-planet model successfully reproduces the pre-industrial Earth after 4.5 Gyr of model evolution. We also generate a corresponding reflected light spectrum to compare against that of a realistic, cloudy Earth. In Table \ref{tab:input_parameters} we report the calibrated interior parameter values following our optimization procedure described in Section \ref{sec:IC}, and model results and a comparison to measured values within uncertainties are summarized in Table \ref{tab:model-validation}.

\begin{table*}[htb!]
\centering
    \begin{tabular}{lc}
       \hline
    \textbf{Parameter} & \textbf{Value} \\
   \hline
    Initial \textsuperscript{40}K power in the mantle & $5.14 \times 10^{13}$ W \\
     Initial \textsuperscript{40}K power in the core & $4.05 \times 10^{13}$ W  \\
     Initial \textsuperscript{232}Th power in the mantle & $7.01 \times 10^{12}$ W  \\
     Initial \textsuperscript{232}Th power in the core & $1.30 \times 10^{11}$ W  \\
    Initial  \textsuperscript{238}U power in the mantle & $1.04 \times 10^{13}$ W  \\
     Initial \textsuperscript{238}U power in the core & $1.0 \times 10^{11}$ W  \\
     Initial \textsuperscript{235}U power in the mantle & $1.66 \times 10^{13}$ W  \\
     Initial \textsuperscript{235}U power in the core & $4.68 \times 10^{11}$ W \\
    Initial mantle temperature & 2700 K \\
    Initial core temperature & 5700 K \\
    Melt eruption efficiency & 0.01 [n.d.] \\
    Core liquidus depression & $4.04 \times 10^{-4}$ K  \\
    Lower/upper mantle viscosity ratio & 1.487 [n.d.]  \\
    Depression of mantle viscosity activation energy due to water & $9.642\times 10^{6}$ K  \\
    Viscosity activation energy & $2.64 \times 10^{5}$ J mol\textsuperscript{-1} \\
    Upper mantle thermal conductivity & 4.24 W/m/K  \\
    \hline
    \end{tabular}
\caption{Initial conditions and constants of the interior Earth model. Core radiogenic power includes BMO contribution (see Section \ref{sec:interior}).}
\label{tab:input_parameters}
\end{table*}

\begin{figure}[htb!]
    \centering
    \includegraphics[width=\linewidth]{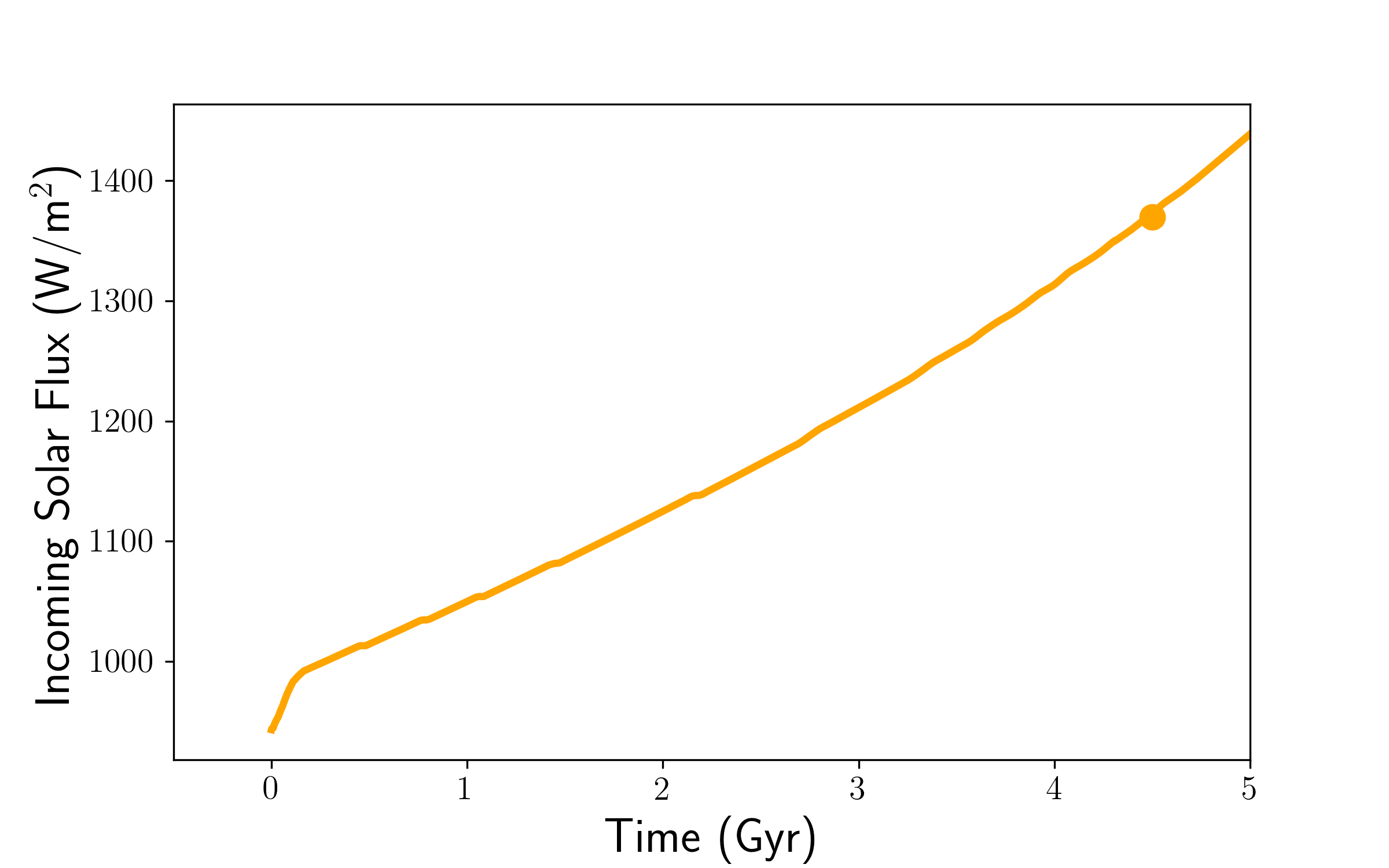}
    \caption{Incoming solar flux evolution with time, where the dot at 4.5 Gyr represents the present-day value.}
    \label{fig:sun}
\end{figure}

In Figure \ref{fig:sun}, we show the incoming solar flux at the planet as a function of time. Using the \citet{baraffe2015new} model, the evolving solar flux starts at 943 W/m$^2$ ($\sim$$0.7 L_{\odot}$) and reaches 1370 W/m$^2$ ($1.0 L_{\odot}$) by 4.5 Gyr. The final incoming stellar flux is thus consistent with its present-day value. Furthermore, the early stellar flux is consistent with predictions of other stellar evolution models used to model the faint young Sun \citep{sagan1972earth, newman1977implications, gough1981solar}.

\begin{figure*}[htb!]
    \centering
    \includegraphics[width=\linewidth]{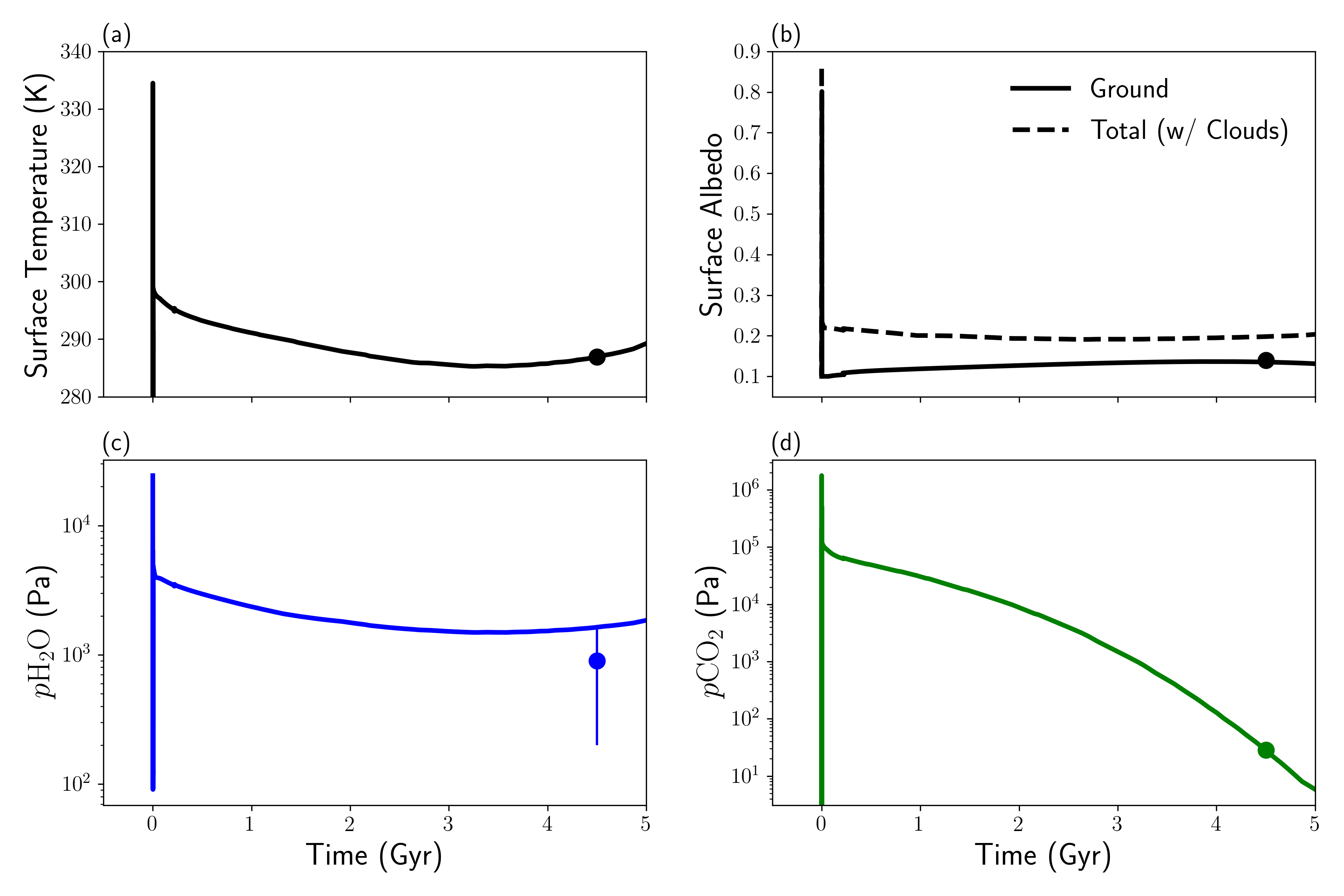}
    \caption{Some bulk properties of the abiotic Earth as a function of time. (a) Global average surface temperature warms as the star gradually brightens. (b) The ground albedo increases as dark ocean water subducts into the mantle and comparatively brighter land becomes exposed. Total albedo also increases as the warming planet's atmosphere can hold more water vapor, and more clouds form. (c) As the planet warms, the atmosphere can hold more water and the \ch{H2O} partial pressure increases. (d) Finally, atmospheric \ch{CO2} spikes early on when the planet is cold, and gradually decreases as weathering becomes more efficient on the warming planet. On all plots, dots represent measurements of PIE (4.5 Gyr) properties with uncertainties.}
    \label{fig:bulk}
\end{figure*}

Figure \ref{fig:bulk} shows the evolution of bulk planetary parameters including surface temperature, total albedo (ground and clouds), atmospheric $p$H$_2$O, and atmospheric $p$CO$_2$. By 4.5 Gyr, the planet surface temperature reaches 286.9 K (a), the ground albedo (solid line) is 0.14 while the total albedo including clouds (dashed line) is 0.198 (b), the atmospheric $p$H$_2$O is 1600 Pa with an assumed relative humidity of 100\% (c), and the atmospheric $p$CO$_2$ is 28.2 Pa (d). The atmospheric water vapor abundance (c) shows a strong relationship with planetary temperature (a) due to the efficiency of OLR absorption and strong temperature-dependence of saturation vapor pressure. After 4.5 Gyr, our model {suggests} that the total degassing flux of \ch{CO2} is 9 Tmol/yr and that of \ch{H2O} is 85 Tmol/yr -- both are reasonably close to the present-day observations of 7.5 and 95 Tmol/yr, respectively \citep{catling2017atmospheric}.

At the start of the evolution, following a brief ($\sim$100 Myr) equilibration period, the planet is relatively warm, with $T_{\mathrm{surf}} = 297$ K. The degassing flux dominates, allowing the atmospheric CO$_2$ abundance to rise to $\sim$1 bar 100,000 years after model initialization. As the planet warms, weathering is enhanced and the carbon sinks begin to overwhelm the sources, resulting in a gradual decrease in $p$CO$_2$. Initial oscillations in \ch{CO2} ($<$5 Myr) occur due to a disequilibrium between \ch{CO2} partitioning and the rapidly cooling mantle and surface temperatures. The final state of the model is ultimately insensitive to these initial model conditions. Furthermore, the transition between magma ocean solidification and the onset of plate tectonics is poorly understood, so modeling this transition falls outside the scope of this work.

The total albedo includes contributions from the ground and clouds, and is therefore higher. Clouds are tuned to match the PIE temperature. From tuning the constant cloud opacity pressure, we find a total albedo of 0.198 corresponding to a cloud fraction of 8\%.  We note that this is likely an unrealistically small cloud fraction for the PIE, most likely due to our simplified cloud implementation (see Section \ref{sec:futurework}).

\begin{figure*}[htb!]
    \centering
    \includegraphics[width=\linewidth]{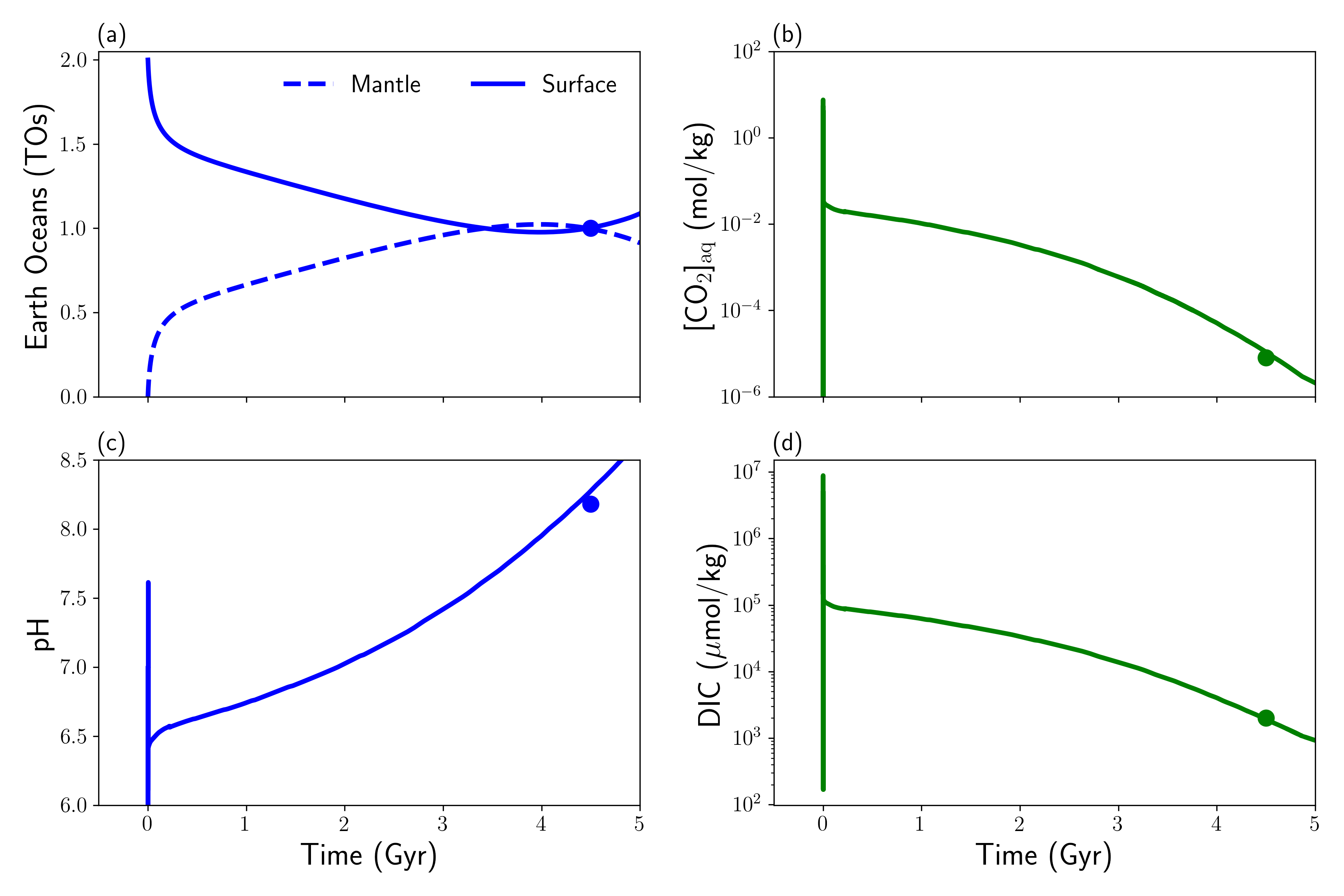}
    \caption{Properties of the ocean as a function of time. (a) After 4.5 Gyr, 1 TO has entered the mantle, ultimately leaving 1 TO on the surface. (b) The concentration of \ch{CO2} in the oceans tracks closely with the atmospheric \ch{CO2} inventory, following  Henry's law. (c) Prior to equilibration, the pH initially spikes because of the initial concentration of carbonate and bicarbonate in the calcite-saturated oceans is also very high. Following equilibration, the pH gradually becomes more alkaline as the abundance of \ch{CO2} in the atmosphere and ocean gradually decreases. (d) The total dissolved inorganic carbon at the ocean's surface, which includes aqueous \ch{CO2}, carbonate, and bicarbonate, tracks closely with the atmospheric \ch{CO2} budget. On all plots, dots represent measurements of PIE (4.5 Gyr) properties with uncertainties.}
    \label{fig:ocean}
\end{figure*}

Figure \ref{fig:ocean} shows the evolution of the surface and mantle water budgets (a), the aqueous CO$_2$ concentration (b), ocean pH (c), and the surface ocean's total dissolved inorganic carbon (DIC) concentration (d). We initialize the surface with 2 TOs, resulting in 1 TO in the mantle after 4.5 Gyr. Using Henry's law, we calculate an aqueous CO$_2$ concentration of 10.8 $\mu$mol/kg by 4.5 Gyr (b). Assuming a constant carbonate concentration based on the abiotic calcite budget, we obtain a pH of 8.28 after 4.5 Gyr (c). Finally, the total dissolved inorganic carbon (DIC) at the ocean's surface is 1919 $\mu$mol/kg after 4.5 Gyr (d). Ocean pH exhibits strong temperature-dependent behavior, while the DIC evolution tracks closely with the atmospheric CO$_2$ evolution. When the atmospheric CO$_2$ abundance is initially high (Figure \ref{fig:bulk}, d), the aqueous CO$_2$ concentration is also high (Figure \ref{fig:bulk}, b). Following model equilibration, as atmospheric CO$_2$ abundances sharply increase prior to 1 Gyr, the aqueous CO$_2$ concentration rises to $2\times10^{-2}$ mol/kg, and the ocean pH is $\sim$6.6, or 30$\times$ more acidic than present-day Earth oceans. As the atmospheric CO$_2$ content gradually wanes to PIE values, the aqueous CO$_2$ concentration and DIC follow suit, resulting in increasingly alkaline oceans.

\begin{figure*}[htb!]
    \centering
    \includegraphics[width=\linewidth, 
    trim=1cm 2.0cm 1cm 2.5cm,clip]{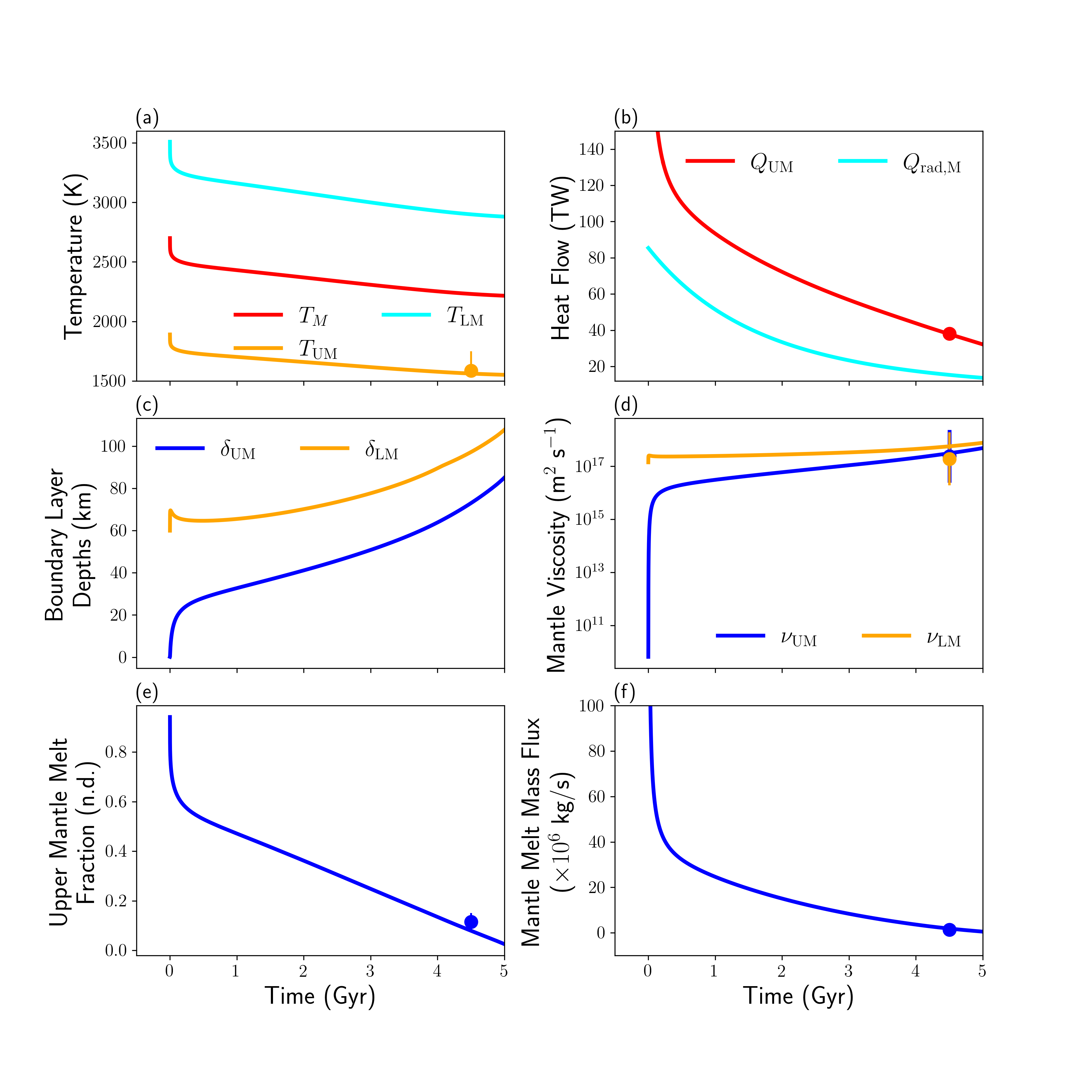}
    \caption{Key properties of the Earth's mantle through time. (a) The mantle and core gradually cool over time. (b) As the core and mantle cool, the heat flow across the upper mantle and the core-mantle boundary and the heat flow due to radiogenics in the core and mantle decrease. (c) Both the upper and lower mantle boundaries grow deeper as the core-mantle system cools and the mantle solidifies downwards. (d) The lower mantle {TBL} is initially less viscous than {that of} the upper mantle. As water subducts into the cooling mantle, both the upper and lower mantle viscosities gradually increase. (e) As the interior cools, the upper mantle melt fraction rapidly decreases. (f) Similarly, the melt flux from the mantle also decreases as the mantle cools. On all plots where applicable, dots represent measurements of PIE (4.5 Gyr) properties with uncertainties.}
    \label{fig:mantle}
\end{figure*}

Figure \ref{fig:mantle} shows the evolution of the mantle. In Figure \ref{fig:mantle}(a) we show the evolution of the average temperature of the mantle ($T_M$), the evolving upper mantle temperature $T_{\mathrm{UM}}$, and the evolving lower mantle temperature $T_{\mathrm{LM}}$. In panel (b) of the same figure, we show the evolution of heat flow in the upper mantle ($Q_{\mathrm{UM}}$), and the radiogenic heat flow in the mantle ($Q_\mathrm{{rad,M}}$). Panel (c) shows the changing boundary layer depths of the upper ($\delta_{\mathrm{UM}}$) and lower mantle ($\delta_{\mathrm{LM}}$). {Note that the thickness of the assumed BMO always remains smaller than the thickness of the lower boundary layer ($<60$--100 km), preventing the radiogenically enriched BMO from mixing into the convecting mantle.} Panel (d) shows the viscosity near the upper {TBL} ($\nu_{\mathrm{UM}}$) and {near the lower TBL of} the mantle ($\nu_{\mathrm{LM}}$) with time. Finally, panel (e) shows the upper mantle melt fraction, and panel (f) shows the melt mass flux over geological time. By 4.5 Gyr, we confirm that $T_{\mathrm{UM}}$,  $Q_{\mathrm{UM}}$, $\nu_{\mathrm{UM}}$, $\nu_{\mathrm{LM}}$, $F_{\mathrm{melt}}$, and mantle melt mass flux all match measurements of Earth properties within measurement uncertainties. 

\begin{figure*}[htb!]
    \centering
    \includegraphics[width=\linewidth, 
    trim=1cm 2cm 1cm 2cm,clip]{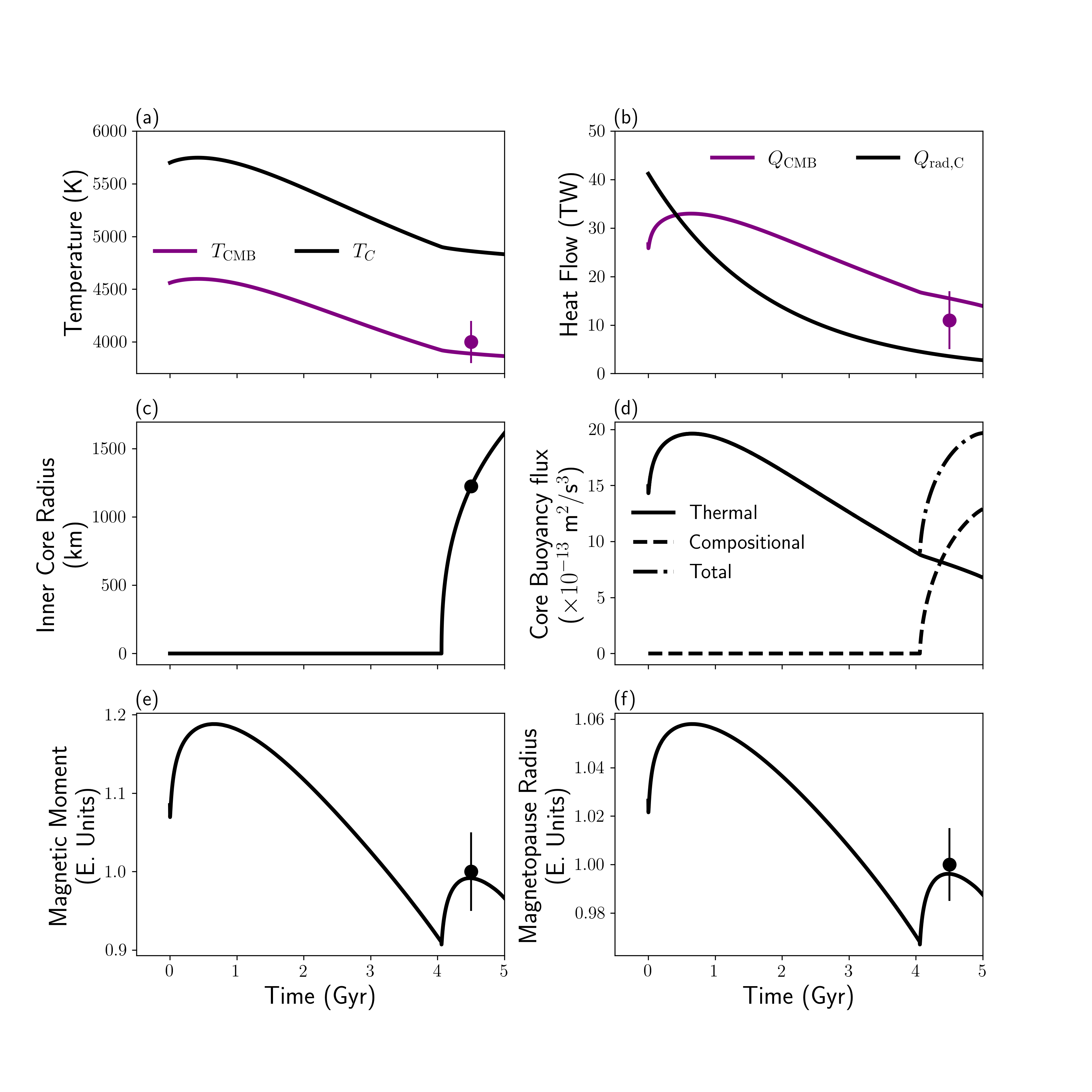}
    \caption{Key properties of the Earth's core through time. (a and b) The temperature and heat flows of the core and core-mantle boundary secularly decrease with time. (c) Once the interior has sufficiently cooled, the inner core rapidly solidifies at $\sim$4 Gyr. (d) The thermal buoyancy flux of the core gradually decreases as the core cools. Once the inner core solidifies and exsolves light elements, the compositional changes inject energy into the outer core and boost the overall core buoyancy flux. (e and f) The late energy increase in the outer core due to inner core solidification causes coinciding discontinuities in the magnetic moment and the magnetopause radius. On all plots where applicable, dots represent measurements of PIE (4.5 Gyr) properties with uncertainties.}
    \label{fig:core}
\end{figure*}

In Figure \ref{fig:core}, we show the evolution of key core properties over time. Panel (a) shows the evolution of the core-mantle boundary ($T_{\mathrm{CMB}}$) and core ($T_C$) temperatures, while Panel (b) shows the evolution of the heat flow at the core-mantle boundary and the heat flow in the core due to radiogenics ($Q_{\mathrm{rad,C}}$). Panel (c) shows the radius of the inner core with time, and panel (d) shows the core buoyancy flux. Finally, panel (e) shows the magnetic moment in Earth units, and panel (f) shows the magnetopause radius in Earth units. 

For all properties shown in this figure, we note a discontinuity that occurs $\sim$4 Gyr. This feature represents inner core nucleation in the model, when light elements begin to be rejected from the solidifying inner core \citep{driscoll2014thermal}. The thermal buoyancy flux of the core gradually decreases as the core cools (b). Once the inner core begins to solidify light elements are injected into the base of the liquid outer core, generating compositional buoyancy and boosting the total core buoyancy flux (b). This buoyancy injection into the outer core boosts the magnetic moment and magnetopause radius (c and d). All four core related properties in Figure \ref{fig:core} match the final expected values within measurement uncertainties.

{\setlength{\tabcolsep}{32pt}
\begin{deluxetable*}{lrr}[htb!]
\tablecaption{Abiotic Earth model values compared to pre-industrial Earth measurements. Model values are reported at 4.5 Gyr.}
\tablehead{
\colhead{Properties} & 
\colhead{Target Value} & 
\colhead{Model Value}
}
\label{tab:model-validation}
\startdata
Atmospheric \ch{{CO2}} ($p$\ch{{CO2}}) & 28.4 $\pm$ 0.4 Pa & 28.2 Pa \\
Atmospheric \ch{{H2O}} ($p$\ch{{H2O}}) & $900 \pm 700$ Pa & 1600 Pa \\
Surface temperature ($T_{{\mathrm{{surf}}}}$) & 286.9 $\pm$ 0.1 K & 286.9 K \\
Ground albedo ($A_g$) & $0.14$ & 0.14 \\
Ocean mass ($w_{{\textrm{{ocean}}}}$) & $1.4 \times 10^{{21}}$ kg & $1.4 \times 10^{21}$ kg \\
Oceanic (\ch{{CO2}})${{\mathrm{{aq}}}}$ & $8 \pm 2$ $\mu$mol/kg & 10.8 $\mu$mol/kg \\
Total Dissolved Inorganic Carbon (DIC) & $2000 \pm 200$ $\mu$mol/kg & 1919 $\mu$mol/kg \\
pH & $8.18 \pm 0.05$ & 8.28 \\
Upper mantle temperature ($T_{{\mathrm{{UM}}}}$) & $1587_{{-34}}^{{+161}}$ K & 1561 K \\
Core-mantle boundary temperature ($T_{{\mathrm{{CMB}}}}$) & $4000 \pm 200$ K & 3889 K \\
Upper mantle heat flow ($Q_{{\mathrm{{UM}}}}$) & $38 \pm 3$ TW & 38 TW \\
Core-mantle boundary heat flow ($Q_{{\mathrm{{CMB}}}}$) & $11 \pm 6$ TW & 15 TW \\
Upper TBL viscosity ($\nu_{\mathrm{UM}}$) & $10^{17.38 \pm 1}$ m$^2$ s$^{-1}$ & $10^{17.46}$ m$^2$ s$^{-1}$ \\
Lower TBL viscosity ($\nu_{\mathrm{LM}}$) & $10^{17.28 \pm 1}$ m$^2$ s$^{-1}$ & $10^{17.73}$ m$^2$ s$^{-1}$ \\
Upper mantle melt fraction ($F_{{\mathrm{{UM}}}}$) & $11.5 \pm 3.5$\% & 10.0\% \\
Melt flux & $(1.3 \pm 0.8) \times 10^6$ kg/s & $1.9 \times 10^6$ kg/s \\
Inner core radius ($R_{{\mathrm{{IC}}}}$) & $1221 \pm 1$ km & 1221 km \\
Magnetic moment & $1.00 \pm 0.05$ E. Unit & 1.01 E. Unit \\
Magnetopause radius & $1.00 \pm 0.02$ E. Unit & 1.00 E. Unit \\
\enddata
\tablecomments{All measured values and uncertainties are detailed in Section \ref{sec:props}, and summarized in Table \ref{tab:earth-meas}.}
\end{deluxetable*}
}

\vspace{-24pt}

Table \ref{tab:model-validation} summarizes our final model values for 19 pre-industrial characteristics. For 17 out of 19 key parameters, our model results are {within the reported measurement uncertainties}, {while oceanic (\ch{CO2})$_{\mathrm{aq}}$ and ocean pH fall slightly outside measurement uncertainties. Our modeled (\ch{CO2})$_{\mathrm{aq}}$ of 10.8~$\mu$mol/kg slightly exceeds the measured range of $8 \pm 2$~$\mu$mol/kg, while our modeled ocean pH of 8.28 is marginally more alkaline than the measured value of $8.18 \pm 0.05$  (an offset of 0.10 pH units, or roughly twice the reported measurement uncertainty). Assuming the symmetric uncertainties of these measurements represent normal underlying distributions, our model is no more than 2-$\sigma$ offset from the mean measurements, and we take this to indicate sufficient model agreement.} Nonetheless, this slight deviation from expectation for both ocean pH and (\ch{CO2})$_{{\mathrm{aq}}}$ can likely be attributed to our simplifying assumptions regarding calcite concentration. We discuss the implications of this assumption in Section \ref{sec:calcite}. 

\begin{figure*}[htb!]
    \centering
    \includegraphics[width=\linewidth]{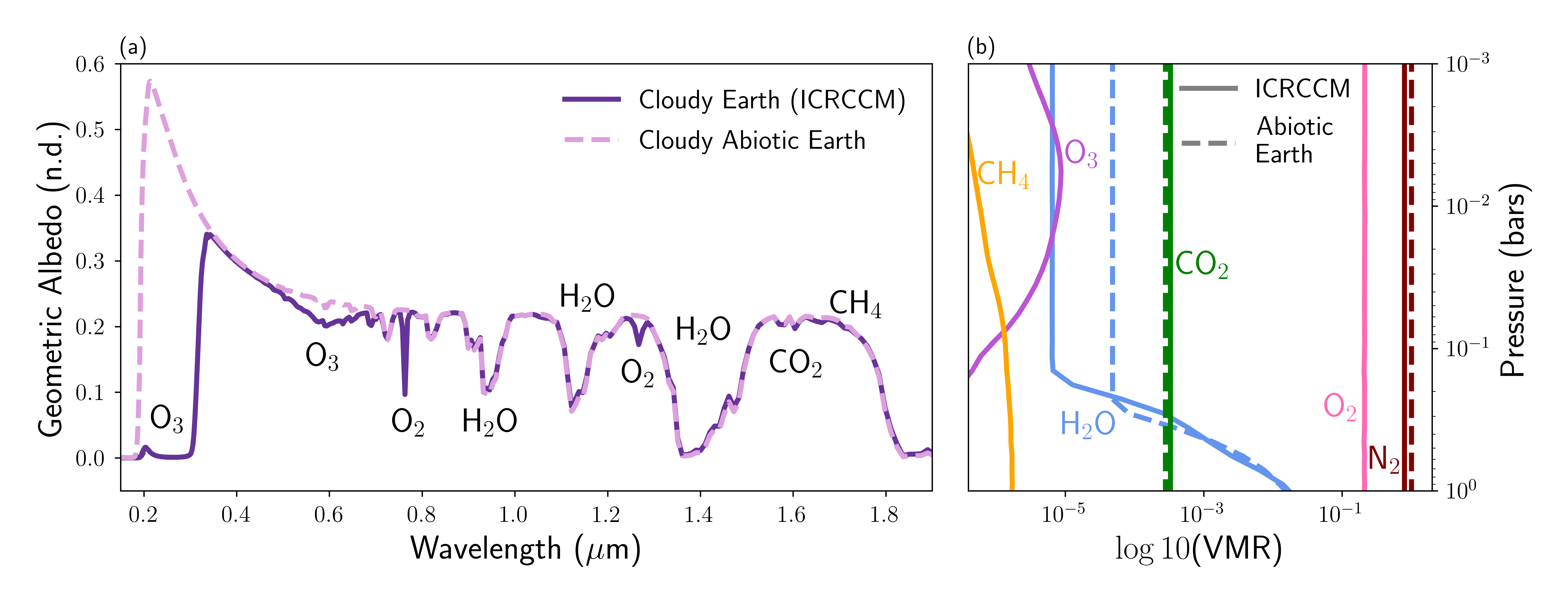}
    \caption{(a) A comparison of reflected light spectra for a realistic, cloudy Earth and the abiotic atmospheric state after 4.5 Gyr. The abiotic atmosphere lacks \ch{O2} (and therefore \ch{O3}) and \ch{CH4}, contributing to differences in absorption. (b) A comparison of the input molecular abundance profiles used to generate each reflected light spectrum. We address minor differences between the \ch{CO2}, \ch{H2O}, and \ch{N2} profiles in the accompanying text. Though the true profiles extend to a top-of-atmosphere pressure of $1 \times 10^{-5}$ bars, we only show the surface to the upper stratosphere to better emphasize the differences in the water vapor profiles given the logarithmic scale.}
    \label{fig:spec}
\end{figure*}

Finally, as a proof of concept, in Figure \ref{fig:spec}, we compare reflected light spectra of a realistic cloudy Earth and a snapshot of the abiotic Earth after 4.5 Gyr (a), as well as their corresponding abundance profiles (b). Volume-mixing ratio profiles are shown in log-log space as a function of decreasing pressure in bars on the y-axis. While both atmospheres include \ch{CO2}, \ch{H2O}, and \ch{N2}, the abiotic atmosphere lacks \ch{O2} (and therefore \ch{O3}) and \ch{CH4}. The lack of \ch{O2} and \ch{O3} contributes to the most significant differences in absorption between the two spectra. For example, in the near-UV $\sim$0.25 $\mu$m, the true Earth spectrum shows absorption due to the \ch{O3} Hartley band. In the visible, the Earth spectrum has absorption features at $\sim$0.6 $\mu$m due to the \ch{O3} Chappuis band, and a prominent narrow-band feature at $\sim$0.76 $\mu$m due to the \ch{O2} A-band. In the near-infrared, the Earth spectrum shows absorption at $\sim$1.27 $\mu$m due to the weak \ch{O2} feature. Finally, the true Earth spectrum has $\sim$2 ppm of \ch{CH4}, which produces an absorption feature at $\sim$1.69 $\mu$m in the near-infrared, resulting in a relatively weak absorption feature that our abiotic planet lacks. 

Figure \ref{fig:spec}b demonstrates some minor differences between the \ch{N2}, \ch{CO2}, and \ch{H2O} profiles. While the ICRCCM atmosphere is 79\% \ch{N2}, the abiotic atmosphere has 1 bar to compensate for the lack of bulk \ch{O2}. Furthermore, the ICRCCM atmosphere was benchmarked to more recent terrestrial atmospheric abundances. As a result, the ICRCCM atmosphere has a higher \ch{CO2} abundance of 330 ppm versus our 280 ppm. Lastly, though both water vapor profiles have similar near-surface abundances, our tropopause appears at a slightly lower pressure (and higher altitude) than the ICRCCM profile. This difference occurs for two reasons. First, the ICRCCM profile represents an average mid-latitude climate during the summer months whereas our profile represents a globally averaged profile. Second, \clima estimates the location of the tropopause by calculating the atmospheric skin temperature rather than explicitly calculating the radiative-convective boundary. Nonetheless, these differences in the atmospheric profiles do not contribute to major differences in the resulting spectra -- the weak \ch{CO2} bands are insensitive to small differences in abundance, and most spectral water vapor absorption occurs below the tropopause.

\section{Discussion}
\label{sec:discussion}

{We have presented a coupled core-mantle-crust-climate evolution model that reproduces 19 properties of the pre-industrial Earth within estimated measurement uncertainties, with two ocean chemistry parameters showing modest offsets attributable to our simplified treatment of calcite saturation.} We have also shown that our model can be used to generate synthetic HWO-like reflected light spectra. Our findings suggest that the abiotic Earth could evolve to a similar state (in terms of global geophysics and climate) to present-day Earth without life, but without abundant atmospheric oxygen. In this section, we discuss the implications of key assumptions in our model, such as neglecting \ch{CH4} and abiotic calcite saturation in the oceans. Finally, we discuss our findings in the context of the Gaia hypothesis and forecast future research directions.

\subsection{Spectroscopic Predictions for HWO Observations}

The atmospheric states produced by our coupled evolution model can generate realistic reflected light spectra suitable for comparison with future HWO observations (see Figure \ref{fig:spec}). The most significant spectral differences arise from the absence of biologically-mediated gases: the modern Earth spectrum exhibits strong absorption features from \ch{O2}, \ch{O3}, and \ch{CH4}, none of which appear in our abiotic atmosphere.

This comparison validates that our model produces realistic atmospheric states that can be propagated through radiative transfer codes to generate synthetic HWO observations. The abiotic spectrum represents the expected appearance of a habitable but lifeless Earth-like planet, establishing a null hypothesis against which to test biosignature detections. While both spectra show prominent \ch{H2O} and \ch{CO2} features, only the biological Earth exhibits the \ch{O2} (or \ch{O3})/\ch{CH4} disequilibrium signature that would constitute compelling evidence for modern Earth-like life \citep{meadows2018exoplanet}. Future parameter space explorations varying planetary properties and host star types will reveal the diversity of reflected light spectra representing abiotic habitable worlds, establishing the range of null hypothesis scenarios that HWO must discriminate from life.

\subsection{\ch{CH4} as a Greenhouse Gas}
\label{sec:discch4}

Our model establishes that the carbonate-silicate cycle can operate effectively to maintain habitable surface temperatures over geological timescales using only \ch{CO2} and \ch{H2O} as greenhouse agents, without requiring additional radiative forcing from trace gases like \ch{CH4}. This finding has important implications for understanding both Earth's climate history and the potential habitability of rocky exoplanets.

Abundant atmospheric \ch{CH4} due to microbial life on the early Earth has been proposed as a solution to the faint young Sun paradox \citep{pavlov2000greenhouse, kharecha2005coupled}. \citet{pavlov2000greenhouse} showed that atmospheric \ch{CH4} mixing ratios of 100--1000 ppmv, combined with elevated \ch{CO2} levels, could maintain above-freezing temperatures despite 20--30\% lower solar luminosity. While biological fluxes of \ch{CH4} may have warmed Earth's climate in certain epochs -- with some models suggesting temperature increases of 5--10$\degree$C from elevated methane during the Archean \citep{roberson2011greenhouse} -- our results suggest that such biological enhancement is not necessary for maintaining habitability. An abiotic Earth {may} remain habitable through the \ch{CO2}-\ch{H2O} greenhouse alone, provided the carbonate-silicate cycle operates.

Future work should nevertheless explore the sensitivity of climate evolution to low levels of abiotic \ch{CH4}. Even modest trace gas abundances could influence atmospheric photochemistry and oxidation state. The atmospheric lifetime of methane depends strongly on photochemically generated hydroxyl radicals (\ch{OH-}) produced through water vapor photolysis \citep{pavlov2001uv}. In an anoxic atmosphere without oxygenic photosynthesis, extended methane lifetimes could potentially allow greater accumulation than our simplified analysis suggests given sufficiently available \ch{OH-}. Additionally, planets orbiting M- and K-dwarf stars may experience different photochemical regimes due to their host stars' UV spectra, potentially affecting methane lifetimes and the viability of methane as a warming agent on abiotic worlds \citep{segura2005biosignatures, arney2016pale}. Incorporating volcanic \ch{CH4} outgassing coupled to mantle redox state evolution would establish a more complete abiotic baseline for interpreting future observations of exoplanets orbiting different stars.

\subsection{Abiotic Marine Calcite Budget}
\label{sec:calcite}

Tuning the calcite saturation and concentration to achieve the observed pre-industrial ocean DIC and approximate pH represents an oversimplification of the terrestrial carbon cycle, as demonstrated by the larger model disagreement for ocean pH and aqueous \ch{CO2} concentration. We tune the calcite saturation and concentration to match the observed total DIC, but this choice then constrains the carbonate ion concentration via the calcite saturation equation (Equation [\ref{eq:carb}]), which in turn determines the partitioning among carbonate species through equilibrium constants (Equations [\ref{eq:k1}], [\ref{eq:k2}], and [\ref{eq:bicarb}]), and ultimately fixes the pH. This cascade of constraints means that solely tuning calcite concentration to match total DIC does not guarantee agreement with other observables that depend on the same carbonate chemistry, including pH and (\ch{CO2})$_{\mathrm{aq}}$. 

The {larger} pH offset highlights a fundamental limitation of our simplified ocean chemistry model: we assume calcite saturation and do not explicitly track total alkalinity as an independent variable, such as in \citet{krissansen2021oxygen}. In a more complete carbonate system, total alkalinity and DIC jointly determine pH and carbonate speciation. By fixing calcite concentration to match DIC, we implicitly constrain alkalinity through the carbonate equilibrium, removing a degree of freedom that would otherwise allow independent adjustment of pH.

The primary distinction between our abiotic model and a biologically mediated carbon cycle lies in how carbon is packaged for gravitational transport from the surface to the seafloor. The biological pump creates heavier, rapidly-sinking aggregates (marine snow, fecal pellets, shells) from organic and inorganic carbon \citep{Passow2014}, allowing more carbon to reach the deep ocean before re-mineralization. Life effectively enhances gravitational export of carbon and influences the depth at which re-mineralization occurs. Organisms migrating vertically through the water column also transport metabolized carbon \citep{le2019pathways}, but gravitational sinking dominates in modern oceans \citep{boyd2019multi, nowicki2022quantifying, siegel2023quantifying}. In an abiotic ocean, gravitational settling of inorganic precipitates would still occur, but the size distribution, sinking velocities, and spatial patterns of carbonate formation would likely differ from the biologically-mediated system.

Furthermore, it is unclear whether abiotic calcite is less thermodynamically stable than its biogenic counterpart. \citet{zhuang2018calcite} suggest that abiotic calcite is less stable than biogenic calcite due to lower crystallinity and activation energy, but \citet{stalport2005search} found that biologically precipitated calcite thermally degrades at temperatures approximately 40$\degree$C cooler than abiotic calcite, suggesting greater thermal stability of abiotic crystals at elevated temperatures. Pessimistically, these differing material properties combined with the orders-of-magnitude slower precipitation kinetics \citep{carpenter1992srmg} imply that achieving high calcite concentrations in an abiotic ocean may require either longer equilibration times or higher degrees of supersaturation than in living systems.

The extent to which {an uninhabited} planet could accumulate and maintain the calcite concentrations necessary to operate an effective carbonate-silicate thermostat remains an open question with implications for the long-term habitability of abiotic, Earth-like exoplanets. The calcite concentration directly impacts carbonate chemistry and thus the efficiency of oceanic carbon sequestration and weathering feedbacks. Future work should incorporate a more complete carbonate system with total alkalinity as a prognostic variable, following approaches like \citet{krissansen2021oxygen}. Additionally, systematic parameter space explorations should assess the sensitivity of ocean chemistry, atmospheric CO$_2$ regulation, and surface temperature to variations in calcite concentration and alkalinity. Such studies would help establish the range of abiotic carbonate cycling rates compatible with maintaining temperate surface conditions on terrestrial planets over Gyr timescales.

\subsection{Implications for the Gaia Hypothesis}

Since its introduction by \citet{lovelock1974atmospheric}, the Gaia hypothesis has spurred long-running debates over life's role in long-term climate stability (i.e., habitability). In evaluating Lovelock's original theory, \citet{kirchner1989gaia} argues for re-categorizing Gaia into ``weak'' and ``strong'' variants -- weak Gaia argues that interactions between living and abiotic processes have some stabilizing effect on the Earth; strong Gaia asserts that such interactions are the \textit{most significant} process stabilizing Earth's climate. Strong Gaia proponents thus argue that life stabilizes the planetary thermostat by regulating temperature, atmospheric composition, and ocean chemistry to maintain conditions favorable to its survival and propagation \citep{lovelock1974atmospheric, lovelock1983gaia, lovelock2000ages, lovelock2016gaia}. 

Generally, criticisms of the Gaia hypothesis acknowledge that life alters its environment, but point out that, despite some compelling examples, life's activities are not always favorable to its long-term survival. For instance, on the modern Earth, marine plankton ``plunder'' surface ocean nutrients, creating widespread biological deserts \citep{volk2002toward, kirchner2003gaia}. Furthermore, anthropogenic climate change is a salient contemporary example of how biological activity can \textit{disrupt} rather than stabilize the climate system \citep{kirchner2002gaia, kirchner2003gaia}. These examples demonstrate that biological feedbacks are not intrinsically homeostatic.

More recently, modeling studies have investigated whether life can actively contribute to maintaining long-term planetary habitability. \citet{nicholson2018gaian} developed the ExoGaia model, which simulates how abiotic geochemistry and microbial metabolisms interact to influence surface temperature. They found that under stable conditions, microbial metabolisms can prevent planets from reaching inhospitable temperatures that would otherwise occur on their lifeless counterparts.  \citet{alcabes2020robustness} extended this work by subjecting planets to three types of global climate perturbations, finding that planets with Gaian feedbacks are most resilient to global climate change, especially when changes are gradual. Together, these ExoGaia studies suggest that if life does emerge on a planet, it may under certain conditions evolve stabilizing feedbacks that enhance the long-term habitability of the host planet.

However, these models investigate whether life \textit{can} stabilize planetary climate under a weak Gaia hypothesis, not whether life is \textit{necessary} for climate stability as in the strong Gaia hypothesis. As further evidence against a strong Gaia mechanism, \citet{kirchner1989gaia} identifies purely geophysical models that produce long-term climate stability without invoking life at all \citep{berner1983carbonate, walker1981negative}. Our results are in agreement with these models. We demonstrate that an Earth-like planet can maintain habitable surface temperatures over 4.5 Gyr of evolution (and longer) through abiotic processes alone, without requiring biological feedbacks for climate regulation. 

The carbonate-silicate cycle, driven by the interplay between silicate weathering, volcanic outgassing, and ocean chemistry, provides a robust negative feedback that keeps atmospheric \ch{CO2} and surface temperature within habitable bounds despite the Sun's luminosity increasing by approximately 30\% over 4.5 Gyr. Critically, our model demonstrates this climate stability while self-consistently evolving all major planetary subsystems: the interior cools and outgassing rates decline; the magnetic dynamo weakens and the magnetopause contracts; ocean chemistry adjusts to changing atmospheric CO$_2$; and the incoming stellar flux steadily increases. Despite these dramatic changes, the coupled system maintains surface temperatures between 285--297 K throughout the evolution. This represents a more complete test of climate stability than previous models that considered the carbon cycle in isolation, as it demonstrates that the carbonate-silicate thermostat functions robustly even when embedded within a fully evolving planetary system with realistic feedbacks between interior, surface, atmosphere, and external forcing from the brightening Sun.

As discussed above, our model does simplify ocean chemistry, which biology enhances on the modern Earth (Section \ref{sec:calcite}). However, the fundamental mechanism of climate stabilization -- the temperature and \ch{CO2} dependence of silicate weathering coupled to carbonate precipitation and subduction -- operates independently of life. While life may affect the \textit{spatial distribution} and \textit{kinetics} of carbonate precipitation, these biological enhancements may not be necessary for long-term climate stability.

That life may not be a requirement of a habitable planet has important implications for understanding Earth's habitability and for predicting conditions on rocky exoplanets. The persistence of habitable conditions over geological time may not require life to actively regulate the climate, rather potentially emerging from volatile cycling between planetary reservoirs. Habitable conditions may have been created on the early Earth by the abiotic carbonate-silicate cycle. Once life emerged, it may have developed stabilizing feedbacks as suggested by ExoGaia-type models \citep{nicholson2018gaian, alcabes2020robustness}, potentially altering weathering rates, ocean chemistry, and atmospheric composition. However, our results suggest that the underlying abiotic stabilization mechanism provided by the carbonate-silicate cycle can maintain habitability independently. While we cannot rule out that life enhances climate stability (weak Gaia), our results suggest that life {may} not be a requirement of a habitable planet.

The implication that life may not be a requirement of a habitable planet raises a complementary question: given a habitable planet, how long does it take for life to emerge? On Earth, mineralogical evidence confirms that liquid water oceans existed as early as 4.4 Ga \citep{cameron2024evidence}, while the oldest widely accepted evidence for life -- stromatolites from the Pilbara Craton -- dates to 3.5 Ga \citep{walter1980stromatolites}. Though more ancient biogenic signatures have been proposed (and continue to be highly disputed), the stromatolite evidence provides a conservative estimate that life emerged within approximately 900 Myr of ocean formation. It has been suggested that this relatively rapid emergence of life on Earth implies that life may also rapidly emerge on habitable exoplanets \citep{kipping2025strong}. Our results support this conjecture by showing how Earth could have been habitable but abiotic for billions of years, yet life still appeared relatively shortly after the appearance of habitable conditions. On the other hand, our model suggests that habitable surface conditions can persist for billions of years through purely abiotic mechanisms, potentially providing a prolonged ``prebiotic window'' during which the chemical building blocks of life could accumulate and undergo the reactions necessary for creating life. 

Potentially habitable exoplanets observed by future missions may be captured at various stages of their prebiotic evolution: some may harbor early life similar to Earth at 3.5 Ga, while others could still be accumulating prebiotic chemistry despite having maintained habitable conditions for comparable or even longer durations. Indeed, our results suggest that rocky planets with plate tectonics, liquid water oceans, and silicate weathering can maintain habitable surface conditions over billion-year timescales without requiring biology. Thus, the mere presence of habitable conditions is not sufficient evidence that a planet is inhabited. Some worlds may remain prebiotic for prolonged periods, or never give rise to life at all. Distinguishing inhabited from uninhabited-but-habitable worlds will therefore require (1) detecting disequilibrium biosignature gases \citep{lovelock1974atmospheric, meadows2018exoplanet} or surface features directly attributable to biological activity \citep{schwieterman2018exoplanet}, but also (2) a well-quantified abiotic baseline for the potential mimics of biosignature gases and surface biosignatures \citep[e.g.,][]{wogan2020abundant, krissansen2021oxygen}. 

\subsection{Comparison to Earth Evolution}
\label{sec:earthcomp}

\begin{figure*}[htbp]
    \centering
    \includegraphics[width=\linewidth]{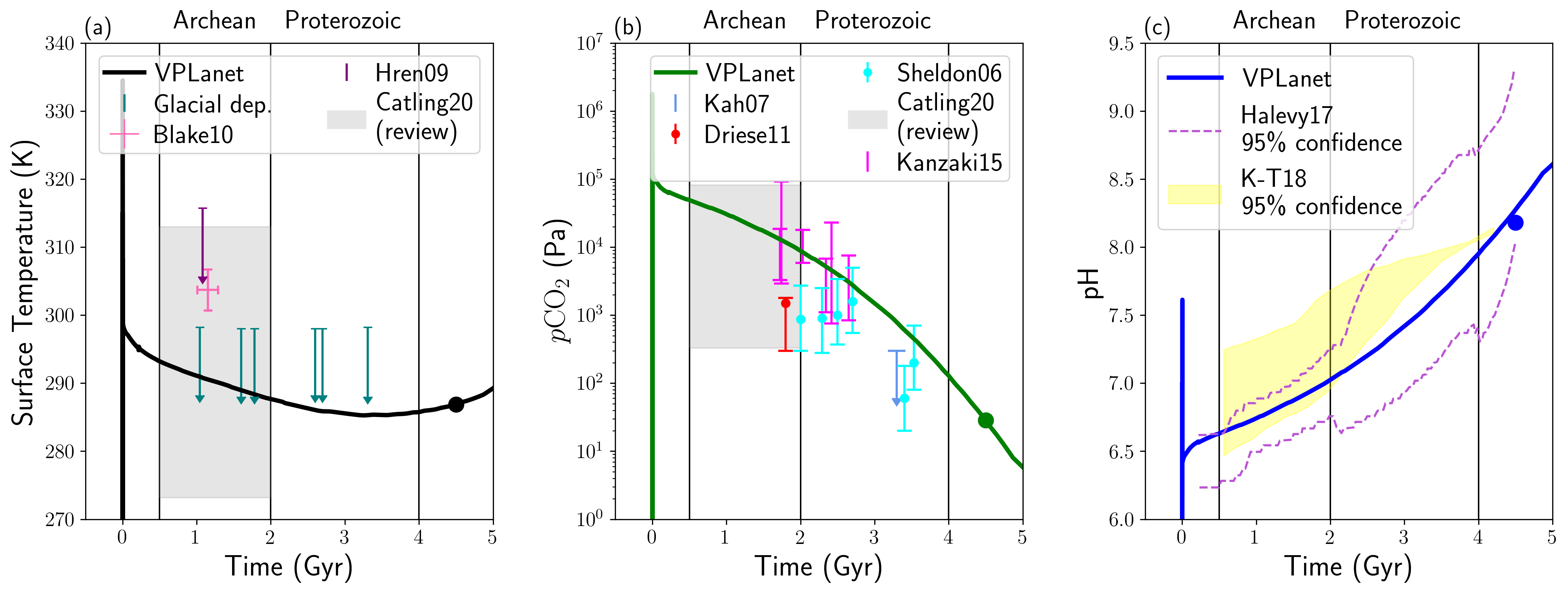}
    \caption{{Comparison of the \vplanet abiotic Earth model evolution (solid curves) against paleo-climate proxy reconstructions from the literature, following the framework of \citet{krissansen2018constraining}. Vertical black lines demarcate the Archean ($\sim$0.5--2.0~Gyr) and Proterozoic ($\sim$2.0--4.0~Gyr) eons. Filled black, green, and blue circles mark pre-industrial Earth values. \textbf{(a)} Surface temperature evolution compared with glacial deposit occurrences \citep[teal arrows;][]{geboy2013re, kuipers2013periglacial,williams2005subglacial, ojakangas2014talya, young1998earth, de20163}, $\delta^{18}$O and organic phosphate constraints \citep[Blake10;][]{blake2010phosphate}, $\delta^{18}$O and deuterium isotopes in 3.42~Ga cherts \citep[Hren09;][]{hren2009oxygen}, and the habitable temperature range reviewed in \citet[Catling20;][]{catling2020archean}. Downward arrows denote upper limits. \textbf{(b)} Atmospheric $p$\ch{CO2} evolution compared with cyanobacterial sheath calcification upper bounds \citep[Kah07;][]{kah2007mesoproterozoic}, paleosol-based estimates from \citet[Driese11;][]{driese2011neoarchean} and \citet[Sheldon06;][]{sheldon2006precambrian}, aqueous soil chemistry models \citep[Kanzaki15;][]{kanzaki2015estimates}, and the habitable range reviewed in Catling20. \textbf{(c)} Ocean pH evolution compared with 95\% confidence intervals from the carbonate-alkalinity model of \citet[Halevy17;][]{halevy2017geologic} and the coupled carbon-cycle model of \citet[K-T18;][]{krissansen2018constraining}.}}
    \label{fig:earth-through-time}
\end{figure*}

{While we did not aim to validate our results on the early Earth we can broadly compare our model predictions to measurements of global average paleo-climate conditions in the literature. Following \citet{krissansen2018constraining}, we compare the evolution of surface temperature, atmospheric $p$\ch{CO2}, and ocean pH against paleo-climate proxy data in Figure \ref{fig:earth-through-time}a, b, and c, respectively. We also qualitatively compare our interior thermal evolution to previous work. Overall, we find that our abiotic model largely reproduces paleo $p$\ch{CO2} and ocean pH estimates across much of Earth's history and yields a mantle thermal history that is consistent with most previous models, while predicting systematically cooler surface temperatures than biotic proxy reconstructions suggest.}

{Across the Archean ($\sim$0.5--2.0 Gyr) and Proterozoic ($\sim$2.0--4.0 Gyr) eons, we find that our predicted $p$\ch{CO2} remains broadly consistent with measurements established via mass balance models of weathering \citep{sheldon2006precambrian} and higher \ch{CO2} abundances predicted by \citet{kanzaki2015estimates} based on modeling of aqueous soil chemistry. Our predicted $p$\ch{CO2} is several times higher than a Precambrian upper bound estimated from calcification of cyanobacterial sheaths \citep{kah2007mesoproterozoic}. Our model also predicts global Archean surface temperatures that are consistent with the lower-end of estimates from measurements of glacial deposits \citep{geboy2013re, kuipers2013periglacial, williams2005subglacial, ojakangas2014talya, young1998earth, de20163}, but several K cooler than estimates from measurements of $\delta^{18}$O and deuterium isotopes in 3.42 Ga cherts \citep{hren2009oxygen} and $\delta^{18}$O and organic phosphate \citep{blake2010phosphate}. During the Proterozoic, our model predicts consistently cooler surface temperatures from measurements of glacial deposits. Lastly, our ocean pH throughout the evolution is consistent with  estimates from \citet{halevy2017geologic}, but is consistently more acidic than estimates from \citet{krissansen2018constraining}. The offset in pH likely reflects our simplified treatment of ocean alkalinity (see Section~\ref{sec:futurework}), while our somewhat cooler surface temperatures are in line with prior work identifying biogenic \ch{CH4} from methanogens as an important greenhouse contributor on the early Earth \citep{pavlov2000greenhouse, kharecha2005coupled}. An abiotic world would lack such elevated \ch{CH4} abundances and, under a faint young Sun, may evolve along a colder trajectory.}

{While proxy evidence gives a validation baseline for paleo-Earth climate, Earth's early interior evolution is harder to constrain. \citet{herzberg2010thermal} compiled petrologic estimates of mantle temperature over Earth history, which reveal more or less monotonic mantle cooling since the Archean.  However, there is significant uncertainty in mantle temperature and cooling rates in the Archean and Hadean, with models predicting both early heating periods \citep{korenaga2008plate, korenaga2008urey} or monotonic cooling \citep{driscoll2014thermal}.}

{Earth's noble gas record provides specific evidence that past escape may have been significant. Atmospheric xenon is strongly depleted and isotopically fractionated relative to other noble gases, and one plausible explanation is enhanced escape in Earth's past -- \citep{zahnle2019strange} vigorous hydrogen escape fueled by the young Sun's elevated XUV flux could have dragged along ionized xenon. Such vigorous atmospheric escape early in Earth's history potentially contributed to the permanent oxidation of our atmosphere \citep{catling2001biogenic}. If xenon fractionation represents continuous escape, Earth may have lost the hydrogen from at least one ocean of water across the Hadean and Archean \citep{zahnle2019strange}, which would make our assumption of a fixed volatile inventory unreasonable (Section~\ref{sec:assumptions}). However, translating the xenon record into a well-constrained escape history remains difficult. The duration and continuity of xenon escape mechanisms are uncertain \citep{zahnle2019strange}, and alternative explanations like trapping of heavy xenon isotopes in organic hazes \citep{hebrard2014coupled, avice2018evolution} and sequestration in the crust \citep{parai2018xenon}  may explain the observed fractionation without requiring enhanced atmospheric escape rates \citep{gronoff2020atmospheric}. Given these open questions, we acknowledge that an explicit treatment of atmospheric escape should be included in future work (Section~\ref{sec:futurework}).}

{A subtler tension in this comparison is that the present-day Earth system parameters against which we tune and validate
our model are themselves the product of ${\sim}4$~Gyr of coupled biological and geological evolution. In calibrating an abiotic model to reproduce these endpoint values, we are implicitly asking whether an abiotic Earth \textit{could} have arrived at a state quantitatively similar to the present-day. Our results indicate that such a solution is achievable in principle, but the broader question of what global properties depend on biotic versus abiotic processes requires future work.}

{Nonetheless, our model demonstrates that many of the present-day global geophysical conditions remain achievable along an abiotic pathway. We emphasize that this solution is not necessarily unique: other plausible evolutionary trajectories could also reproduce modern conditions, and our results likely represent one of a suite of Earth-like solutions. More broadly, our findings suggest that the presence or absence of life may meaningfully contribute to the thermal trajectory of an Earth-like planet — an abiotic Earth plausibly evolves through a substantially cooler climate history than the biotic Earth inferred from proxy data, even while converging on the same present-day state. This work is a step toward systematically mapping the broader space of both biotic and abiotic evolutionary histories consistent with present-day Earth.}

\subsection{Future Work}
\label{sec:futurework}

Our core-mantle-crust-climate model reproduces the conditions of the pre-industrial Earth without life, and provides a foundation for future studies aiming to simulate the evolving habitability of rocky exoplanets as well as the observable signatures of these states. However, we found that our simple cloud parameterization is likely insufficient to model realistic radiative forcing from clouds. We tuned a wavelength-independent cloud opacity pressure to match pre-industrial Earth surface temperature, resulting in an unrealistically low cloud fraction of 8\% compared to Earth's observed 67\% average cloud coverage \citep{wielicki1996clouds}. This discrepancy arises because we ``paint'' clouds onto the surface rather than self-consistently calculating cloud formation and radiative effects within the atmospheric column \citep{fauchez2018explicit}. Future work should reconcile this inconsistency. 

Our treatment of atmospheric processes spans a spectrum of physical fidelity. For surface temperature, atmospheric water vapor partitioning, and radiative transfer, we employ first-principles physics: \clima solves the radiative transfer equations with wavelength-dependent opacities, while water vapor is constrained by the Clausius-Clapeyron relation at each atmospheric level. By contrast, clouds are parameterized -- the cloud opacity scale and reflectance parameters are calibrated to reproduce the PIE's cloud radiative effect, and are not derived from first-principles cloud microphysics. 

While more sophisticated 1-D cloud treatments exist \citep{ackerman2001precipitating, zsom20121d, fauchez2018explicit, helling2019exoplanet, windsor2023radiative}, implementing them would largely trade our current free parameters for others (e.g., sedimentation efficiency in microphysical models), without fundamentally resolving the uncertainties inherent to representing 3-D cloud processes in 1-D globally-averaged models. Clouds remain the dominant source of uncertainty in 3-D general circulation models due to the complex interplay of microphysics, planet-wide dynamics, and radiative transfer \citep{mauritsen2012tuning, vial2013interpretation, way2018climates}. For the purposes of investigating long-term planetary evolution across diverse parameter spaces, a more pragmatic approach is to acknowledge this uncertainty explicitly and marginalize over plausible cloud parameter ranges when conducting broad explorations of Earth-like exoplanets. By varying our cloud opacity pressure and cloud albedo parameters across physically motivated ranges in future studies, we can assess how cloud radiative forcing uncertainties affect predicted habitability outcomes without requiring self-consistent calculations of cloud microphysics.

Further improvements to the model would ensure that it can capture additional feedback mechanisms of planets on diverse evolutionary paths. These improvements include incorporating additional volatile cycles and early Earth processes that affect the planet's initial conditions. Below we discuss potential model improvements and describe applications to exoplanet studies, which may include coupling our model outputs to 3-D general circulation models (GCMs) to better capture the complexity of surface processes like clouds and ice coverage.

Our model currently only handles \ch{CO2} and \ch{H2O} cycling, which is a subset of Earth's geologically active volatiles. Following \citet{krissansen2021oxygen}, we assume that \ch{N2} remains a bulk gas by setting it to a constant of 1 bar throughout the evolution. Future work should incorporate abiotic nitrogen cycling into our whole-planet model, as the abundance of atmospheric \ch{N2} affects surface pressure and impacts the surface temperature of the planet via collision-induced absorption \citep{schwieterman2015detecting} and pressure broadening of \ch{CO2} and \ch{H2O} \citep{goldblatt2009nitrogen}. Furthermore, \citet{laneuville2018earth} showed that, in an abiotic context, the abundance of atmospheric nitrogen is highly dependent on surface and interior processes, suggesting this form of volatile cycling is ripe for inclusion in a whole-planet modeling framework. 

{Future work should explore the sensitivity of long-term climate evolution to different assumptions about the efficiency of continental weathering on an abiotic planet without land vegetation. Previous habitability studies have suggested that land plants enhance continental weathering rates beyond purely abiotic levels, potentially driving abiotic planets to equilibrate at a higher $p$\ch{CO2} \citep{haqq2016limit} and enhancing the likelihood of living planets experiencing global glaciation events \citep{kadoya2019outer}. However, the quantitative contribution of biology to silicate weathering rates remains poorly constrained. Field estimates of the biotic enhancement factor span nearly two orders of magnitude \citep[e.g.,][]{dessert2003basalt, cochran1996promotion}, and no true abiotic control site exists on Earth today \citep{wild2022contribution}. Marginalizing over a range of weathering assumptions would help quantify the effect of this uncertainty and bracket the range of plausible $p$\ch{CO2} trajectories and surface temperatures for abiotic Earth-like planets, establishing more robust predictions for HWO observations.}

While low abiotic \ch{CH4} abundances likely do not directly contribute to greenhouse warming, methane nevertheless plays an important role in atmospheric chemistry and redox balance \citep{pavlov2001uv}. The atmospheric abundance and lifetime of \ch{CH4} depends on the availability of photochemically generated oxidants, which in turn depends on the planet's oxidation state, which is itself influenced by hydrogen escape to space \citep{catling2001biogenic, zahnle2019strange}. Atmospheric escape is therefore critical for determining both the feasibility of detecting abiotic methane and for interpreting \ch{CH4} observations as potential biosignatures. 

More broadly, atmospheric escape processes -- particularly XUV-driven hydrodynamic and thermal escape -- can substantially alter volatile inventories, atmospheric composition, and planetary oxidation state over Gyr timescales \citep{catling2001biogenic, luger2015extreme, gialluca2024implications}. {On Earth itself, predictive models suggest that as the Sun brightens by ${\sim}10\%$ {$\sim$1} Gyr, hydrogen escape will effectively desiccate the surface, leaving small habitable oases only at the poles \citep{kasting1988runaway, catling2009planetary}.} This possibility is especially important for planets orbiting lower mass stars like M- and K-dwarfs, which experience extended pre-main-sequence phases compared to G- and F-dwarfs, with high XUV fluxes that can drive significant atmospheric loss \citep{chadney2015xuv, atri2021stellar}. For example, on M-dwarf planets, preferential hydrogen escape oxidizes the atmosphere and potentially the mantle, affecting volcanic outgassing composition (the \ch{CH4}/\ch{CO2} ratio), ocean chemistry, and the operation of the carbonate-silicate cycle \citep{gaillard2022redox, schaefer2016predictions}.

{The planetary magnetic field also helps determine which atmospheric escape channels operate. While a magnetosphere shields the atmosphere from stellar wind sputtering and ion pickup, it simultaneously enables ion outflow through the polar caps and cusps, and models spanning a range of magnetizations find that intermediate magnetic moments can enhance, rather than suppress, total escape \citep{gunell2018intrinsic, gronoff2020atmospheric}. Furthermore, some atmospheric constituents may be more amenable to magnetic-limited escape, which would imply that certain types of atmospheres (e.g. low molecular mass) are more vulnerable to escape \citep{gronoff2020atmospheric}. Coupling such magnetization-dependent, non-thermal escape processes to the dynamo evolution predicted by interior thermal models in a whole-planet framework like ours would offer a path toward addressing this question self-consistently.}

Giant impacts during planetary formation and early bombardment can also drive atmospheric loss while simultaneously creating transient reduced atmospheres enriched in \ch{H2} and potentially \ch{CH4} \citep{zahnle2020creation}. Future work should couple atmospheric escape and impact-driven atmospheric evolution to our model to track how early volatile loss shapes long-term habitability, particularly for planets around active K-dwarfs, which will be prime targets for atmospheric characterization with HWO \citep{arney2019k, habworldstargets}. Incorporating volcanic CH$_4$ outgassing, escape-driven oxidation, and impact-driven reducing conditions will establish a complete abiotic baseline against which to interpret methane detections in exoplanet atmospheres.

Our model is initialized after the magma ocean has crystallized. However, the magma ocean phase shapes the atmospheric composition and surface temperature of the early Earth. Volatiles incompatible with the magma ocean are released into the atmosphere \citep{elkins2012magma}, and the rate at which the magma ocean cools may also contribute to whether or not a planet develops plate tectonics \citep{schaefer2018magma}. Thus, our model potentially underestimates the initial surface temperature and atmospheric volatile inventory of the planet, and does not account for how this early evolutionary process ultimately shapes Earth's habitability. 

Though this work does not couple to the magma ocean, \vplanet already includes a magma ocean model. \citet{carone2025co2} updated \magmoc, \vplanet's magma ocean module, demonstrating good agreement with previous magma ocean models \citep{elkins2008linked, nikolaou2019factors, lichtenberg2021vertically}. Future work could integrate the \magmoc module into this whole-planet modeling framework, and continuously model the planet from the magma ocean phase into the solid mantle phase \citep[e.g.,][]{krissansen2021oxygen}. 

Finally, our 1-D model is not designed to account for complicated 3-D planetary phenomena like transitions to global glaciation states. Modeling a realistic ice-albedo feedback and snowball Earth transitions likely requires latitude-dependent insolation and heat transport that 1-D globally-averaged models cannot wholly capture \citep{hoffman2017snowball, abbot2013robust, rose2017ice, wilhelm2022ice}. Similarly, substantial discrepancies exist between 1-D and 3-D models regarding the height of cloud formation, cloud optical properties, and the climate impacts of clouds \citep{yang2016differences, way2018climates, komacek2019atmospheric, helling2019exoplanet}. 

Rather than attempting to fully incorporate these 3-D phenomena into our 1-D framework, we envision a complementary modeling approach. Our evolutionary model provides realistic time-dependent boundary conditions for GCM simulations: atmospheric composition, volatile inventories, interior heat flux, and incoming stellar flux at snapshots throughout planetary evolution. GCMs can then simulate the spatial climate patterns, ice coverage, cloud distributions, and circulation for these evolved atmospheric states \citep{wolf2022exocam}.  Conversely, GCM-derived quantities -- such as a spatially-averaged albedo that includes realistic cloud distributions, ice surface fractions, and meridional heat transport efficiencies -- can inform improved parameterizations in 1-D evolutionary models without sacrificing the computational speed required to cover large parameter sweeps over Gyr-timescales. This iterative dialogue between complementary modeling approaches leverages the strengths of each: 1-D models are appropriate for long-term interior-atmosphere coupling, while 3-D models are useful for simulating spatially resolved climate states with realistic cloud and ice feedbacks. The integration of different modeling techniques will likely be essential for interpreting future observations of exoplanet atmospheres, and for predicting which planets are most likely to maintain habitable conditions over geological time.

With the additional model developments described above, a whole-planet framework could be used to assess abiotic mimics of life on Earth-like exoplanets. For example, an improved version of the model could build on the work of \citet{krissansen2021oxygen} by assessing the evolution of abiotic \ch{O2} in the atmospheres of exoplanets around Sun-like stars with a fully coupled core model. Our model would include the effect of an evolving core on the planet's magnetosphere and the efficiency of hydrogen escape, which regulates the accumulation of abiotic \ch{O2}. The improved model could also be used to investigate the abiotic generation of other biosignature gases, like \ch{CH4}. This would expand on previous investigations of abiotic mimics by considering the evolution of the core-mantle system in addition to the star and planetary atmosphere. 

In its current form, our model is well-poised for large parameter sweep explorations of Earth-like exoplanets throughout the HZs of FGK stars, both for the exploration of habitable planet properties as well as the simulation of variations in observable signatures of habitability. For this Earth validation study, we solely show our model coupled to solar evolution, but the \vplanet stellar model can simulate main sequence stars with masses up to 1.4$M_{\odot}$. This mass range encompasses the Sun-like FGK stars that HWO will eventually target for observation. 

Using preliminary HWO stellar target lists from \citet{habworldstargets} and \citet{tuchow2025hwo}, future work could simulate planets with the mass, radius, and tectonic regime of the Earth at various separations from FGK host stars. Since our coupled climate model performs full ASR and OLR calculations, it can be used to simulate the conditions at both the inner and outer edges of the HZ. At the inner edge, planets cease to be habitable when they enter the runaway greenhouse phase and at the outer edge, \ch{CO2} scattering sets the maximum greenhouse limit \citep{kopparapu2013habitable}. To explore a wide range of planets, we can vary the initial \ch{CO2} and \ch{H2O} inventories, as well as interior parameters like the radiogenics budget and the initial mantle and core temperatures. The unique luminosity evolution of each target star combined with various model parameters will shape the climate of a given planet over time, ultimately determining its potential long-term habitability. 

Importantly, our evolutionary framework allows us to generate atmospheric states and synthetic HWO observations not just at a single endpoint, but at multiple snapshots throughout a planet's history. A single Earth-like planet may exhibit dramatically different spectroscopic signatures at 1 Gyr, 3 Gyr, and 5 Gyr as its atmosphere, interior, and surface evolve in response to changing stellar forcing and volatile cycling. This application of our model could reveal how the stability of the Earth system is affected by different stellar types. For example, the comparatively slow luminosity evolution of K-dwarfs may promote long-term climate stability over the extended main sequence lifetime of these stars, while the relatively rapid brightening of F-dwarfs may challenge the climate stability of Earth-like exoplanets. The time-resolved atmospheric states across different evolutionary epochs can then be used to produce simulated HWO observations, providing insight on the range of spectral characteristics consistent with purely abiotic planetary conditions. This temporal dimension is crucial: the diversity of abiotic habitable worlds HWO will sample reflects not only variations in initial conditions and host star properties, but also the natural evolution of planetary systems captured at different ages.

\section{Conclusions}
\label{sec:conclusions}

 Here we have presented a coupled core-mantle-crust-climate model that reproduces geophysical, climate, and ocean properties of the pre-industrial Earth. Our model predicts that life {may} not {be} required to maintain a habitable planet resembling Earth-like conditions. This work thus contributes to the ongoing debate regarding the Gaia hypothesis by suggesting that life and habitability may be somewhat decoupled.

In terms of future applications, our model is intended to be applied to the evolution of rocky exoplanets. Our model shows that the evolution of the planetary interior plays an important role in determining the long-term habitability of Earth-like exoplanets. Thus, including realistically evolving planetary interiors is a crucial aspect of whole planet modeling and should account for potentially non-Earth-like compositions in order to illuminate the potential equilibrium states of rocky planets. 

Though exoplanet interiors themselves will likely never be directly observable, the effects of interior evolution are apparent in the composition of the atmosphere and the albedo of the surface environment, both of which HWO will observe on Earth-like planets around Sun-like stars via reflected light spectroscopy. We have demonstrated that atmospheric states from our model can generate realistic reflected light spectra that capture the key differences between inhabited and uninhabited worlds, establishing an abiotic baseline for biosignature interpretation. The model presented here is compatible with modeling the long-term habitability of theoretical exoplanets around key HWO targets. For quantifying the null hypothesis and establishing an abiotic baseline, additional model development is required. Nevertheless, our core-mantle-crust-ocean-atmosphere-stellar model represents significant progress towards the ambitious goal of whole-planet modeling, which itself is a key step towards advancing our search for life in the universe.

\begin{acknowledgments}
\nolinenumbers
    We thank Joshua Krissansen-Totton, 
    Jacob Lustig-Yaeger, Thomas Quinn, and two anonymous reviewers for their helpful comments, which improved this manuscript. Funding for S. G-J and A.M.M. was provided by the Exoplanet Spectroscopy Technologies Team at NASA Goddard, which is supported through the NASA Astrophysics Division Internal Scientist Funding Model. S. G-J was also supported by NASA award No. 80NSSC24M0049. R.K.B. was supported by NASA award No. 80NSSC24K0856, and R.G. was supported by NASA award No. 80NSSC23K0261. N.F.W. was supported by the NASA Postdoctoral Program.  L.C. acknowledges support by the DFG priority programme SP1833 ``Building a habitable Earth'' Grant CA 1795/3.
\end{acknowledgments}

\software{\vplanet \citep{barnes2020vplanet}, \clima \citep{wogan2025open}, \SciPy \citep{Virtanen2019scipy, SciPy2020}}

\clearpage
 
\appendix
\restartappendixnumbering 

\section{Constants}
\label{sec:constants}
\nopagebreak

\begin{longtblr}[
label = {tab:constants}, 
caption = {All constants appear here in alphabetical order, with Greek letters appearing last. Values without units are denoted as non-dimensional (n.d.).}]
{colspec = {|c|c|c|},
  rowhead = 1,
  hlines} 

Symbol & Parameter & Value \\
$a_\mathrm{grad}$ & best-fit deep ocean temperature/global mean surface temperature gradient & 1.02 [n.d.] \\
$a_1$ & saturation vapor pressure scaling parameter for silicate weathering & 0.3 [n.d.] \\
$a_2$ & plate velocity scaling parameter & 35.0 [n.d.] \\
$A_b$ & bond albedo &  0.29 [n.d.] \\
$A_i$ & ice albedo &  0.6 [n.d.] \\
$A_{oc}$ & ocean albedo &  0.1 [n.d.] \\
$A_{r}$ & rock albedo &  0.17 [n.d.] \\
$A_{\textnormal{Earth}}$ & surface area of the Earth & $5.1 \times 10^{14}$ m$^2$ \\
$b_\mathrm{int}$ & best-fit deep ocean temperature/global mean surface temperature intercept & -16.7 [n.d.] \\
$B$ & viscosity-melt reduction coefficient & 2.5 [n.d.] \\
$D_{\textnormal{H}_2 \textnormal{O}}$ & bulk \ch{H2O} distribution coefficient & 0.01 [n.d.] \\
$E_a$ & silicate weathering activation energy & $42 \times 10^3$ J/mol \\
$E_{a,\textnormal{man}}$ & mantle viscosity activation energy & $3 \times 10^5$ J/mol \\
$E_{bas}$ & basalt weathering activation energy & $92 \times 10^3$ J/mol \\
$E_\textnormal{max}$ & maximum erosion rate & $3.2 \times 10^{-10}$ m/s \\
$f_{cc}$ & fraction of Mg, Ca, K, and Na in the continental crust & $0.08$ [n.d.] \\
$f_h$ & mass fraction of water in the serpentinized layer & 0.03 [n.d.] \\
$F_{\textnormal{sfw},0}$ & present-day seafloor weathering flux & $55,454$ mol/s \\
$F_{\textnormal{weather},0}$ & present-day silicate weathering flux & $380,257$ mol/s \\
$g$ & surface gravity of the Earth & $9.8$ m/s$^2$ \\ 
$k$ & thermal conductivity & 4.2 W/m/K \\
$L$ & present-day length of ocean trenches & $6 \times 10^7$ m \\
$L_w$ & latent heat of water & $2.469 \times 10^{6}$ J/kg \\
$\bar{m}_c$ & molar mass of CO$_2$ & $44 \times 10^{-3}$ kg/mol \\
$\bar{m}_{cc}$ & average molar mass of Mg, Ca, K, and Na & $32 \times 10^{-3}$ kg/mol \\
$\bar{m}_w$ & molar mass of water & $18 \times 10^{-3}$ kg/mol \\
$M_\textnormal{man}$ & mass of the mantle & $4 \times 10^{24}$ kg \\ 
$p$\ch{CO2}$_{,0}$ & present-day partial pressure of atmospheric CO$_2$ & $33$ Pa \\
$P_{\textnormal{sat},0}$ & present-day saturation vapor pressure & $1391$ Pa \\
$P_{\textnormal{sat},\textnormal{ref}}$ & reference saturation vapor pressure & $610$ Pa \\
$R_c$ & radius of Earth's core & $3481 \times 10^3$ m \\ 
$R_g$ & universal gas constant & $8.314$ J/K/mol \\
$RH$ & relative humidity & 100\% \\
$R_\mathrm{man}$ & radius of Earth's mantle & $6371 \times 10^3$ m \\
$Ra_{\textnormal{crit}}$ & critical Rayleigh number & 660 [n.d.]\\
$S$ & salinity & 35 ppt \\
$S_0$ & solar constant & 1369 W/m$^2$  \\
$T_{\textnormal{pore},0}$ & present-day pore-space temperature & $283.03$ K \\
$T_{\textnormal{sat},\textnormal{ref}}$ & reference saturation vapor temperature & $273$ K \\
$T_{\textnormal{surf},0}$ & present-day surface temperature & $285$ K \\ 
$v_{0}$ & present-day plate speed & $1.58 \times 10^{-9}$ m/s \\
$V_{\textnormal{man}}$ & volume of Earth's mantle & $9.1 \times 10^{20}$ m$^3$ \\ 
$\nu_{\textnormal{ref}}$ & reference kinematic mantle viscosity & $4.8272 \times 10^9$ m$^2$/s \\
$x_{i,\mathrm{max}}$ & maximum ice surface fraction & 0.12 [n.d.] \\
$x_{\textnormal{l},0}$ & present-day land fraction & 0.29 [n.d.] \\
$x_{\textnormal{oc},0}$ & present-day ocean fraction & 0.71 [n.d.] \\
$\alpha$ & strength of dependence of basalt carbonation on atmospheric CO$_2$ & 0.25 [n.d.] \\
$\beta_1$ & partial pressure of CO$_2$ scaling parameter for silicate weathering & 0.55 [n.d.] \\
$\beta_2$ & scaling exponent of Rayleigh number for plate speed & 1/3 [n.d.] \\
$\gamma$ & viscosity-melt reduction coefficient & 6.0 [n.d.] \\
$\delta$ & viscosity-melt reduction exponent & 6.0 [n.d.] \\
$\Delta E_{\textnormal{H$_2$O}}$ & depression of mantle viscosity activation energy due to water & $2.22 \times 10^6$ K/wt. frac. \\
$\kappa$ & thermal diffusivity & $1 \times 10^{-6}$ m$^2$/s \\
$\xi$ & viscosity-melt reduction coefficient & $5 \times 10^{-4}$ [n.d.] \\
$\rho_{\mathrm{man}}$ & density of the mantle ($M_{\mathrm{man}}/V_{\mathrm{man}}$) & 4395 kg/m$^3$ \\
$\rho_r$ & density of regolith & 2500 kg/m$^3$\\
$\sigma_{\textrm{SB}}$ & Stefan-Boltzmann constant & $5.67 \times 10^{-8}$ W/m$^2$/K \\
$\tau_{cl}$ & cloud opacity pressure & 4050 Pa \\ 
$\chi_{\mathrm{degas,c}}$ & \ch{CO2} degassing efficiency & $0.3$ [n.d.] \\
$\chi_{\mathrm{sub,c}}$ & \ch{CO2} subduction efficiency & $0.2868$ [n.d.] \\
$\chi_{\mathrm{degas,w}}$ & water degassing efficiency & 1.0 [n.d.] \\
$\chi_{\mathrm{regas,w}}$ & water regassing efficiency & 0.15508 [n.d.] \\ 
$\Omega_{\textnormal{cal}}/[\ch{Ca^2+}]$ & marine calcite concentration & 408.1 kg/mol \\
\end{longtblr}

\section{BMO Radiogenic Heating Calculation}
\label{sec:BMOcalc}

Here we provide a detailed derivation of the effective core radiogenic heating rate, which includes contributions from an enriched basal magma ocean (BMO) layer. We calculate the $^{40}$K enrichment needed to generate up to 3.5 TW of heat in the core, which our model requires to match observed upper mantle heat flow.

\subsection{Derivation of Effective Core Heating Rate}

The core and BMO energy balances may be written as 

$$ M_C c_C \dot{T_C} = Q_{\mathrm{rad,}C} + Q_{IC} - Q_{\mathrm{CMB}},$$
\noindent and 
$$ M_{\mathrm{BMO}} c_{\mathrm{BMO}} \dot{T_B} = Q_{\mathrm{CMB}} + Q_{\mathrm{rad, BMO}} - Q_{\mathrm{BMO}},$$

\noindent respectively, where $M$ denotes mass, $c$ denotes specific heat, $\dot{T}$ denotes the time derivative of reservoir temperature, and $Q$ denotes heat flow. The subscripts indicate the relevant reservoir: $C$ and $IC$ denotes the core and inner core respectively, $CMB$ gives the core-mantle boundary, and $BMO$ is the basal magma ocean layer. Thus, $Q_{rad, BMO}$ represents the radiogenic heat flow in the BMO, and $Q_{rad,C}$ represents the radiogenic heat flow in the core. We neglect tidal heating, as it is not a significant source of internal heating for Earth \citep{ray1996detection}. The negative sign on the left-hand side represents secular cooling. Adding these equations yields

\begin{equation*}
M_C c_C \dot{T_C} + M_{\mathrm{BMO}} c_{\mathrm{BMO}} \dot{T_{\mathrm{BMO}}} = Q_{\mathrm{rad, BMO}} + Q_{\mathrm{rad,}C} + Q_{IC} - Q_{\mathrm{BMO}}.
\end{equation*}

We assume that the core and BMO cool at the same rate such that $\dot{T_C} \approx \dot{T_{\mathrm{BMO}}}$. This assumption is justified if the more viscous solid mantle regulates the cooling rate of both reservoirs. This allows us to write

\begin{equation}
\dot{T_C}(M_C c_C + M_{\mathrm{BMO}} c_{\mathrm{BMO}}) =  Q_{\mathrm{rad, BMO}} + Q_{\mathrm{rad,}C} + Q_{IC} - Q_{\mathrm{BMO}}.
\label{eq:deriv}
\end{equation}

The heat released at the inner core boundary is given by

\begin{equation}
    Q_{IC} =  \dot{T_C} \frac{dM_{IC}}{dT_C} (L_{H} + E_G),
\end{equation}
\noindent where ${dM_{IC}}/{dT_C}$ represents the change in inner core mass with temperature, and $L_{H}$ and $E_G$ are the latent and gravitational energy released per unit mass at the inner core boundary. Substituting into Equation (\ref{eq:deriv}) yields

\begin{equation}
\dot{T_C} = \frac{ (Q_{rad,BMO} + Q_{rad,C})- Q_{\mathrm{BMO}}}{(M_{\mathrm{BMO}}c_{\mathrm{BMO}} + M_cc_C) - \frac{dM_{IC}}{dT_C} (L_{H} + E_G)}.
\label{eq:deriv2}
\end{equation}

{The mass of the BMO is likely much smaller than that of the core. To show this, we can compare the density of each, finding that}

\begin{equation}
    \frac{M_{\mathrm{BMO}}}{M_C} = \left[\left(\frac{R_{\mathrm{BMO}}}{R_C}\right)^3 -1 \right] \frac{\rho_\mathrm{BMO}}{\rho_C},
\end{equation}

{\noindent where $R_{\mathrm{BMO}} = R_C + L_{\mathrm{BMO}}$ with $R_C = R_{\mathrm{CMB}} = 3481$ km the radius of the core-mantle boundary (i.e., the outer surface of the core), and $L_{\mathrm{BMO}}$ is the thickness of the BMO layer. Consistent with seismic observations of the density jump at the CMB \citep{dziewonski1981preliminary} and with high-pressure constraints showing that iron-enriched silicate melts are less dense than the metallic core \citep{labrosse2007crystallizing}, we assume that the BMO is half as dense as the core, $\rho_{\mathrm{BMO}}/\rho_C \approx 0.5$. Using this conservative estimate of the density ratio, we find that $M_{\mathrm{BMO}}$ is maximally 20\% that of the core for an initial BMO thickness of 400 km, as in \citet{labrosse2007crystallizing}. The rapidly-thinning BMO is even less massive than the core -- for example, $M_{\mathrm{BMO}}/M_C \approx 0.044$ for a 100 km-thick BMO, and $M_{\mathrm{BMO}}/M_C \approx 0.009$ for a 20 km-thick BMO layer. Thus, for the majority of the evolution $M_{\mathrm{BMO}} \ll M_C$, and} we neglect the $M_{\mathrm{BMO}} c_{\mathrm{BMO}}$ term in Equation (\ref{eq:deriv2}). Combining the radiogenic heat from the BMO and core into one term, we find the final expression given in Equation (\ref{eq:derivfinal}) in Section \ref{sec:interior}:

\begin{equation}
\dot{T_C} = \frac{ Q_{\mathrm{rad,BMO}+C} - Q_{\mathrm{BMO}}}{M_Cc_C - \frac{dM_{IC}}{dT_C} (L_{H} + E_G)}.
\end{equation}

\subsection{Required Radiogenic Enrichment}

To estimate the possible heating power generated by a radiogenically enriched BMO, {we account for heating from the four radioactive isotopes included in our model: $^{40}$K, $^{232}$Th, $^{235}$U, and $^{238}$U. We define the present-day heating contribution from each isotope as,} 

\begin{equation}
     Q_{\mathrm{rad,BMO},i,0} =  M_{\mathrm{BMO}} x_{\mathrm{BMO}} c_{i} h_{i},
     \label{eq:heatk40}
\end{equation}

\noindent {where $M_{\mathrm{BMO}}$ is the mass of the BMO, $x_{BMO}$ is the concentration of the parent species in the BMO, $c_i$ is the relative abundance of each isotope, and $h_i$ is the total radiogenic heating power per unit mass of the isotope. Table \ref{tab:radiogenic} provides the values of $x_{\mathrm{BMO}}$, $c_{i}$, and $h_{i}$ for each species. To solve for the mass of the BMO, we can write}

\begin{equation}
    M_{\mathrm{BMO}} = \rho_{\mathrm{BMO}} \frac{4\pi}{3} \left((R_{\mathrm{CMB}} + L_{\mathrm{BMO}})^3 -(R_{\mathrm{CMB}})^3\right)
\end{equation}

\noindent {where $\rho_{\mathrm{BMO}}$ is the density of the BMO, and we solve for the volume of a $L_{\mathrm{BMO}}$-thick BMO using the radius of the core mantle boundary, $R_{\mathrm{CMB}}\approx3481$ km.}

\begin{table}[htb!]
    \centering
    \begin{tabular}{llcl}
    \hline
        \textbf{Parameter} & \textbf{Species} & \textbf{Value} & \textbf{Reference}\\
    \hline
    \multirow{3}{*}{Bulk Silicate Earth (BSE) concentration} 
        & K  & 260 ppmw   & \multirow{3}{*}{\citet{mcdonough2025earths}} \\
        & Th & 77.6 ppbw  & \\
        & U  & 20.6 ppbw  & \\
    \hline
    BMO enrichment factor ($x_{\mathrm{BMO}}$) & all & $200\times$ BSE & \citet[][their Figure 4]{labrosse2007crystallizing} \\
    \hline
    \multirow{4}{*}{Isotopic abundance ($c_i$)} 
        & $^{40}$K   & $1.28 \times 10^{-4}$ & \multirow{4}{*}{\citet{korenaga2006archean}} \\
        & $^{232}$Th & 1                     & \\
        & $^{235}$U  & $7.2 \times 10^{-3}$  & \\
        & $^{238}$U  & 0.9927                & \\
    \hline
    \multirow{4}{*}{Radiogenic heat per unit mass ($h_i$, $\mu$W/kg)} 
        & $^{40}$K   & 27.9  & \multirow{4}{*}{\citet{turcotte2002geodynamics}} \\
        & $^{232}$Th & 26.9  & \\
        & $^{235}$U  & 569   & \\
        & $^{238}$U  & 93.7  & \\
    \hline
    \end{tabular}
    \caption{Physical and Earth properties used for radiogenic heating calculations in the basal magma ocean (BMO).}
    \label{tab:radiogenic}
\end{table}

{\citet{labrosse2007crystallizing} modeled a primordial BMO of $L_{\mathrm{BMO}}\sim$400 km that fractionally crystallizes to $\sim$8 km over 4.5 Gyr. However, as the BMO solidifies and thins, incompatible elements are continuously expelled into the residual liquid, such that a thinner present-day BMO is expected to be more concentrated than a thicker earlier one.} We assume a present-day BMO thickness of $L_{\mathrm{BMO}} = 4$ km, {which we interpret as the highly enriched residual of a larger primordial BMO}. Finally, we assume a BMO density of $\rho_{\mathrm{BMO}} = 6000$ kg m$^{-3}$, which is slightly denser than the solid mantle due to higher pressures. {Substituting into Equation \ref{eq:heatk40} to solve for the individual radiogenic heat contributions for each of the 4 species, we find that the present-day total $Q_{\mathrm{rad,BMO},0}=3.67$ TW for a 4 km-thick BMO.} This calculation demonstrates that attributing 3.5 TW of core heating to a radiogenically enriched, thin present-day BMO is physically plausible.

\bibliography{abioticearth}{}

@article{foley2015role,
  title={The role of plate tectonic--climate coupling and exposed land area in the development of habitable climates on rocky planets},
  author={Foley, Bradford J},
  journal={The Astrophysical Journal},
  volume={812},
  number={1},
  pages={36},
  year={2015},
  publisher={IOP Publishing}
}

@article{seales2020deep,
  title={Deep Water Cycling and the Multi-Stage Cooling of the Earth},
  author={Seales, Johnny and Lenardic, Adrian},
  journal={Geochemistry, Geophysics, Geosystems},
  volume={21},
  number={10},
  pages={e2020GC009106},
  year={2020},
  publisher={Wiley Online Library}
}

@article{driscoll2013divergent,
  title={Divergent evolution of Earth and Venus: influence of degassing, tectonics, and magnetic fields},
  author={Driscoll, P and Bercovici, D},
  journal={Icarus},
  volume={226},
  number={2},
  pages={1447--1464},
  year={2013},
  publisher={Elsevier}
}

@article{barnes2020vplanet,
  title={VPLanet: the virtual planet simulator},
  author={Barnes, Rory and Luger, Rodrigo and Deitrick, Russell and Driscoll, Peter and Quinn, Thomas R and Fleming, David P and Smotherman, Hayden and McDonald, Diego V and Wilhelm, Caitlyn and Garcia, Rodolfo and others},
  journal={Publications of the Astronomical Society of the Pacific},
  volume={132},
  number={1008},
  pages={024502},
  year={2020},
  publisher={IOP Publishing}
}

@article{schwieterman2019rethinking,
  title={Rethinking CO antibiosignatures in the search for life beyond the solar system},
  author={Schwieterman, Edward W and Reinhard, Christopher T and Olson, Stephanie L and Ozaki, Kazumi and Harman, Chester E and Hong, Peng K and Lyons, Timothy W},
  journal={The Astrophysical Journal},
  volume={874},
  number={1},
  pages={9},
  year={2019},
  publisher={IOP Publishing}
}

@article{krissansen2021oxygen,
  title={Oxygen false positives on habitable zone planets around sun-like stars},
  author={Krissansen-Totton, Joshua and Fortney, Jonathan J and Nimmo, Francis and Wogan, Nicholas},
  journal={AGU Advances},
  volume={2},
  number={2},
  pages={e2020AV000294},
  year={2021},
  publisher={Wiley Online Library}
}

@article{halevy2017geologic,
  title={The geologic history of seawater pH},
  author={Halevy, I and Bachan, A},
  journal={Science},
  volume={355},
  number={6329},
  pages={1069--1071},
  year={2017},
  publisher={American Association for the Advancement of Science}
}

@book{national2021pathways,
  title={Pathways to Discovery in Astronomy and Astrophysics for the 2020s},
  author={{National Academies of Sciences, Engineering, and Medicine}},
  publisher={The National Academies Press},
  year={2021}
}

@article{arney2016pale,
  title={The pale orange dot: the spectrum and habitability of hazy {Archean Earth}},
  author={Arney, Giada and Domagal-Goldman, Shawn D and Meadows, Victoria S and Wolf, Eric T and Schwieterman, Edward and Charnay, Benjamin and Claire, Mark and H{\'e}brard, Eric and Trainer, Melissa G},
  journal={Astrobiology},
  volume={16},
  number={11},
  pages={873--899},
  year={2016},
  publisher={Mary Ann Liebert, Inc. 140 Huguenot Street, 3rd Floor New Rochelle, NY 10801 USA}
}

@article{catling2020archean,
  title={The {Archean} atmosphere},
  author={Catling, David C and Zahnle, Kevin J},
  journal={Science advances},
  volume={6},
  number={9},
  pages={eaax1420},
  year={2020},
  publisher={American Association for the Advancement of Science}
}

@article{wogan2022rapid,
  title={Rapid timescale for an oxic transition during the Great Oxidation Event and the instability of low atmospheric {O$_2$}},
  author={Wogan, Nicholas F and Catling, David C and Zahnle, Kevin J and Claire, Mark W},
  journal={Proceedings of the National Academy of Sciences},
  volume={119},
  number={37},
  pages={e2205618119},
  year={2022},
  publisher={National Acad Sciences}
}

@article{arney2017pale,
  title={Pale orange dots: the impact of organic haze on the habitability and detectability of {Earthlike} exoplanets},
  author={Arney, Giada N and Meadows, Victoria S and Domagal-Goldman, Shawn D and Deming, Drake and Robinson, Tyler D and Tovar, Guadalupe and Wolf, Eric T and Schwieterman, Edward},
  journal={The Astrophysical Journal},
  volume={836},
  number={1},
  pages={49},
  year={2017},
  publisher={IOP Publishing}
}

@article{arney2019k,
  title={The {K dwarf} advantage for biosignatures on directly imaged exoplanets},
  author={Arney, Giada N},
  journal={The Astrophysical Journal Letters},
  volume={873},
  number={1},
  pages={L7},
  year={2019},
  publisher={IOP Publishing}
}

@article{young2023inferring,
  title={Inferring Chemical Disequilibrium Biosignatures for {Proterozoic Earth-Like} Exoplanets},
  author={Young, Amber V and Robinson, Tyler D and Krissansen-Totton, Joshua and Schwieterman, Edward W and Wogan, Nicholas F and Way, Michael J and Sohl, Linda E and Arney, Giada N and Reinhard, Christopher T and Line, Michael R and others},
  journal={arXiv preprint arXiv:2311.06083},
  year={2023}
}

@article{foley2016whole,
  title={Whole planet coupling between climate, mantle, and core: Implications for rocky planet evolution},
  author={Foley, Bradford J and Driscoll, Peter E},
  journal={Geochemistry, Geophysics, Geosystems},
  volume={17},
  number={5},
  pages={1885--1914},
  year={2016},
  publisher={Wiley Online Library}
}

@book{catling2017atmospheric,
  title={Atmospheric evolution on inhabited and lifeless worlds},
  author={Catling, David C and Kasting, James F},
  year={2017},
  publisher={Cambridge University Press}
}

@article{coogan2015alteration,
  title={Alteration of ocean crust provides a strong temperature dependent feedback on the geological carbon cycle and is a primary driver of the Sr-isotopic composition of seawater},
  author={Coogan, Laurence A and Dosso, Stan E},
  journal={Earth and Planetary Science Letters},
  volume={415},
  pages={38--46},
  year={2015},
  publisher={Elsevier}
}

@article{driscoll2014thermal,
  title={On the thermal and magnetic histories of {Earth and Venus}: Influences of melting, radioactivity, and conductivity},
  author={Driscoll, P and Bercovici, D},
  journal={Physics of the Earth and Planetary Interiors},
  volume={236},
  pages={36--51},
  year={2014},
  publisher={Elsevier}
}

@article{pavlov2001uv,
  title={{UV} shielding of {NH$_3$} and {O$_2$} by organic hazes in the {Archean} atmosphere},
  author={Pavlov, Alexander A and Brown, Lisa L and Kasting, James F},
  journal={Journal of Geophysical Research: Planets},
  volume={106},
  number={E10},
  pages={23267--23287},
  year={2001},
  publisher={Wiley Online Library}
}

@article{krissansen2018constraining,
  title={Constraining the climate and ocean {pH} of the early {Earth} with a geological carbon cycle model},
  author={Krissansen-Totton, Joshua and Arney, Giada N and Catling, David C},
  journal={Proceedings of the National Academy of Sciences},
  volume={115},
  number={16},
  pages={4105--4110},
  year={2018},
  publisher={National Acad Sciences}
}

@article{kopparapu2013habitable,
  title={Habitable zones around main-sequence stars: new estimates},
  author={Kopparapu, Ravi Kumar and Ramirez, Ramses and Kasting, James F and Eymet, Vincent and Robinson, Tyler D and Mahadevan, Suvrath and Terrien, Ryan C and Domagal-Goldman, Shawn and Meadows, Victoria and Deshpande, Rohit},
  journal={The Astrophysical Journal},
  volume={765},
  number={2},
  pages={131},
  year={2013},
  publisher={IOP Publishing}
}

@techreport{habworldstargets,
   author = {Mamajek, Eric and Stapelfeldt, Karl},
   title = {{NASA} {ExEP} Mission Star List for the {Habitable Worlds Observatory}},
   year = {2023},
   institution = {NASA Jet Propulsion Laboratory and California Institute of Technology},
}

@article{sleep2001carbon,
  title={Carbon dioxide cycling and implications for climate on ancient Earth},
  author={Sleep, Norman H and Zahnle, Kevin},
  journal={Journal of Geophysical Research: Planets},
  volume={106},
  number={E1},
  pages={1373--1399},
  year={2001},
  publisher={Wiley Online Library}
}

@article{walker1981negative,
  title={A negative feedback mechanism for the long-term stabilization of {Earth's} surface temperature},
  author={Walker, James CG and Hays, PB and Kasting, James F},
  journal={Journal of Geophysical Research: Oceans},
  volume={86},
  number={C10},
  pages={9776--9782},
  year={1981},
  publisher={Wiley Online Library}
}

@article{segura2010effect,
  title={The effect of a strong stellar flare on the atmospheric chemistry of an {Earth-like} planet orbiting an {M} dwarf},
  author={Segura, Ant{\'\i}gona and Walkowicz, Lucianne M and Meadows, Victoria and Kasting, James and Hawley, Suzanne},
  journal={Astrobiology},
  volume={10},
  number={7},
  pages={751--771},
  year={2010},
  publisher={Mary Ann Liebert, Inc. 140 Huguenot Street, 3rd Floor New Rochelle, NY 10801 USA}
}

@article{segura2005biosignatures,
  title={Biosignatures from {Earth-like} planets around {M} dwarfs},
  author={Segura, Ant{\'\i}gona and Kasting, James F and Meadows, Victoria and Cohen, Martin and Scalo, John and Crisp, David and Butler, Rebecca AH and Tinetti, Giovanna},
  journal={Astrobiology},
  volume={5},
  number={6},
  pages={706--725},
  year={2005},
  publisher={Mary Ann Liebert, Inc. 2 Madison Avenue Larchmont, NY 10538 USA}
}

@article{wolf2017constraints,
  title={Constraints on climate and habitability for Earth-like exoplanets determined from a general circulation model},
  author={Wolf, Eric T and Shields, Aomawa L and Kopparapu, Ravi K and Haqq-Misra, Jacob and Toon, Owen B},
  journal={The Astrophysical Journal},
  volume={837},
  number={2},
  pages={107},
  year={2017},
  publisher={IOP Publishing}
}

@article{rodler2014feasibility,
  title={Feasibility Studies for the Detection of {O$_2$} in an {Earth-like} Exoplanet},
  author={Rodler, Florian and L{\'o}pez-Morales, Mercedes},
  journal={The Astrophysical Journal},
  volume={781},
  number={1},
  pages={54},
  year={2014},
  publisher={IOP Publishing}
}

@article{fan2019earth,
  title={Earth as an exoplanet: A two-dimensional alien map},
  author={Fan, Siteng and Li, Cheng and Li, Jia-Zheng and Bartlett, Stuart and Jiang, Jonathan H and Natraj, Vijay and Crisp, David and Yung, Yuk L},
  journal={The Astrophysical Journal Letters},
  volume={882},
  number={1},
  pages={L1},
  year={2019},
  publisher={IOP Publishing}
}

@article{shim2014earth,
  title={Un-{Earth}-Like Interiors of {Earth}-Like Exoplanets},
  author={Shim, S-HD},
  journal={Search for Life Beyond the Solar System. Exoplanets, Biosignatures \& Instruments},
  pages={2--19},
  year={2014}
}

@article{santerne2018earth,
  title={An {Earth-sized} exoplanet with a {Mercury}-like composition},
  author={Santerne, A and Brugger, B and Armstrong, DJ and Adibekyan, V and Lillo-Box, J and Gosselin, H and Aguichine, A and Almenara, J-M and Barrado, D and Barros, SCC and others},
  journal={Nature Astronomy},
  volume={2},
  number={5},
  pages={393--400},
  year={2018},
  publisher={Nature Publishing Group UK London}
}

@article{vladilo2013habitable,
  title={The habitable zone of {Earth-like} planets with different levels of atmospheric pressure},
  author={Vladilo, Giovanni and Murante, Giuseppe and Silva, Laura and Provenzale, Antonello and Ferri, Gaia and Ragazzini, Gregorio},
  journal={The Astrophysical Journal},
  volume={767},
  number={1},
  pages={65},
  year={2013},
  publisher={IOP Publishing}
}

@article{wordsworth2014abiotic,
  title={Abiotic oxygen-dominated atmospheres on terrestrial habitable zone planets},
  author={Wordsworth, Robin and Pierrehumbert, Raymond},
  journal={The Astrophysical Journal Letters},
  volume={785},
  number={2},
  pages={L20},
  year={2014},
  publisher={IOP Publishing}
}

@article{lopez2019detecting,
  title={Detecting {Earth-like} biosignatures on rocky exoplanets around nearby stars with ground-based extremely large telescopes},
  author={L{\'o}pez-Morales, Mercedes and Currie, Thayne and Teske, Johanna and Gaidos, Eric and Kempton, Eliza and Males, Jared and Lewis, Nikole and Rackham, Benjamin V and Ben-Ami, Sagi and Birkby, Jayne and others},
  journal={arXiv preprint arXiv:1903.09523},
  year={2019}
}

@article{rugheimer2018spectra,
  title={Spectra of {Earth-like} planets through geological evolution around {FGKM} stars},
  author={Rugheimer, Sarah and Kaltenegger, Lisa},
  journal={The Astrophysical Journal},
  volume={854},
  number={1},
  pages={19},
  year={2018},
  publisher={IOP Publishing}
}

@article{herzberg2010thermal,
  title={Thermal history of the Earth and its petrological expression},
  author={Herzberg, Claude and Condie, Kent and Korenaga, Jun},
  journal={Earth and Planetary Science Letters},
  volume={292},
  number={1-2},
  pages={79--88},
  year={2010},
  publisher={Elsevier}
}

@article{wilhelm2022ice,
  title={The Ice Coverage of {Earth-like} Planets Orbiting {FGK} Stars},
  author={Wilhelm, Caitlyn and Barnes, Rory and Deitrick, Russell and Mellman, Rachel},
  journal={The Planetary Science Journal},
  volume={3},
  number={1},
  pages={13},
  year={2022},
  publisher={IOP Publishing}
}

@article{godolt20153d,
  title={{3D} climate modeling of {Earth-like} extrasolar planets orbiting different types of host stars},
  author={Godolt, M and Grenfell, JL and Hamann-Reinus, A and Kitzmann, Daniel and Kunze, M and Langematz, Ulrike and Von Paris, P and Patzer, ABC and Rauer, H and Stracke, Barbara},
  journal={Planetary and Space Science},
  volume={111},
  pages={62--76},
  year={2015},
  publisher={Elsevier}
}

@article{godolt2016assessing,
  title={Assessing the habitability of planets with {Earth-like} atmospheres with {1D} and {3D} climate modeling},
  author={Godolt, M and Grenfell, JL and Kitzmann, Daniel and Kunze, M and Langematz, U and Patzer, ABC and Rauer, H and Stracke, Barbara},
  journal={Astronomy \& Astrophysics},
  volume={592},
  pages={A36},
  year={2016},
  publisher={EDP Sciences}
}

@article{frank2014radiogenic,
  title={A radiogenic heating evolution model for cosmochemically {Earth-like} exoplanets},
  author={Frank, Elizabeth A and Meyer, Bradley S and Mojzsis, Stephen J},
  journal={Icarus},
  volume={243},
  pages={274--286},
  year={2014},
  publisher={Elsevier}
}

@article{gebauer2017evolution,
  title={Evolution of {Earth-like} extrasolar planetary atmospheres: assessing the atmospheres and biospheres of early {Earth} analog planets with a coupled atmosphere biogeochemical model},
  author={Gebauer, S and Grenfell, JL and Stock, Joachim Wolfgang and Lehmann, Ralph and Godolt, M and von Paris, Philip and Rauer, H},
  journal={Astrobiology},
  volume={17},
  number={1},
  pages={27--54},
  year={2017},
  publisher={Mary Ann Liebert, Inc. 140 Huguenot Street, 3rd Floor New Rochelle, NY 10801 USA}
}

@article{meadows2018exoplanet,
  title={Exoplanet biosignatures: understanding oxygen as a biosignature in the context of its environment},
  author={Meadows, Victoria S and Reinhard, Christopher T and Arney, Giada N and Parenteau, Mary N and Schwieterman, Edward W and Domagal-Goldman, Shawn D and Lincowski, Andrew P and Stapelfeldt, Karl R and Rauer, Heike and DasSarma, Shiladitya and others},
  journal={Astrobiology},
  volume={18},
  number={6},
  pages={630--662},
  year={2018},
  publisher={Mary Ann Liebert, Inc. 140 Huguenot Street, 3rd Floor New Rochelle, NY 10801 USA}
}

@article{kharecha2005coupled,
  title={A coupled atmosphere--ecosystem model of the early {Archean Earth}},
  author={Kharecha, P and Kasting, James and Siefert, J},
  journal={Geobiology},
  volume={3},
  number={2},
  pages={53--76},
  year={2005},
  publisher={Wiley Online Library}
}

@article{charnay2013exploring,
  title={Exploring the faint young {Sun} problem and the possible climates of the {Archean} Earth with a {3-D GCM}},
  author={Charnay, Benjamin and Forget, Fran{\c{c}}ois and Wordsworth, Robin and Leconte, J{\'e}r{\'e}my and Millour, Ehouarn and Codron, Francis and Spiga, Aymeric},
  journal={Journal of Geophysical Research: Atmospheres},
  volume={118},
  number={18},
  pages={10--414},
  year={2013},
  publisher={Wiley Online Library}
}

@article{barley2005late,
  title={Late {Archean} to Early {Paleoproterozoic} global tectonics, environmental change and the rise of atmospheric oxygen},
  author={Barley, Mark E and Bekker, Andrey and Krape{\v{z}}, Bryan},
  journal={Earth and Planetary Science Letters},
  volume={238},
  number={1-2},
  pages={156--171},
  year={2005},
  publisher={Elsevier}
}

@article{zerkle2012bistable,
  title={A bistable organic-rich atmosphere on the {Neoarchaean Earth}},
  author={Zerkle, Aubrey L and Claire, Mark W and Domagal-Goldman, Shawn D and Farquhar, James and Poulton, Simon W},
  journal={Nature Geoscience},
  volume={5},
  number={5},
  pages={359--363},
  year={2012},
  publisher={Nature Publishing Group UK London}
}

@article{schwieterman2019limited,
  title={A limited habitable zone for complex life},
  author={Schwieterman, Edward W and Reinhard, Christopher T and Olson, Stephanie L and Harman, Chester E and Lyons, Timothy W},
  journal={The Astrophysical Journal},
  volume={878},
  number={1},
  pages={19},
  year={2019},
  publisher={IOP Publishing}
}

@article{wogan2020abundant,
  title={Abundant atmospheric methane from volcanism on terrestrial planets is unlikely and strengthens the case for methane as a biosignature},
  author={Wogan, Nicholas and Krissansen-Totton, Joshua and Catling, David C},
  journal={The Planetary Science Journal},
  volume={1},
  number={3},
  pages={58},
  year={2020},
  publisher={IOP Publishing}
}

@article{baraffe2015new,
  title={New evolutionary models for pre-main sequence and main sequence low-mass stars down to the hydrogen-burning limit},
  author={Baraffe, Isabelle and Homeier, Derek and Allard, France and Chabrier, Gilles},
  journal={Astronomy \& Astrophysics},
  volume={577},
  pages={A42},
  year={2015},
  publisher={EDP Sciences}
}

@article{toon1989rapid,
  title={Rapid calculation of radiative heating rates and photodissociation rates in inhomogeneous multiple scattering atmospheres},
  author={Toon, Owen B and McKay, CP and Ackerman, TP and Santhanam, K},
  journal={Journal of Geophysical Research: Atmospheres},
  volume={94},
  number={D13},
  pages={16287--16301},
  year={1989},
  publisher={Wiley Online Library}
}

@article{wolf2022exocam,
  title={ExoCAM: a 3D climate model for exoplanet atmospheres},
  author={Wolf, Eric T and Kopparapu, Ravi and Haqq-Misra, Jacob and Fauchez, Thomas J},
  journal={The Planetary Science Journal},
  volume={3},
  number={1},
  pages={7},
  year={2022},
  publisher={IOP Publishing}
}

@article{meadows1996ground,
  title={Ground-based near-infrared observations of the Venus nightside: The thermal structure and water abundance near the surface},
  author={Meadows, Victoria S and Crisp, David},
  journal={Journal of Geophysical Research: Planets},
  volume={101},
  number={E2},
  pages={4595--4622},
  year={1996},
  publisher={Wiley Online Library}
}

@article{kaltenegger2007spectral,
  title={Spectral evolution of an Earth-like planet},
  author={Kaltenegger, Lisa and Traub, Wesley A and Jucks, Kenneth W},
  journal={The Astrophysical Journal},
  volume={658},
  number={1},
  pages={598},
  year={2007},
  publisher={IOP Publishing}
}

@book{de2010planetary,
  title={Planetary Sciences},
  author={de Pater, Imke and Lissauer, J},
  year={2010},
  publisher={Cambridge University Press,}
}

@book{pilson2012introduction,
  title={An Introduction to the Chemistry of the Sea},
  author={Pilson, Michael EQ},
  year={2012},
  publisher={cambridge university press}
}

@book{brent2013algorithms,
  title={Algorithms for minimization without derivatives},
  author={Brent, Richard P},
  year={2013},
  publisher={Courier Corporation}
}

@article{hashimoto2001predictions,
  title={Predictions of a simple cloud model for water vapor cloud albedo feedback on Venus},
  author={Hashimoto, George L and Abe, Yutaka},
  journal={Journal of Geophysical Research: Planets},
  volume={106},
  number={E7},
  pages={14675--14690},
  year={2001},
  publisher={Wiley Online Library}
}

@article{hirose2013composition,
  title={Composition and state of the core},
  author={Hirose, Kei and Labrosse, St{\'e}phane and Hernlund, John},
  journal={Annual Review of Earth and Planetary Sciences},
  volume={41},
  number={1},
  pages={657--691},
  year={2013},
  publisher={Annual Reviews}
}

@misc{jaupart2015treatise,
  title={Treatise on geophysics. Temperatures, heat and energy in the mantle of the Earth},
  author={Jaupart, C and Labrosse, S and Lucazeau, F and Mareschal, JC},
  year={2015},
  publisher={Elsevier, Oxford}
}

@article{cogne2004temporal,
  title={Temporal variation of oceanic spreading and crustal production rates during the last 180 My},
  author={Cogn{\'e}, Jean-Pascal and Humler, Eric},
  journal={Earth and Planetary Science Letters},
  volume={227},
  number={3-4},
  pages={427--439},
  year={2004},
  publisher={Elsevier}
}

@article{li2015seismic,
  title={Seismic observation of an extremely magmatic accretion at the ultraslow spreading Southwest Indian Ridge},
  author={Li, Jiabiao and Jian, Hanchao and Chen, Yongshun John and Singh, Satish C and Ruan, Aiguo and Qiu, Xuelin and Zhao, Minghui and Wang, Xianguang and Niu, Xiongwei and Ni, Jianyu and others},
  journal={Geophysical Research Letters},
  volume={42},
  number={8},
  pages={2656--2663},
  year={2015},
  publisher={Wiley Online Library}
}

@article{kinzler1992primary,
  title={Primary magmas of mid-ocean ridge basalts 1. Experiments and methods},
  author={Kinzler, Rosamond J and Grove, Timothy L},
  journal={Journal of Geophysical Research: Solid Earth},
  volume={97},
  number={B5},
  pages={6885--6906},
  year={1992},
  publisher={Wiley Online Library}
}

@article{katsura2004olivine,
  title={Olivine-wadsleyite transition in the system (Mg, Fe) 2SiO4},
  author={Katsura, Tomoo and Yamada, Hitoshi and Nishikawa, Osamu and Song, Maoshuang and Kubo, Atsushi and Shinmei, Toru and Yokoshi, Sho and Aizawa, Yoshitaka and Yoshino, Takashi and Walter, Michael J and others},
  journal={Journal of Geophysical Research: Solid Earth},
  volume={109},
  number={B2},
  year={2004},
  publisher={Wiley Online Library}
}

@article{katsura2003post,
  title={Post-spinel transition in Mg2SiO4 determined by high P--T in situ X-ray diffractometry},
  author={Katsura, Tomoo and Yamada, Hitoshi and Shinmei, Toru and Kubo, Atsushi and Ono, Shigeaki and Kanzaki, Masami and Yoneda, Akira and Walter, Michael J and Ito, Eiji and Urakawa, Satoru and others},
  journal={Physics of the Earth and Planetary Interiors},
  volume={136},
  number={1-2},
  pages={11--24},
  year={2003},
  publisher={Elsevier}
}

@article{alfe2002composition,
  title={Composition and temperature of the Earth’s core constrained by combining ab initio calculations and seismic data},
  author={Alf{\`e}, D and Gillan, MJ and Price, G David},
  journal={Earth and Planetary Science Letters},
  volume={195},
  number={1-2},
  pages={91--98},
  year={2002},
  publisher={Elsevier}
}

@article{labrosse2003thermal,
  title={Thermal and magnetic evolution of the Earth’s core},
  author={Labrosse, St{\'e}phane},
  journal={Physics of the Earth and Planetary Interiors},
  volume={140},
  number={1-3},
  pages={127--143},
  year={2003},
  publisher={Elsevier}
}

@article{paulson2005modelling,
  title={Modelling post-glacial rebound with lateral viscosity variations},
  author={Paulson, Archie and Zhong, Shijie and Wahr, John},
  journal={Geophysical Journal International},
  volume={163},
  number={1},
  pages={357--371},
  year={2005},
  publisher={Blackwell Publishing Ltd Oxford, UK}
}

@incollection{kivelson2014planetary,
  title={Planetary magnetospheres},
  author={Kivelson, Margaret Galland and Bagenal, Fran},
  booktitle={Encyclopedia of the solar system},
  pages={137--157},
  year={2014},
  publisher={Elsevier}
}

@book{united1976us,
  title={US standard atmosphere, 1976},
  author={{United States Committee on Extension to the Standard Atmosphere}},
  year={1976},
  publisher={National Oceanic and Atmospheric Administration}
}

@article{goldblatt2009nitrogen,
  title={Nitrogen-enhanced greenhouse warming on early Earth},
  author={Goldblatt, Colin and Claire, Mark W and Lenton, Timothy M and Matthews, Adrian J and Watson, Andrew J and Zahnle, Kevin J},
  journal={Nature Geoscience},
  volume={2},
  number={12},
  pages={891--896},
  year={2009},
  publisher={Nature Publishing Group UK London}
}

@article{etheridge1996natural,
  title={Natural and anthropogenic changes in atmospheric CO2 over the last 1000 years from air in Antarctic ice and firn},
  author={Etheridge, David M and Steele, LP and Langenfelds, R Ll and Francey, Roger J and Barnola, J-M and Morgan, VI},
  journal={Journal of Geophysical Research: Atmospheres},
  volume={101},
  number={D2},
  pages={4115--4128},
  year={1996},
  publisher={Wiley Online Library}
}

@article{hawkins2017estimating,
  title={Estimating changes in global temperature since the preindustrial period},
  author={Hawkins, Ed and Ortega, Pablo and Suckling, Emma and Schurer, Andrew and Hegerl, Gabi and Jones, Phil and Joshi, Manoj and Osborn, Timothy J and Masson-Delmotte, Val{\'e}rie and Mignot, Juliette and others},
  journal={Bulletin of the American Meteorological Society},
  volume={98},
  number={9},
  pages={1841--1856},
  year={2017}
}

@article{hawkins2016connecting,
  title={Connecting climate model projections of global temperature change with the real world},
  author={Hawkins, Ed and Sutton, Rowan},
  journal={Bulletin of the American Meteorological Society},
  volume={97},
  number={6},
  pages={963--980},
  year={2016}
}

@article{labrosse2007crystallizing,
  title={A crystallizing dense magma ocean at the base of the Earth’s mantle},
  author={Labrosse, St{\'e}phane and Hernlund, JW and Coltice, Nicolas},
  journal={Nature},
  volume={450},
  number={7171},
  pages={866--869},
  year={2007},
  publisher={Nature Publishing Group UK London}
}

@article{kump2008rise,
  title={The rise of atmospheric oxygen},
  author={Kump, Lee R},
  journal={Nature},
  volume={451},
  number={7176},
  pages={277--278},
  year={2008},
  publisher={Nature Publishing Group UK London}
}

@article{lyons2014rise,
  title={The rise of oxygen in {Earth’s} early ocean and atmosphere},
  author={Lyons, Timothy W and Reinhard, Christopher T and Planavsky, Noah J},
  journal={Nature},
  volume={506},
  number={7488},
  pages={307--315},
  year={2014},
  publisher={Nature Publishing Group UK London}
}

@article{tyndall1861xxiii,
  title={XXIII. On the absorption and radiation of heat by gases and vapours, and on the physical connexion of radiation, absorption, and conduction.—The bakerian lecture},
  author={Tyndall, John},
  journal={The London, Edinburgh, and Dublin Philosophical Magazine and Journal of Science},
  volume={22},
  number={146},
  pages={169--194},
  year={1861},
  publisher={Taylor \& Francis}
}

@book{fleming1998historical,
  title={Historical perspectives on climate change},
  author={Fleming, James Rodger},
  year={1998},
  publisher={Oxford University Press}
}

@article{trenberth1987global,
  title={Global atmospheric mass, surface pressure, and water vapor variations},
  author={Trenberth, Kevin E and Christy, Johan R and Olson, Jerry G},
  journal={Journal of Geophysical Research: Atmospheres},
  volume={92},
  number={D12},
  pages={14815--14826},
  year={1987},
  publisher={Wiley Online Library}
}

@article{held2000water,
  title={Water vapor feedback and global warming},
  author={Held, Isaac M and Soden, Brian J},
  journal={Annual review of energy and the environment},
  volume={25},
  number={1},
  pages={441--475},
  year={2000},
  publisher={Annual reviews 4139 El Camino Way, PO Box 10139, Palo Alto, CA 94303-0139, USA}
}

@article{mockler1995water,
  title={Water vapor in the climate system},
  author={Mockler, SB},
  journal={Special Report, American, Geophysical Union},
  year={1995}
}

@article{allan2022global,
  title={Global changes in water vapor 1979--2020},
  author={Allan, Richard P and Willett, Kate M and John, Viju O and Trent, Tim},
  journal={Journal of Geophysical Research: Atmospheres},
  volume={127},
  number={12},
  pages={e2022JD036728},
  year={2022},
  publisher={Wiley Online Library}
}

@article{patel2023increase,
  title={Increase in tropospheric water vapor amplifies global warming and climate change},
  author={Patel, Vikas Kumar and Kuttippurath, Jayanarayanan},
  journal={Ocean-Land-Atmosphere Research},
  volume={2},
  pages={0015},
  year={2023},
  publisher={AAAS}
}

@article{brewer1949evidence,
  title={Evidence for a world circulation provided by the measurements of helium and water vapour distribution in the stratosphere},
  author={Brewer, AW},
  journal={Quarterly Journal of the Royal Meteorological Society},
  volume={75},
  number={326},
  pages={351--363},
  year={1949},
  publisher={Wiley Online Library}
}

@article{randel2019diagnosing,
  title={Diagnosing observed stratospheric water vapor relationships to the cold point tropical tropopause},
  author={Randel, William and Park, Mijeong},
  journal={Journal of Geophysical Research: Atmospheres},
  volume={124},
  number={13},
  pages={7018--7033},
  year={2019},
  publisher={Wiley Online Library}
}

@article{robinson2011earth,
  title={Earth as an extrasolar planet: Earth model validation using EPOXI Earth observations},
  author={Robinson, Tyler D and Meadows, Victoria S and Crisp, David and Deming, Drake and A'hearn, Michael F and Charbonneau, David and Livengood, Timothy A and Seager, Sara and Barry, Richard K and Hearty, Thomas and others},
  journal={Astrobiology},
  volume={11},
  number={5},
  pages={393--408},
  year={2011},
  publisher={Mary Ann Liebert, Inc. 140 Huguenot Street, 3rd Floor New Rochelle, NY 10801 USA}
}

@article{lustig2023earth,
  title={Earth as a transiting exoplanet: A validation of transmission spectroscopy and atmospheric retrieval methodologies for terrestrial exoplanets},
  author={Lustig-Yaeger, Jacob and Meadows, Victoria S and Crisp, David and Line, Michael R and Robinson, Tyler D},
  journal={The Planetary Science Journal},
  volume={4},
  number={9},
  pages={170},
  year={2023},
  publisher={IOP Publishing}
}

@article{gunell2018intrinsic,
  title={Why an intrinsic magnetic field does not protect a planet against atmospheric escape},
  author={Gunell, Herbert and Maggiolo, Romain and Nilsson, Hans and Wieser, Gabriella Stenberg and Slapak, Rikard and Lindkvist, Jesper and Hamrin, Maria and De Keyser, Johan},
  journal={Astronomy \& Astrophysics},
  volume={614},
  pages={L3},
  year={2018},
  publisher={EDP Sciences}
}

@article{tian2013atmosphere,
  title={Atmosphere escape and climate evolution of terrestrial planets},
  author={Tian, Feng and Chassefi{\`e}re, Eric and Leblanc, Fran{\c{c}}ois and Brain, DA},
  journal={Comparative climatology of terrestrial planets},
  pages={chapitre--23},
  year={2013},
  publisher={University of Arizona Press}
}

@article{gronoff2020atmospheric,
  title={Atmospheric escape processes and planetary atmospheric evolution},
  author={Gronoff, Guillaume and Arras, Phil and Baraka, S and Bell, Jared M and Cessateur, Ga{\"e}l and Cohen, Ofer and Curry, Shannon M and Drake, Jeremy J and Elrod, M and Erwin, J and others},
  journal={Journal of Geophysical Research: Space Physics},
  volume={125},
  number={8},
  pages={e2019JA027639},
  year={2020},
  publisher={Wiley Online Library}
}

@article{zahnle2019strange,
  title={Strange messenger: A new history of hydrogen on Earth, as told by Xenon},
  author={Zahnle, Kevin J and Gacesa, Marko and Catling, David C},
  journal={Geochimica et Cosmochimica Acta},
  volume={244},
  pages={56--85},
  year={2019},
  publisher={Elsevier}
}

@article{kasting1988runaway,
  title={Runaway and moist greenhouse atmospheres and the evolution of Earth and Venus},
  author={Kasting, James F},
  journal={Icarus},
  volume={74},
  number={3},
  pages={472--494},
  year={1988},
  publisher={Elsevier}
}

@article{catling2009planetary,
  title={The planetary air leak},
  author={Catling, David C and Zahnle, Kevin J},
  journal={Scientific American},
  volume={300},
  number={5},
  pages={36--43},
  year={2009},
  publisher={JSTOR}
}

@article{genda2016origin,
  title={Origin of Earth’s oceans: An assessment of the total amount, history and supply of water},
  author={Genda, Hidenori},
  journal={Geochemical Journal},
  volume={50},
  number={1},
  pages={27--42},
  year={2016},
  publisher={GEOCHEMICAL SOCIETY OF JAPAN}
}

@article{charette2010volume,
  title={The volume of Earth's ocean},
  author={Charette, Matthew A and Smith, Walter HF},
  journal={Oceanography},
  volume={23},
  number={2},
  pages={112--114},
  year={2010},
  publisher={JSTOR}
}

@misc{aster_spectral_library_1999,
  author = {{Jet Propulsion Laboratory}},
  title = {{ASTER} Spectral Library},
  year = {1999},
  organization = {California Institute of Technology},
  address = {Pasadena, California},
  url = {http://speclib.jpl.nasa.gov},
  note = {Copyright 1999, California Institute of Technology}
}

@techreport{clark_usgs_spectral_2003,
  author = {Clark, R. N. and Swayze, G. A. and Wise, R. and Livo, K. E. and Hoefen, T. M. and Kokaly, R. F. and Sutley, S. J.},
  title = {{USGS} Digital Spectral Library splib05a},
  institution = {U.S. Geological Survey},
  year = {2003},
  type = {Open File Report},
  number = {03-395},
  url = {http://pubs.usgs.gov/of/2003/ofr-03-395/datatable.html}
}

@article{feely2001uptake,
  title={Uptake and storage of carbon dioxide in the ocean: The global co\~{} 2 survey},
  author={Feely, Richard A and Sabine, Christopher L and Takahashi, Taro and Wanninkhof, Rik and others},
  journal={Oceanography},
  volume={14},
  number={4},
  pages={18--32},
  year={2001},
  publisher={The Oceanographic Society; 1999}
}

@article{eide2017global,
  title={A global ocean climatology of preindustrial and modern ocean $\delta$13C},
  author={Eide, Marie and Olsen, Are and Ninnemann, Ulysses S and Johannessen, Truls},
  journal={Global Biogeochemical Cycles},
  volume={31},
  number={3},
  pages={515--534},
  year={2017},
  publisher={Wiley Online Library}
}

@article{carroll2022attribution,
  title={Attribution of space-time variability in global-ocean dissolved inorganic carbon},
  author={Carroll, Dustin and Menemenlis, Dimitris and Dutkiewicz, Stephanie and Lauderdale, Jonathan M and Adkins, Jess F and Bowman, Kevin W and Brix, Holger and Fenty, Ian and Gierach, Michelle M and Hill, Chris and others},
  journal={Global Biogeochemical Cycles},
  volume={36},
  number={3},
  pages={e2021GB007162},
  year={2022},
  publisher={Wiley Online Library}
}

@article{le2019pathways,
  title={Pathways of organic carbon downward transport by the oceanic biological carbon pump},
  author={Le Moigne, Fr{\'e}d{\'e}ric AC},
  journal={Frontiers in Marine Science},
  volume={6},
  pages={634},
  year={2019},
  publisher={Frontiers Media SA}
}

@article{o1981carbon,
  title={Carbon isotope fractionation in plants},
  author={O'Leary, Marion H},
  journal={Phytochemistry},
  volume={20},
  number={4},
  pages={553--567},
  year={1981},
  publisher={Elsevier}
}

@article{quay2003changes,
  title={Changes in the 13C/12C of dissolved inorganic carbon in the ocean as a tracer of anthropogenic CO2 uptake},
  author={Quay, P and Sonnerup, R and Westby, T and Stutsman, J and McNichol, A},
  journal={Global Biogeochemical Cycles},
  volume={17},
  number={1},
  pages={4--1},
  year={2003},
  publisher={Wiley Online Library}
}

@article{jiang2019surface,
  title={Surface ocean pH and buffer capacity: past, present and future},
  author={Jiang, Li-Qing and Carter, Brendan R and Feely, Richard A and Lauvset, Siv K and Olsen, Are},
  journal={Scientific reports},
  volume={9},
  number={1},
  pages={18624},
  year={2019},
  publisher={Nature Publishing Group UK London}
}

@article{catling2001biogenic,
  title={Biogenic methane, hydrogen escape, and the irreversible oxidation of early Earth},
  author={Catling, David C and Zahnle, Kevin J and McKay, Christopher P},
  journal={Science},
  volume={293},
  number={5531},
  pages={839--843},
  year={2001},
  publisher={American Association for the Advancement of Science}
}

@book{kargel2014global,
  title={Global land ice measurements from space},
  author={Kargel, Jeffrey S and Leonard, Gregory J and Bishop, Michael P and K{\"a}{\"a}b, Andreas and Raup, Bruce H},
  year={2014},
  publisher={Springer}
}

@article{solomon2007ipcc,
  title={IPCC fourth assessment report (AR4)},
  author={Solomon, Scott and Qin, D and Manning, M and Chen, Z and Marquis, M and Averyt, K and Tignor, M and Miller, H},
  journal={Climate change},
  volume={374},
  year={2007}
}

@article{kasting1984response,
  title={Response of Earth's atmosphere to increases in solar flux and implications for loss of water from Venus},
  author={Kasting, James F and Pollack, James B and Ackerman, Thomas P},
  journal={Icarus},
  volume={57},
  number={3},
  pages={335--355},
  year={1984},
  publisher={Elsevier}
}

@article{gough1981solar,
  title={Solar interior structure and luminosity variations},
  author={Gough, DO},
  journal={Solar Physics},
  volume={74},
  number={1},
  pages={21--34},
  year={1981},
  publisher={Springer}
}

@article{sagan1972earth,
  title={Earth and Mars: Evolution of atmospheres and surface temperatures},
  author={Sagan, Carl and Mullen, George},
  journal={Science},
  volume={177},
  number={4043},
  pages={52--56},
  year={1972},
  publisher={American Association for the Advancement of Science}
}

@article{luger2015extreme,
  title={Extreme water loss and abiotic O2 buildup on planets throughout the habitable zones of M dwarfs},
  author={Luger, Rodrigo and Barnes, Rory},
  journal={Astrobiology},
  volume={15},
  number={2},
  pages={119--143},
  year={2015},
  publisher={Mary Ann Liebert, Inc. 140 Huguenot Street, 3rd Floor New Rochelle, NY 10801 USA}
}

@article{gialluca2024implications,
  title={The Implications of Thermal Hydrodynamic Atmospheric Escape on the TRAPPIST-1 Planets},
  author={Gialluca, Megan T and Barnes, Rory and Meadows, Victoria S and Garcia, Rodolfo and Birky, Jessica and Agol, Eric},
  journal={The Planetary Science Journal},
  volume={5},
  number={6},
  pages={137},
  year={2024},
  publisher={IOP Publishing}
}

@article{newman1977implications,
  title={Implications of solar evolution for the Earth's early atmosphere},
  author={Newman, Michael J and Rood, Robert T},
  journal={Science},
  volume={198},
  number={4321},
  pages={1035--1037},
  year={1977},
  publisher={American Association for the Advancement of Science}
}

@article{wielicki1996clouds,
  title={Clouds and the Earth's Radiant Energy System (CERES): An earth observing system experiment},
  author={Wielicki, Bruce A and Barkstrom, Bruce R and Harrison, Edwin F and Lee III, Robert B and Smith, G Louis and Cooper, John E},
  journal={Bulletin of the American Meteorological Society},
  volume={77},
  number={5},
  pages={853--868},
  year={1996},
  publisher={American Meteorological Society}
}

@article{loeb2009toward,
  title={Toward optimal closure of the Earth's top-of-atmosphere radiation budget},
  author={Loeb, Norman G and Wielicki, Bruce A and Doelling, David R and Smith, G Louis and Keyes, Dennis F and Kato, Seiji and Manalo-Smith, Natividad and Wong, Takmeng},
  journal={Journal of Climate},
  volume={22},
  number={3},
  pages={748--766},
  year={2009}
}

@article{ramanathan1989cloud,
  title={Cloud-radiative forcing and climate: Results from the Earth Radiation Budget Experiment},
  author={Ramanathan, VLRD and Cess, RD and Harrison, EF and Minnis, P and Barkstrom, BR and Ahmad, E and Hartmann, D},
  journal={Science},
  volume={243},
  number={4887},
  pages={57--63},
  year={1989},
  publisher={American Association for the Advancement of Science}
}

@article{harrison1990seasonal,
  title={Seasonal variation of cloud radiative forcing derived from the Earth Radiation Budget Experiment},
  author={Harrison, Edwin F and Minnis, Patrick and Barkstrom, BR and Ramanathan, V and Cess, RD and Gibson, GG},
  journal={Journal of Geophysical Research: Atmospheres},
  volume={95},
  number={D11},
  pages={18687--18703},
  year={1990},
  publisher={Wiley Online Library}
}

@article{kasting1993habitable,
  title={Habitable zones around main sequence stars},
  author={Kasting, James F and Whitmire, Daniel P and Reynolds, Ray T},
  journal={Icarus},
  volume={101},
  number={1},
  pages={108--128},
  year={1993},
  publisher={Elsevier}
}

@article{watson1981dynamics,
  title={The dynamics of a rapidly escaping atmosphere: applications to the evolution of Earth and Venus},
  author={Watson, Andrew J and Donahue, Thomas M and Walker, James CG},
  journal={Icarus},
  volume={48},
  number={2},
  pages={150--166},
  year={1981},
  publisher={Elsevier}
}

@article{gillmann2022long,
  title={The Long-Term Evolution of the Atmosphere of Venus: Processes and Feedback Mechanisms: Interior-Exterior Exchanges},
  author={Gillmann, Cedric and Way, Michael J and Avice, Guillaume and Breuer, Doris and Golabek, Gregor J and H{\"o}ning, Dennis and Krissansen-Totton, Joshua and Lammer, Helmut and O’Rourke, Joseph G and Persson, Moa and others},
  journal={Space Science Reviews},
  volume={218},
  number={7},
  pages={56},
  year={2022},
  publisher={Springer}
}

@article{armann2012simulating,
  title={Simulating the thermochemical magmatic and tectonic evolution of Venus's mantle and lithosphere: Two-dimensional models},
  author={Armann, Marina and Tackley, Paul J},
  journal={Journal of Geophysical Research: Planets},
  volume={117},
  number={E12},
  year={2012},
  publisher={Wiley Online Library}
}

@article{condie2016great,
  title={A great thermal divergence in the mantle beginning 2.5 Ga: Geochemical constraints from greenstone basalts and komatiites},
  author={Condie, Kent C and Aster, Richard C and Van Hunen, Jeroen},
  journal={Geoscience Frontiers},
  volume={7},
  number={4},
  pages={543--553},
  year={2016},
  publisher={Elsevier}
}

@article{siegel2023quantifying,
  title={Quantifying the ocean's biological pump and its carbon cycle impacts on global scales},
  author={Siegel, David A and DeVries, Timothy and Cetini{\'c}, Ivona and Bisson, Kelsey M},
  journal={Annual review of marine science},
  volume={15},
  number={1},
  pages={329--356},
  year={2023},
  publisher={Annual Reviews}
}

@article{boyd2019multi,
  title={Multi-faceted particle pumps drive carbon sequestration in the ocean},
  author={Boyd, Philip W and Claustre, Herv{\'e} and Levy, Marina and Siegel, David A and Weber, Thomas},
  journal={Nature},
  volume={568},
  number={7752},
  pages={327--335},
  year={2019},
  publisher={Nature Publishing Group UK London}
}

@article{nowicki2022quantifying,
  title={Quantifying the carbon export and sequestration pathways of the ocean's biological carbon pump},
  author={Nowicki, Michael and DeVries, Tim and Siegel, David A},
  journal={Global Biogeochemical Cycles},
  volume={36},
  number={3},
  pages={e2021GB007083},
  year={2022},
  publisher={Wiley Online Library}
}

@article{zhuang2018calcite,
  title={Calcite precipitation induced by Bacillus cereus MRR2 cultured at different Ca2+ concentrations: Further insights into biotic and abiotic calcite},
  author={Zhuang, Dingxiang and Yan, Huaxiao and Tucker, Maurice E and Zhao, Hui and Han, Zuozhen and Zhao, Yanhong and Sun, Bin and Li, Dan and Pan, Juntong and Zhao, Yanyang and others},
  journal={Chemical Geology},
  volume={500},
  pages={64--87},
  year={2018},
  publisher={Elsevier}
}

@article{carpenter1992srmg,
  title={SrMg ratios of modern marine calcite: Empirical indicators of ocean chemistry and precipitation rate},
  author={Carpenter, Scott J and Lohmann, Kyger C},
  journal={Geochimica et Cosmochimica Acta},
  volume={56},
  number={5},
  pages={1837--1849},
  year={1992},
  publisher={Elsevier}
}

@article{stalport2005search,
  title={Search for past life on Mars: Physical and chemical characterization of minerals of biotic and abiotic origin: Part 1-calcite},
  author={Stalport, Fabien and Coll, Patrice and Cabane, Michel and Person, Alain and Gonz{\'a}lez, Rafael Navarro and Raulin, Francois and Vaulay, Marie Jo and Ausset, Patrick and McKay, Chris P and Szopa, Cyril and others},
  journal={Geophysical Research Letters},
  volume={32},
  number={23},
  year={2005},
  publisher={Wiley Online Library}
}

@article{blanchard2017solubility,
  title={The solubility of heat-producing elements in Earth’s core},
  author={Blanchard, Ingrid and Siebert, Julien and Borensztajn, Stephan and Badro, James},
  journal={Geochemical Perspectives Letters},
  volume={5},
  number={5},
  pages={1--5},
  year={2017}
}

@article{watanabe2014abundance,
  title={The abundance of potassium in the Earth’s core},
  author={Watanabe, Kosui and Ohtani, Eiji and Kamada, Seiji and Sakamaki, Tatsuya and Miyahara, Masaaki and Ito, Yoshinori},
  journal={Physics of the Earth and Planetary Interiors},
  volume={237},
  pages={65--72},
  year={2014},
  publisher={Elsevier}
}

@article{jackson2010evidence,
  title={Evidence for the survival of the oldest terrestrial mantle reservoir},
  author={Jackson, Matthew G and Carlson, Richard W and Kurz, Mark D and Kempton, Pamela D and Francis, Don and Blusztajn, Jerzy},
  journal={Nature},
  volume={466},
  number={7308},
  pages={853--856},
  year={2010},
  publisher={Nature Publishing Group UK London}
}

@article{campbell2012evidence,
  title={Evidence against a chondritic Earth},
  author={Campbell, Ian H and O’Neill, Hugh St C.},
  journal={Nature},
  volume={483},
  number={7391},
  pages={553--558},
  year={2012},
  publisher={Nature Publishing Group UK London}
}

@article{carlson2014did,
  title={How did early Earth become our modern world?},
  author={Carlson, Richard W and Garnero, Edward and Harrison, T Mark and Li, Jie and Manga, Michael and McDonough, William F and Mukhopadhyay, Sujoy and Romanowicz, Barbara and Rubie, David and Williams, Quentin and others},
  journal={Annual Review of Earth and Planetary Sciences},
  volume={42},
  number={1},
  pages={151--178},
  year={2014},
  publisher={Annual Reviews}
}

@article{Chave1970,
  author = {Chave, K. E. and Suess, E.},
  title = {Calcium carbonate saturation in seawater: effects of dissolved organic matter},
  journal = {Limnology and Oceanography},
  volume = {15},
  number = {4},
  pages = {633--637},
  year = {1970}
}

@article{Bialik2022,
  author = {Bialik, Or M. and Sisma-Ventura, Guy and Vogt-Vincent, Noam and Silverman, Jacob and Katz, Timor},
  title = {Role of oceanic abiotic carbonate precipitation in future atmospheric {CO}$_2$ regulation},
  journal = {Scientific Reports},
  volume = {12},
  pages = {15970},
  year = {2022},
  doi = {10.1038/s41598-022-20446-7}
}

@article{Grotzinger1989,
  author = {Grotzinger, John P.},
  title = {Facies and evolution of {Precambrian} carbonate depositional systems: Emergence of the modern platform archetype},
  journal = {SEPM Special Publication},
  volume = {44},
  pages = {79--106},
  year = {1989}
}

@article{Grotzinger1993,
  author = {Grotzinger, John P. and Kasting, James F.},
  title = {New constraints on {Precambrian} ocean composition},
  journal = {Journal of Geology},
  volume = {101},
  pages = {235--243},
  year = {1993},
  doi = {10.1086/648218}
}

@incollection{Grotzinger2000,
  author = {Grotzinger, John P. and James, Noel P.},
  title = {Precambrian carbonates: evolution of understanding},
  booktitle = {Carbonate Sedimentation and Diagenesis in the Evolving Precambrian World},
  editor = {Grotzinger, John P. and James, Noel P.},
  publisher = {SEPM Society for Sedimentary Geology},
  pages = {3--20},
  year = {2000}
}

@article{Sumner1996,
  author = {Sumner, Dawn Y. and Grotzinger, John P.},
  title = {Were kinetics of {Archean} calcium carbonate precipitation related to oxygen concentration?},
  journal = {Geology},
  volume = {24},
  number = {2},
  pages = {119--122},
  year = {1996},
  doi = {10.1130/0091-7613(1996)024<0119:WKOACC>2.3.CO;2}
}

@article{Passow2014,
  author = {Passow, Uta and Carlson, Craig A.},
  title = {The biological pump in a high {CO}$_2$ world},
  journal = {Marine Ecology Progress Series},
  volume = {470},
  pages = {249--271},
  year = {2012},
  doi = {10.3354/meps09985}
}

@article{lovelock1974atmospheric,
  title={Atmospheric homeostasis by and for the biosphere: the Gaia hypothesis},
  author={Lovelock, James E and Margulis, Lynn},
  journal={Tellus},
  volume={26},
  number={1-2},
  pages={2--10},
  year={1974},
  publisher={Taylor \& Francis}
}

@inproceedings{lovelock1983gaia,
  title={Gaia as seen through the atmosphere},
  author={Lovelock, James E},
  booktitle={Biomineralization and biological metal accumulation: biological and geological perspectives papers presented at the fourth international symposium on biomineralization, renesse, The Netherlands, june 2--5, 1982},
  pages={15--25},
  year={1983},
  organization={Springer}
}

@book{lovelock2016gaia,
  title={Gaia: A new look at life on earth},
  author={Lovelock, James},
  year={2016},
  publisher={Oxford University Press}
}

@book{lovelock2000ages,
  title={The ages of Gaia: A biography of our living earth},
  author={Lovelock, James},
  year={2000},
  publisher={Oxford University Press, USA}
}

@article{kirchner2002gaia,
  title={The Gaia hypothesis: fact, theory, and wishful thinking},
  author={Kirchner, James W},
  journal={Climatic change},
  volume={52},
  number={4},
  pages={391--408},
  year={2002},
  publisher={Springer}
}

@article{kirchner2003gaia,
  title={The Gaia hypothesis: conjectures and refutations},
  author={Kirchner, James W},
  journal={Climatic Change},
  volume={58},
  number={1},
  pages={21--45},
  year={2003},
  publisher={Springer}
}

@article{thompson2022case,
  title={The case and context for atmospheric methane as an exoplanet biosignature},
  author={Thompson, Maggie A and Krissansen-Totton, Joshua and Wogan, Nicholas and Telus, Myriam and Fortney, Jonathan J},
  journal={Proceedings of the National Academy of Sciences},
  volume={119},
  number={14},
  pages={e2117933119},
  year={2022},
  publisher={National Academy of Sciences}
}

@article{etiope2013abiotic,
  title={Abiotic methane on Earth},
  author={Etiope, Giuseppe and Sherwood Lollar, Barbara},
  journal={Reviews of Geophysics},
  volume={51},
  number={2},
  pages={276--299},
  year={2013},
  publisher={Wiley Online Library}
}

@article{roberson2011greenhouse,
  title={Greenhouse warming by nitrous oxide and methane in the Proterozoic Eon},
  author={Roberson, April L and Roadt, J and Halevy, I and Kasting, JF},
  journal={Geobiology},
  volume={9},
  number={4},
  pages={313--320},
  year={2011},
  publisher={Wiley Online Library}
}

@article{pavlov2000greenhouse,
  title={Greenhouse warming by CH4 in the atmosphere of early Earth},
  author={Pavlov, Alexander A and Kasting, James F and Brown, Lisa L and Rages, Kathy A and Freedman, Richard},
  journal={Journal of Geophysical Research: Planets},
  volume={105},
  number={E5},
  pages={11981--11990},
  year={2000},
  publisher={Wiley Online Library}
}

@article{cockell2014habitable,
  title={Habitable worlds with no signs of life},
  author={Cockell, Charles S},
  journal={Philosophical Transactions of the Royal Society A: Mathematical, Physical and Engineering Sciences},
  volume={372},
  number={2014},
  pages={20130082},
  year={2014},
  publisher={The Royal Society Publishing}
}

@article{berner1983carbonate,
  title={Carbonate-silicate geochemical cycle and its effect on atmospheric carbon dioxide over the past 100 million years},
  author={Berner, Robert A and Lasaga, Antonio C and Garrels, Robert M},
  journal={Am. J. Sci.;(United States)},
  volume={283},
  number={7},
  year={1983},
  publisher={Yale Univ., New Haven, CT}
}

@article{boehler1996melting,
  title={Melting temperature of the Earth's mantle and core: Earth's thermal structure},
  author={Boehler, Reinhard},
  journal={Annual Review of Earth and Planetary Sciences},
  volume={24},
  number={1},
  pages={15--40},
  year={1996},
  publisher={Annual Reviews 4139 El Camino Way, PO Box 10139, Palo Alto, CA 94303-0139, USA}
}

@article{jeanloz1986temperature,
  title={Temperature distribution in the crust and mantle},
  author={Jeanloz, Raymond and Morris, S},
  journal={IN: Annual review of earth and planetary sciences. Volume 14 (A87-13190 03-46). Palo Alto, CA, Annual Reviews, Inc., 1986, p. 377-415. NSF-NASA-supported research.},
  volume={14},
  pages={377--415},
  year={1986}
}

@article{nicholson2018gaian,
  title={Gaian bottlenecks and planetary habitability maintained by evolving model biospheres: The ExoGaia model},
  author={Nicholson, Arwen E and Wilkinson, David M and Williams, Hywel TP and Lenton, Timothy M},
  journal={Monthly Notices of the Royal Astronomical Society},
  volume={477},
  number={1},
  pages={727--740},
  year={2018},
  publisher={Oxford University Press}
}

@article{volk2002toward,
  title={Toward a future for Gaia theory},
  author={Volk, Tyler},
  journal={Climatic Change},
  volume={52},
  number={4},
  pages={423},
  year={2002},
  publisher={Springer Nature BV}
}

@article{laneuville2018earth,
  title={Earth without life: A systems model of a global abiotic nitrogen cycle},
  author={Laneuville, Matthieu and Kameya, Masafumi and Cleaves, H James},
  journal={Astrobiology},
  volume={18},
  number={7},
  pages={897--914},
  year={2018},
  publisher={Mary Ann Liebert, Inc. 140 Huguenot Street, 3rd Floor New Rochelle, NY 10801 USA}
}

@article{schwieterman2015detecting,
  title={Detecting and constraining N2 abundances in planetary atmospheres using collisional pairs},
  author={Schwieterman, Edward W and Robinson, Tyler D and Meadows, Victoria S and Misra, Amit and Domagal-Goldman, Shawn},
  journal={The Astrophysical Journal},
  volume={810},
  number={1},
  pages={57},
  year={2015},
  publisher={IOP Publishing}
}

@book{millero2005chemical,
  title={Chemical oceanography},
  author={Millero, Frank J},
  volume={30},
  year={2005},
  publisher={CRC press}
}

@article{tuchow2025hwo,
  title={HWO Target Stars and Systems: A Prioritized Community List of Potential Stellar Targets for the Habitable Worlds Observatory’s ExoEarth Survey},
  author={Tuchow, Noah W and Harada, Caleb K and Mamajek, Eric E and Tanner, Angelle and Hinkel, Natalie R and Belikov, Ruslan and Sirbu, Dan and Ciardi, David R and Stark, Christopher C and Morgan, Rhonda M and others},
  journal={Publications of the Astronomical Society of the Pacific},
  volume={137},
  number={10},
  pages={104402},
  year={2025},
  publisher={IOP Publishing}
}

@article{kirchner1989gaia,
  title={The Gaia hypothesis: Can it be tested?},
  author={Kirchner, James W},
  journal={Reviews of Geophysics},
  volume={27},
  number={2},
  pages={223--235},
  year={1989},
  publisher={Wiley Online Library}
}

@article{schwieterman2018exoplanet,
  title={Exoplanet biosignatures: a review of remotely detectable signs of life},
  author={Schwieterman, Edward W and Kiang, Nancy Y and Parenteau, Mary N and Harman, Chester E and DasSarma, Shiladitya and Fisher, Theresa M and Arney, Giada N and Hartnett, Hilairy E and Reinhard, Christopher T and Olson, Stephanie L and others},
  journal={Astrobiology},
  volume={18},
  number={6},
  pages={663--708},
  year={2018},
  publisher={Mary Ann Liebert, Inc. 140 Huguenot Street, 3rd Floor New Rochelle, NY 10801 USA}
}

@article{yang2016differences,
  title={Differences in water vapor radiative transfer among 1D models can significantly affect the inner edge of the habitable zone},
  author={Yang, Jun and Leconte, J{\'e}r{\'e}my and Wolf, Eric T and Goldblatt, Colin and Feldl, Nicole and Merlis, Timothy and Wang, Yuwei and Koll, Daniel DB and Ding, Feng and Forget, Fran{\c{c}}ois and others},
  journal={The Astrophysical Journal},
  volume={826},
  number={2},
  pages={222},
  year={2016},
  publisher={IOP Publishing}
}

@article{way2018climates,
  title={Climates of warm Earth-like planets. I. 3D model simulations},
  author={Way, Michael J and Del Genio, Anthony D and Aleinov, Igor and Clune, Thomas L and Kelley, Maxwell and Kiang, Nancy Y},
  journal={The Astrophysical Journal Supplement Series},
  volume={239},
  number={2},
  pages={24},
  year={2018},
  publisher={IOP Publishing}
}

@article{komacek2019atmospheric,
  title={The atmospheric circulation and climate of terrestrial planets orbiting Sun-like and M dwarf stars over a broad range of planetary parameters},
  author={Komacek, Thaddeus D and Abbot, Dorian S},
  journal={The Astrophysical Journal},
  volume={871},
  number={2},
  pages={245},
  year={2019},
  publisher={IOP Publishing}
}

@article{hoffman2017snowball,
  title={Snowball Earth climate dynamics and Cryogenian geology-geobiology},
  author={Hoffman, Paul F and Abbot, Dorian S and Ashkenazy, Yosef and Benn, Douglas I and Brocks, Jochen J and Cohen, Phoebe A and Cox, Grant M and Creveling, Jessica R and Donnadieu, Yannick and Erwin, Douglas H and others},
  journal={Science Advances},
  volume={3},
  number={11},
  pages={e1600983},
  year={2017},
  publisher={American Association for the Advancement of Science}
}

@article{abbot2013robust,
  title={Robust elements of Snowball Earth atmospheric circulation and oases for life},
  author={Abbot, Dorian S and Voigt, Aiko and Li, Dawei and Hir, Guillaume Le and Pierrehumbert, Raymond T and Branson, Mark and Pollard, David and B. Koll, Daniel D},
  journal={Journal of Geophysical Research: Atmospheres},
  volume={118},
  number={12},
  pages={6017--6027},
  year={2013},
  publisher={Wiley Online Library}
}

@article{rose2017ice,
  title={Ice Caps and Ice Belts: The Effects of Obliquity on Ice- Albedo Feedback},
  author={Rose, Brian EJ and Cronin, Timothy W and Bitz, Cecilia M},
  journal={The Astrophysical Journal},
  volume={846},
  number={1},
  pages={28},
  year={2017},
  publisher={IOP Publishing}
}

@article{schaefer2018magma,
  title={Magma oceans as a critical stage in the tectonic development of rocky planets},
  author={Schaefer, Laura and Elkins-Tanton, Linda T},
  journal={Philosophical Transactions of the Royal Society A: Mathematical, Physical and Engineering Sciences},
  volume={376},
  number={2132},
  pages={20180109},
  year={2018},
  publisher={The Royal Society Publishing}
}

@article{elkins2012magma,
  title={Magma oceans in the inner solar system},
  author={Elkins-Tanton, Linda T},
  journal={Annual Review of Earth and Planetary Sciences},
  volume={40},
  number={1},
  pages={113--139},
  year={2012},
  publisher={Annual Reviews}
}

@article{carone2025co2,
  title={From CO2-to H2O-dominated atmospheres and back-How mixed outgassing changes the volatile distribution in magma oceans around M dwarf stars},
  author={Carone, Ludmila and Barnes, Rory and Noack, Lena and Chubb, K and Barth, Patrick and Bitsch, Bertram and Thamm, Alexander and Balduin, Alexander and Garcia, Rodolfo and Helling, Ch},
  journal={Astronomy \& Astrophysics},
  volume={693},
  pages={A303},
  year={2025},
  publisher={EDP Sciences}
}

@article{elkins2008linked,
  title={Linked magma ocean solidification and atmospheric growth for Earth and Mars},
  author={Elkins-Tanton, Linda T},
  journal={Earth and Planetary Science Letters},
  volume={271},
  number={1-4},
  pages={181--191},
  year={2008},
  publisher={Elsevier}
}

@article{chadney2015xuv,
  title={XUV-driven mass loss from extrasolar giant planets orbiting active stars},
  author={Chadney, JM and Galand, Marina and Unruh, YC and Koskinen, TT and Sanz-Forcada, Jorge},
  journal={Icarus},
  volume={250},
  pages={357--367},
  year={2015},
  publisher={Elsevier}
}

@article{atri2021stellar,
  title={Stellar flares versus luminosity: XUV-induced atmospheric escape and planetary habitability},
  author={Atri, Dimitra and Mogan, Shane R Carberry},
  journal={Monthly Notices of the Royal Astronomical Society: Letters},
  volume={500},
  number={1},
  pages={L1--L5},
  year={2021},
  publisher={Oxford University Press}
}

@article{zahnle2020creation,
  title={Creation and evolution of impact-generated reduced atmospheres of early Earth},
  author={Zahnle, Kevin J and Lupu, Roxana and Catling, David C and Wogan, Nick},
  journal={The Planetary Science Journal},
  volume={1},
  number={1},
  pages={11},
  year={2020},
  publisher={IOP Publishing}
}

@article{gaillard2022redox,
  title={Redox controls during magma ocean degassing},
  author={Gaillard, Fabrice and Bernadou, Fabien and Roskosz, Mathieu and Bouhifd, Mohamed Ali and Marrocchi, Yves and Iacono-Marziano, Giada and Moreira, Manuel and Scaillet, Bruno and Rogerie, Gregory},
  journal={Earth and Planetary Science Letters},
  volume={577},
  pages={117255},
  year={2022},
  publisher={Elsevier}
}

@article{schaefer2016predictions,
  title={Predictions of the atmospheric composition of GJ 1132b},
  author={Schaefer, Laura and Wordsworth, Robin D and Berta-Thompson, Zachory and Sasselov, Dimitar},
  journal={The Astrophysical Journal},
  volume={829},
  number={2},
  pages={63},
  year={2016},
  publisher={IOP Publishing}
}

@article{chao2021lava,
  title={Lava worlds: From early earth to exoplanets},
  author={Chao, Keng-Hsien and deGraffenried, Rebecca and Lach, Mackenzie and Nelson, William and Truax, Kelly and Gaidos, Eric},
  journal={Geochemistry},
  volume={81},
  number={2},
  pages={125735},
  year={2021},
  publisher={Elsevier}
}

@article{lammer2018origin,
  title={Origin and evolution of the atmospheres of early Venus, Earth and Mars},
  author={Lammer, Helmut and Zerkle, Aubrey L and Gebauer, Stefanie and Tosi, Nicola and Noack, Lena and Scherf, Manuel and Pilat-Lohinger, Elke and G{\"u}del, Manuel and Grenfell, John Lee and Godolt, Mareike and others},
  journal={The Astronomy and Astrophysics Review},
  volume={26},
  number={1},
  pages={2},
  year={2018},
  publisher={Springer}
}

@article{stueeken2020mission,
  title={Mission to planet Earth: the first two billion years},
  author={Stueeken, Eva E and Som, SM and Claire, Mark and Rugheimer, Sarah and Scherf, M and Spro{\ss}, L and Tosi, Nicola and Ueno, Y and Lammer, H},
  journal={Space Science Reviews},
  volume={216},
  number={2},
  pages={31},
  year={2020},
  publisher={Springer}
}

@article{Nelder1965,
    author = {Nelder, John A. and Mead, Roger},
    title = {A Simplex Method for Function Minimization},
    journal = {The Computer Journal},
    volume = {7},
    number = {4},
    pages = {308--313},
    year = {1965},
    doi = {10.1093/comjnl/7.4.308}
}

@misc{jones2001scipy,
  title={SciPy: Open source scientific tools for Python},
  author={Jones, Eric and Oliphant, Travis and Peterson, Pearu and others},
  year={2001},
  publisher={https://www. scipy. org}
}

@techreport{bartman1980time,
  title={A time variable model of earth's albedo},
  author={Bartman, Fred L},
  year={1980},
  institution={NASA}
}

@incollection{chu2019fractional,
  title={Fractional vegetation cover},
  author={Chu, Duo},
  booktitle={Remote sensing of land use and land cover in mountain region: a comprehensive study at the central Tibetan Plateau},
  pages={195--207},
  year={2019},
  publisher={Springer}
}

@article{fauchez2018explicit,
  title={Explicit cloud representation in the Atmos 1D climate model for Earth and rocky planet applications},
  author={Fauchez, Thomas and Arney, Giada and Kopparapu, Ravi Kumar and Goldman, Shawn Domagal},
  journal={arXiv preprint arXiv:1811.10759},
  year={2018}
}

@article{farquhar2011geological,
  title={Geological constraints on the origin of oxygenic photosynthesis},
  author={Farquhar, James and Zerkle, Aubrey L and Bekker, Andrey},
  journal={Photosynthesis research},
  volume={107},
  number={1},
  pages={11--36},
  year={2011},
  publisher={Springer}
}

@article{helling2019exoplanet,
  title={Exoplanet clouds},
  author={Helling, Christiane},
  journal={Annual Review of Earth and Planetary Sciences},
  volume={47},
  number={1},
  pages={583--606},
  year={2019},
  publisher={Annual Reviews}
}

@article{zsom20121d,
  title={A 1D microphysical cloud model for Earth, and Earth-like exoplanets: Liquid water and water ice clouds in the convective troposphere},
  author={Zsom, Andras and Kaltenegger, Lisa and Goldblatt, Colin},
  journal={Icarus},
  volume={221},
  number={2},
  pages={603--616},
  year={2012},
  publisher={Elsevier}
}

@article{ackerman2001precipitating,
  title={Precipitating condensation clouds in substellar atmospheres},
  author={Ackerman, Andrew S and Marley, Mark S},
  journal={The Astrophysical Journal},
  volume={556},
  number={2},
  pages={872},
  year={2001},
  publisher={IOP Publishing}
}

@article{windsor2023radiative,
  title={A radiative-convective model for terrestrial planets with self-consistent patchy clouds},
  author={Windsor, James D and Robinson, Tyler D and kumar Kopparapu, Ravi and Young, Amber V and Trilling, David E and LLama, Joe},
  journal={The Planetary Science Journal},
  volume={4},
  number={5},
  pages={94},
  year={2023},
  publisher={IOP Publishing}
}

@ARTICLE{SciPy2020,
  author  = {Virtanen, Pauli and Gommers, Ralf and Oliphant, Travis E. and
            Haberland, Matt and Reddy, Tyler and Cournapeau, David and
            Burovski, Evgeni and Peterson, Pearu and Weckesser, Warren and
            Bright, Jonathan and {van der Walt}, St{\'e}fan J. and
            Brett, Matthew and Wilson, Joshua and Millman, K. Jarrod and
            Mayorov, Nikolay and Nelson, Andrew R. J. and Jones, Eric and
            Kern, Robert and Larson, Eric and Carey, C J and
            Polat, {\.I}lhan and Feng, Yu and Moore, Eric W. and
            {VanderPlas}, Jake and Laxalde, Denis and Perktold, Josef and
            Cimrman, Robert and Henriksen, Ian and Quintero, E. A. and
            Harris, Charles R. and Archibald, Anne M. and
            Ribeiro, Ant{\^o}nio H. and Pedregosa, Fabian and
            {van Mulbregt}, Paul and {SciPy 1.0 Contributors}},
  title   = {{{SciPy} 1.0: Fundamental Algorithms for Scientific
            Computing in Python}},
  journal = {Nature Methods},
  year    = {2020},
  volume  = {17},
  pages   = {261--272},
  adsurl  = {https://rdcu.be/b08Wh},
  doi     = {10.1038/s41592-019-0686-2},
}

@article{Virtanen2019scipy,
	Adsurl = {https://ui.adsabs.harvard.edu/abs/2019arXiv190710121V},
	Archiveprefix = {arXiv},
	Author = {{Virtanen}, Pauli and {Gommers}, Ralf and {Oliphant}, Travis E. and {Haberland}, Matt and {Reddy}, Tyler and {Cournapeau}, David and {Burovski}, Evgeni and {Peterson}, Pearu and {Weckesser}, Warren and {Bright}, Jonathan and {van der Walt}, St{\'e}fan J. and {Brett}, Matthew and {Wilson}, Joshua and {Jarrod Millman}, K. and {Mayorov}, Nikolay and {Nelson}, Andrew R.~J. and {Jones}, Eric and {Kern}, Robert and {Larson}, Eric and {Carey}, CJ and {Polat}, {\.I}lhan and {Feng}, Yu and {Moore}, Eric W. and {Vand erPlas}, Jake and {Laxalde}, Denis and {Perktold}, Josef and {Cimrman}, Robert and {Henriksen}, Ian and {Quintero}, E.~A. and {Harris}, Charles R and {Archibald}, Anne M. and {Ribeiro}, Ant{\^o}nio H. and {Pedregosa}, Fabian and {van Mulbregt}, Paul and {Contributors}, SciPy 1. 0},
	Eid = {arXiv:1907.10121},
	Eprint = {1907.10121},
	Journal = {arXiv e-prints},
	Month = {Jul},
	Pages = {arXiv:1907.10121},
	Primaryclass = {cs.MS},
	Title = {{SciPy 1.0--Fundamental Algorithms for Scientific Computing in Python}},
	Year = {2019}}

@article{alcabes2020robustness,
  title={Robustness of Gaian feedbacks to climate perturbations},
  author={Alcabes, Olivia DN and Olson, Stephanie and Abbot, Dorian S},
  journal={Monthly Notices of the Royal Astronomical Society},
  volume={492},
  number={2},
  pages={2572--2577},
  year={2020},
  publisher={Oxford University Press}
}

@article{wogan2025open,
  title={The Open-source Photochem Code: A General Chemical and Climate Model for Interpreting (Exo) Planet Observations},
  author={Wogan, Nicholas F and Batalha, Natasha E and Zahnle, Kevin and Krissansen-Totton, Joshua and Catling, David C and Wolf, Eric T and Robinson, Tyler D and Meadows, Victoria and Arney, Giada and Domagal-Goldman, Shawn},
  journal={The Planetary Science Journal},
  volume={6},
  number={11},
  pages={256},
  year={2025},
  publisher={IOP Publishing}
}

@article{vial2013interpretation,
  title={On the interpretation of inter-model spread in CMIP5 climate sensitivity estimates},
  author={Vial, Jessica and Dufresne, Jean-Louis and Bony, Sandrine},
  journal={Climate Dynamics},
  volume={41},
  number={11},
  pages={3339--3362},
  year={2013},
  publisher={Springer}
}

@article{mauritsen2012tuning,
  title={Tuning the climate of a global model},
  author={Mauritsen, Thorsten and Stevens, Bjorn and Roeckner, Erich and Crueger, Traute and Esch, Monika and Giorgetta, Marco and Haak, Helmuth and Jungclaus, Johann and Klocke, Daniel and Matei, Daniela and others},
  journal={Journal of advances in modeling Earth systems},
  volume={4},
  number={3},
  year={2012},
  publisher={Wiley Online Library}
}

@article{cameron2024evidence,
  title={Evidence for oceans pre-4300 Ma confirmed by preserved igneous compositions in Hadean zircon},
  author={Cameron, Emilia M and Blum, Tyler B and Cavosie, Aaron J and Kitajima, Kouki and Nasdala, Lutz and Orland, Ian J and Bonamici, Chloe E and Valley, John W},
  journal={American Mineralogist},
  volume={109},
  number={10},
  pages={1670--1681},
  year={2024},
  publisher={Mineralogical Society of America}
}

@article{gale2013mean,
  title={The mean composition of ocean ridge basalts},
  author={Gale, Allison and Dalton, Colleen A and Langmuir, Charles H and Su, Yongjun and Schilling, Jean-Guy},
  journal={Geochemistry, Geophysics, Geosystems},
  volume={14},
  number={3},
  pages={489--518},
  year={2013},
  publisher={Wiley Online Library}
}

@article{klein1987global,
  title={Global correlations of ocean ridge basalt chemistry with axial depth and crustal thickness},
  author={Klein, Emily M and Langmuir, Charles H},
  journal={Journal of Geophysical Research: Solid Earth},
  volume={92},
  number={B8},
  pages={8089--8115},
  year={1987},
  publisher={Wiley Online Library}
}

@article{driscoll2015tidal,
  title={Tidal heating of Earth-like exoplanets around M stars: thermal, magnetic, and orbital evolutions},
  author={Driscoll, Peter E and Barnes, Rory},
  journal={Astrobiology},
  volume={15},
  number={9},
  pages={739--760},
  year={2015},
  publisher={Mary Ann Liebert, Inc. 140 Huguenot Street, 3rd Floor New Rochelle, NY 10801 USA}
}

@article{nimmo2002does,
  title={Why does Venus lack a magnetic field?},
  author={Nimmo, Francis},
  journal={Geology},
  volume={30},
  number={11},
  pages={987--990},
  year={2002},
  publisher={Geological Society of America}
}

@article{boukare2025solidification,
  title={Solidification of Earth’s mantle led inevitably to a basal magma ocean},
  author={Boukar{\'e}, Charles-{\'E}douard and Badro, James and Samuel, Henri},
  journal={Nature},
  pages={1--6},
  year={2025},
  publisher={Nature Publishing Group UK London}
}

@article{lichtenberg2021vertically,
  title={Vertically resolved magma ocean--protoatmosphere evolution: H2, H2O, CO2, CH4, CO, O2, and N2 as primary absorbers},
  author={Lichtenberg, Tim and Bower, Dan J and Hammond, Mark and Boukrouche, Ryan and Sanan, Patrick and Tsai, Shang-Min and Pierrehumbert, Raymond T},
  journal={Journal of Geophysical Research: Planets},
  volume={126},
  number={2},
  pages={e2020JE006711},
  year={2021},
  publisher={Wiley Online Library}
}

@article{nikolaou2019factors,
  title={What factors affect the duration and outgassing of the terrestrial magma ocean?},
  author={Nikolaou, Athanasia and Katyal, Nisha and Tosi, Nicola and Godolt, Mareike and Grenfell, John Lee and Rauer, Heike},
  journal={The Astrophysical Journal},
  volume={875},
  number={1},
  pages={11},
  year={2019},
  publisher={IOP Publishing}
}

@article{lincowski2018evolved,
  title={Evolved climates and observational discriminants for the TRAPPIST-1 planetary system},
  author={Lincowski, Andrew P and Meadows, Victoria S and Crisp, David and Robinson, Tyler D and Luger, Rodrigo and Lustig-Yaeger, Jacob and Arney, Giada N},
  journal={The Astrophysical Journal},
  volume={867},
  number={1},
  pages={76},
  year={2018},
  publisher={IOP Publishing}
}

@article{batalha2019exoplanet,
  title={Exoplanet reflected-light spectroscopy with PICASO},
  author={Batalha, Natasha E and Marley, Mark S and Lewis, Nikole K and Fortney, Jonathan J},
  journal={The Astrophysical Journal},
  volume={878},
  number={1},
  pages={70},
  year={2019},
  publisher={IOP Publishing}
}

@misc{Opacities2025,
  author       = {Wogan, Nicholas and
                  Lopez, Jaden and
                  Batalha, Natasha and
                  Fortney, Jonathan},
  title        = {Resampled Opacity Databases for PICASO from the Photochem model},
  month        = oct,
  year         = 2025,
  publisher    = {Zenodo},
  version      = {v1.0.0},
  doi          = {10.5281/zenodo.17381172},
  url          = {https://doi.org/10.5281/zenodo.17381172},
}

@article{shaik2025advanced,
  title={Advanced deep learning technique for estimating global surface ocean calcium carbonate saturation ($\Omega$cal)},
  author={Shaik, Ibrahim and Nagamani, PV and Yadav, Sandesh and Manmode, Yash and Rao, G Srinivasa},
  journal={Marine Chemistry},
  volume={268},
  pages={104483},
  year={2025},
  publisher={Elsevier}
}

@article{nakagawa2012influence,
  title={Influence of magmatism on mantle cooling, surface heat flow and Urey ratio},
  author={Nakagawa, Takashi and Tackley, Paul J},
  journal={Earth and Planetary Science Letters},
  volume={329},
  pages={1--10},
  year={2012},
  publisher={Elsevier}
}

@misc{feinberg2026habitableworldsobservatorysconcept,
      title={Habitable Worlds Observatory's Concept and Technology Maturation: Initial Feasibility and Trade Space Exploration}, 
      author={Lee D. Feinberg and Breann N. Sitarski and Michael W. McElwain and Giada Arney and Caleb Baker and Matthew R. Bolcar and Marie Levine and Alice Liu and Bertrand Mennesson and Aki Roberge and J. Scott Smith and Feng Zhao and John Ziemer},
      year={2026},
      eprint={2601.11803},
      archivePrefix={arXiv},
      primaryClass={astro-ph.IM},
      url={https://arxiv.org/abs/2601.11803}, 
}

@article{rothman2010hitemp,
  title={HITEMP, the high-temperature molecular spectroscopic database},
  author={Rothman, Laurence S and Gordon, IE and Barber, RJ and Dothe, H and Gamache, Robert R and Goldman, A and Perevalov, VI and Tashkun, SA and Tennyson, J},
  journal={Journal of Quantitative Spectroscopy and Radiative Transfer},
  volume={111},
  number={15},
  pages={2139--2150},
  year={2010},
  publisher={Elsevier}
}

@article{gordon2017hitran2016,
  title={The HITRAN2016 molecular spectroscopic database},
  author={Gordon, Iouli E and Rothman, Laurence S and Hill, Christian and Kochanov, Roman V and Tan, Y and Bernath, Peter F and Birk, Manfred and Boudon, V and Campargue, Alain and Chance, KV and others},
  journal={Journal of quantitative spectroscopy and radiative transfer},
  volume={203},
  pages={3--69},
  year={2017},
  publisher={Elsevier}
}

@article{ray1996detection,
  title={Detection of tidal dissipation in the solid Earth by satellite tracking and altimetry},
  author={Ray, Richard D and Eanes, RJ and Chao, Benjamin F},
  journal={Nature},
  volume={381},
  number={6583},
  pages={595--597},
  year={1996},
  publisher={Nature Publishing Group UK London}
}

@article{walter1980stromatolites,
  title={Stromatolites 3,400--3,500 Myr old from the North pole area, Western Australia},
  author={Walter, MR and Buick, Roger and Dunlop, JSR},
  journal={Nature},
  volume={284},
  number={5755},
  pages={443--445},
  year={1980},
  publisher={Nature Publishing Group UK London}
}

@article{kipping2025strong,
  title={Strong Evidence that Abiogenesis Is a Rapid Process on Earth Analogs},
  author={Kipping, David},
  journal={Astrobiology},
  volume={25},
  number={5},
  pages={323--326},
  year={2025},
  publisher={Mary Ann Liebert, Inc., publishers 140 Huguenot Street, 3rd Floor New~…}
}

@article{constantinou2025comparative,
  title={Comparative Biosignatures},
  author={Constantinou, Tereza and Shorttle, Oliver and Cranmer, Miles and Rimmer, Paul B},
  journal={arXiv preprint arXiv:2505.01512},
  year={2025}
}

@article{garcia2026venus, 
title={Investigation of Venus' thermal history, crustal evolution, and core dynamics with a coupled interior-lithosphere-atmosphere model},
  author={Garcia, Rodolfo and Barnes, Rory and Driscoll, Peter E and Meadows, Victoria S and Gialluca, Megan},
  journal={arXiv preprint arXiv:2603.18070},
  year={2026}
}

@article{costa2009model,
  title={A model for the rheology of particle-bearing suspensions and partially molten rocks},
  author={Costa, Antonio and Caricchi, Luca and Bagdassarov, Nickolai},
  journal={Geochemistry, Geophysics, Geosystems},
  volume={10},
  number={3},
  year={2009},
  publisher={Wiley Online Library}
}

@article{griessmeier2004effect,
  title={The effect of tidal locking on the magnetospheric and atmospheric evolution of “Hot Jupiters”},
  author={Grie{\ss}meier, J-M and Stadelmann, A and Penz, T and Lammer, H and Selsis, F and Ribas, I and Guinan, EF and Motschmann, U and Biernat, HK and Weiss, WW},
  journal={Astronomy \& Astrophysics},
  volume={425},
  number={2},
  pages={753--762},
  year={2004},
  publisher={EDP Sciences}
}

@article{shue1998magnetopause,
  title={Magnetopause location under extreme solar wind conditions},
  author={Shue, J-H and Song, P and Russell, CT and Steinberg, JT and Chao, JK and Zastenker, G and Vaisberg, OL and Kokubun, S and Singer, HJ and Detman, TR and others},
  journal={Journal of Geophysical Research: Space Physics},
  volume={103},
  number={A8},
  pages={17691--17700},
  year={1998},
  publisher={Wiley Online Library}
}

@article{maruyama2018initiation,
  title={Initiation of plate tectonics in the Hadean: Eclogitization triggered by the ABEL Bombardment},
  author={Maruyama, Shigenori and Santosh, M and Azuma, Shintaro},
  journal={Geoscience Frontiers},
  volume={9},
  number={4},
  pages={1033--1048},
  year={2018},
  publisher={Elsevier}
}

@article{al2024coupled,
  title={Coupled fates of Earth’s mantle and core: Early sluggish-lid tectonics and a long-lived geodynamo},
  author={Al Asad, Manar and Lau, Harriet CP},
  journal={Science Advances},
  volume={10},
  number={31},
  pages={eadp1991},
  year={2024},
  publisher={American Association for the Advancement of Science}
}

@article{korenaga2006archean,
  title={Archean geodynamics and the thermal evolution of Earth},
  author={Korenaga, Jun},
  journal={Geophysical Monograph-American Geophysical Union},
  volume={164},
  pages={7},
  year={2006},
  publisher={AGU AMERICAN GEOPHYSICAL UNION}
}

@article{west2012thickness,
  title={Thickness of the chemical weathering zone and implications for erosional and climatic drivers of weathering and for carbon-cycle feedbacks},
  author={West, A Joshua},
  journal={Geology},
  volume={40},
  number={9},
  pages={811--814},
  year={2012},
  publisher={Geological Society of America}
}

@incollection{robinson2018characterizing,
  title={Characterizing exoplanet habitability},
  author={Robinson, Tyler D},
  booktitle={Handbook of Exoplanets},
  pages={3137--3157},
  year={2018},
  publisher={Springer International Publishing}
}

@article{dziewonski1981preliminary,
  title={Preliminary reference Earth model},
  author={Dziewonski, Adam M and Anderson, Don L},
  journal={Physics of the earth and planetary interiors},
  volume={25},
  number={4},
  pages={297--356},
  year={1981},
  publisher={Elsevier}
}

@article{dessert2003basalt,
  title={Basalt weathering laws and the impact of basalt weathering on the global carbon cycle},
  author={Dessert, C{\'e}line and Dupr{\'e}, Bernard and Gaillardet, J{\'e}r{\^o}me and Fran{\c{c}}ois, Louis M and All{\`e}gre, Claude J},
  journal={Chemical Geology},
  volume={202},
  number={3-4},
  pages={257--273},
  year={2003},
  publisher={Elsevier}
}

@article{wild2022contribution,
  title={The contribution of living organisms to rock weathering in the critical zone},
  author={Wild, Bastien and Gerrits, Ruben and Bonneville, Steeve},
  journal={npj Materials degradation},
  volume={6},
  number={1},
  pages={98},
  year={2022},
  publisher={Nature Publishing Group UK London}
}

@article{cochran1996promotion,
  title={Promotion of chemical weathering by higher plants: field observations on Hawaiian basalts},
  author={Cochran, M Ford and Berner, Robert A},
  journal={Chemical Geology},
  volume={132},
  number={1-4},
  pages={71--77},
  year={1996},
  publisher={Elsevier}
}

@article{kadoya2019outer,
  title={Outer limits of the habitable zones in terms of climate mode and climate evolution of earth-like planets},
  author={Kadoya, Shintaro and Tajika, E},
  journal={The Astrophysical Journal},
  volume={875},
  number={1},
  pages={7},
  year={2019},
  publisher={The American Astronomical Society}
}

@article{haqq2016limit,
  title={Limit cycles can reduce the width of the habitable zone},
  author={Haqq-Misra, Jacob and Kopparapu, Ravi Kumar and Batalha, Natasha E and Harman, Chester E and Kasting, James F},
  journal={The Astrophysical Journal},
  volume={827},
  number={2},
  pages={120},
  year={2016},
  publisher={The American Astronomical Society}
}

@article{mcdonough2025earths,
  title={Earths composition: origin, evolution and energy budget},
  author={McDonough, William F},
  journal={arXiv preprint arXiv:2505.02641},
  year={2025}
}

@book{turcotte2002geodynamics,
  title={Geodynamics},
  author={Turcotte, Donald L and Schubert, Gerald},
  year={2002},
  publisher={Cambridge university press}
}

@article{korenaga2008plate,
  title={Plate tectonics, flood basalts and the evolution of Earth’s oceans},
  author={Korenaga, Jun},
  journal={Terra Nova},
  volume={20},
  number={6},
  pages={419--439},
  year={2008},
  publisher={Wiley Online Library}
}

@article{korenaga2008urey,
  title={Urey ratio and the structure and evolution of Earth's mantle},
  author={Korenaga, Jun},
  journal={Reviews of Geophysics},
  volume={46},
  number={2},
  year={2008},
  publisher={Wiley Online Library}
}

@article{blake2010phosphate,
  title={Phosphate oxygen isotopic evidence for a temperate and biologically active Archaean ocean},
  author={Blake, Ruth E and Chang, Sae Jung and Lepland, Aivo},
  journal={Nature},
  volume={464},
  number={7291},
  pages={1029--1032},
  year={2010},
  publisher={Nature Publishing Group UK London}
}

@article{hren2009oxygen,
  title={Oxygen and hydrogen isotope evidence for a temperate climate 3.42 billion years ago},
  author={Hren, MT and Tice, MM and Chamberlain, CP},
  journal={Nature},
  volume={462},
  number={7270},
  pages={205--208},
  year={2009},
  publisher={Nature Publishing Group UK London}
}

@article{geboy2013re,
  title={Re--Os age constraints and new observations of Proterozoic glacial deposits in the Vazante Group, Brazil},
  author={Geboy, Nicholas J and Kaufman, Alan J and Walker, Richard J and Misi, Aroldo and de Oliviera, Tolentino Flavio and Miller, Kristen E and Azmy, Karem and Kendall, Brian and Poulton, Simon W},
  journal={Precambrian Research},
  volume={238},
  pages={199--213},
  year={2013},
  publisher={Elsevier}
}

@article{kuipers2013periglacial,
  title={Periglacial evidence for a 1.91--1.89 Ga old glacial period at low latitude, Central Sweden},
  author={Kuipers, Gerrit and Beunk, Frank F and van der Wateren, Frederik M},
  journal={Geology Today},
  volume={29},
  number={6},
  pages={218--221},
  year={2013},
  publisher={Wiley Online Library}
}

@article{williams2005subglacial,
  title={Subglacial meltwater channels and glaciofluvial deposits in the Kimberley Basin, Western Australia: 1.8 Ga low-latitude glaciation coeval with continental assembly},
  author={Williams, George E},
  journal={Journal of the Geological Society},
  volume={162},
  number={1},
  pages={111--124},
  year={2005},
  publisher={Geological Society of London}
}

@article{ojakangas2014talya,
  title={The Talya conglomerate: An Archean (~ 2.7 Ga) glaciomarine formation, Western Dharwar Craton, Southern India},
  author={Ojakangas, Richard W and Srinivasan, R and Hegde, VS and Chandrakant, SM and Srikantia, SV},
  journal={Current Science},
  pages={387--396},
  year={2014},
  publisher={JSTOR}
}

@article{young1998earth,
  title={Earth's oldest reported glaciation: physical and chemical evidence from the Archean Mozaan Group (~ 2.9 Ga) of South Africa},
  author={Young, Grant M and Brunn, Victor Von and Gold, Digby JC and Minter, WEL},
  journal={The Journal of Geology},
  volume={106},
  number={5},
  pages={523--538},
  year={1998},
  publisher={The University of Chicago Press}
}

@article{de20163,
  title={3.5-Ga hydrothermal fields and diamictites in the Barberton Greenstone Belt—Paleoarchean crust in cold environments},
  author={De Wit, Maarten J and Furnes, Harald},
  journal={Science advances},
  volume={2},
  number={2},
  pages={e1500368},
  year={2016},
  publisher={American Association for the Advancement of Science}
}

@article{kanzaki2015estimates,
  title={Estimates of atmospheric CO2 in the Neoarchean--Paleoproterozoic from paleosols},
  author={Kanzaki, Yoshiki and Murakami, Takashi},
  journal={Geochimica et Cosmochimica Acta},
  volume={159},
  pages={190--219},
  year={2015},
  publisher={Elsevier}
}

@article{kah2007mesoproterozoic,
  title={Mesoproterozoic carbon dioxide levels inferred from calcified cyanobacteria},
  author={Kah, Linda C and Riding, Robert},
  journal={Geology},
  volume={35},
  number={9},
  pages={799--802},
  year={2007},
  publisher={Geological Society of America}
}

@article{driese2011neoarchean,
  title={Neoarchean paleoweathering of tonalite and metabasalt: Implications for reconstructions of 2.69 Ga early terrestrial ecosystems and paleoatmospheric chemistry},
  author={Driese, Steven G and Jirsa, Mark A and Ren, Minghua and Brantley, Susan L and Sheldon, Nathan D and Parker, Don and Schmitz, Mark},
  journal={Precambrian Research},
  volume={189},
  number={1-2},
  pages={1--17},
  year={2011},
  publisher={Elsevier}
}

@article{sheldon2006precambrian,
  title={Precambrian paleosols and atmospheric CO2 levels},
  author={Sheldon, Nathan D},
  journal={Precambrian Research},
  volume={147},
  number={1-2},
  pages={148--155},
  year={2006},
  publisher={Elsevier}
}

@article{dole1964planets,
  title={Planets for man},
  author={Dole, Stephen H and Asimov, Isaac},
  journal={New York},
  year={1964}
}

@article{hamano2013emergence,
  title={Emergence of two types of terrestrial planet on solidification of magma ocean},
  author={Hamano, Keiko and Abe, Yutaka and Genda, Hidenori},
  journal={Nature},
  volume={497},
  number={7451},
  pages={607--610},
  year={2013},
  publisher={Nature Publishing Group UK London}
}

@article{faure2020uranium,
  title={Uranium and thorium partitioning in the bulk silicate Earth and the oxygen content of Earth’s core},
  author={Faure, P and Bouhifd, Mohamed Ali and Boyet, Maud and Manthilake, Geeth and Clesi, Vincent and Devidal, J-L},
  journal={Geochimica et Cosmochimica Acta},
  volume={275},
  pages={83--98},
  year={2020},
  publisher={Elsevier}
}

@article{lacis2010atmospheric,
  title={Atmospheric CO2: Principal control knob governing Earth’s temperature},
  author={Lacis, Andrew A and Schmidt, Gavin A and Rind, David and Ruedy, Reto A},
  journal={Science},
  volume={330},
  number={6002},
  pages={356--359},
  year={2010},
  publisher={American Association for the Advancement of Science}
}

@article{trenberth1998atmospheric,
  title={Atmospheric moisture residence times and cycling: Implications for rainfall rates and climate change},
  author={Trenberth, Kevin E},
  journal={Climatic change},
  volume={39},
  number={4},
  pages={667--694},
  year={1998},
  publisher={Springer}
}

@article{keller2019continuum,
  title={A continuum model of multi-phase reactive transport in igneous systems},
  author={Keller, Tobias and Suckale, Jenny},
  journal={Geophysical Journal International},
  volume={219},
  number={1},
  pages={185--222},
  year={2019},
  publisher={Oxford University Press}
}

@article{hirth2003rheology,
  title={Rheology of the upper mantle and the mantle wedge: A view from the experimentalists},
  author={Hirth, Greg and Kohlstedt, David},
  journal={Geophysical monograph series},
  volume={138},
  pages={83--105},
  year={2003}
}

@article{takei2009viscous,
  title={Viscous constitutive relations of solid-liquid composites in terms of grain boundary contiguity: 1. Grain boundary diffusion control model},
  author={Takei, Yasuko and Holtzman, Benjamin K},
  journal={Journal of Geophysical Research: Solid Earth},
  volume={114},
  number={B6},
  year={2009},
  publisher={Wiley Online Library}
}

@article{rudge2018viscosities,
  title={The viscosities of partially molten materials undergoing diffusion creep},
  author={Rudge, John F},
  journal={Journal of Geophysical Research: Solid Earth},
  volume={123},
  number={12},
  pages={10--534},
  year={2018},
  publisher={Wiley Online Library}
}

@article{hebrard2014coupled,
  title={Coupled noble gas--hydrocarbon evolution of the early Earth atmosphere upon solar UV irradiation},
  author={H{\'e}brard, Eric and Marty, B},
  journal={Earth and Planetary Science Letters},
  volume={385},
  pages={40--48},
  year={2014},
  publisher={Elsevier}
}

@article{avice2018evolution,
  title={Evolution of atmospheric xenon and other noble gases inferred from Archean to Paleoproterozoic rocks},
  author={Avice, Guillaume and Marty, Bernard and Burgess, Raymond and Hofmann, A and Philippot, Pascal and Zahnle, Kevin and Zakharov, David},
  journal={Geochimica et Cosmochimica Acta},
  volume={232},
  pages={82--100},
  year={2018},
  publisher={Elsevier}
}

@article{parai2018xenon,
  title={Xenon isotopic constraints on the history of volatile recycling into the mantle},
  author={Parai, Rita and Mukhopadhyay, Sujoy},
  journal={Nature},
  volume={560},
  number={7717},
  pages={223--227},
  year={2018},
  publisher={Nature Publishing Group UK London}
}

@article{lau2016inferences,
  title={Inferences of mantle viscosity based on ice age data sets: Radial structure},
  author={Lau, Harriet CP and Mitrovica, Jerry X and Austermann, Jacqueline and Crawford, Ophelia and Al-Attar, David and Latychev, Konstantin},
  journal={Journal of Geophysical Research: Solid Earth},
  volume={121},
  number={10},
  pages={6991--7012},
  year={2016},
  publisher={Wiley Online Library}
}

@article{mitrovica2004new,
  title={A new inference of mantle viscosity based upon joint inversion of convection and glacial isostatic adjustment data},
  author={Mitrovica, JX and Forte, AM},
  journal={Earth and Planetary Science Letters},
  volume={225},
  number={1-2},
  pages={177--189},
  year={2004},
  publisher={Elsevier}
}

@article{steinberger2006models,
  title={Models of large-scale viscous flow in the Earth's mantle with constraints from mineral physics and surface observations},
  author={Steinberger, Bernhard and Calderwood, Arthur R},
  journal={Geophysical Journal International},
  volume={167},
  number={3},
  pages={1461--1481},
  year={2006},
  publisher={Blackwell Publishing Ltd Oxford, UK}
}

@article{patovcka2020minimum,
  title={Minimum heat flow from the core and thermal evolution of the Earth},
  author={Pato{\v{c}}ka, V and {\v{S}}r{\'a}mek, O and Tosi, Nicola},
  journal={Physics of the Earth and Planetary Interiors},
  volume={305},
  pages={106457},
  year={2020},
  publisher={Elsevier}
}
\bibliographystyle{aasjournal}

\end{document}